\documentclass{amsart}

\usepackage{amssymb,amscd}

\newtheorem{theorem}{Theorem}
\newtheorem{lemma}{Lemma}
\newtheorem{proposition}{Proposition}

\theoremstyle{definition}
\newtheorem{example}{Example}

\theoremstyle{definition}
\newtheorem{remark}{Remark}

\numberwithin{equation}{section}

\newfont{\germ}{eufm10}

\newcommand\Bc{B^\vee}
\newcommand\et[1]{\tilde{e}_{#1}}
\newcommand\ft[1]{\tilde{f}_{#1}}
\newcommand\aff[1]{\hbox{Aff}(#1)}

\newcommand\ot{\otimes}
\newcommand\ol{\overline}
\newcommand\cd{\cdots}

\newcommand\veps{\varepsilon}
\newcommand\vphi{\varphi}

\newcommand\Z{\mathbb{Z}}

\newcommand\vac{\mbox{vac}}
\newcommand\s{\!\!&\!\!}
\newcommand\sa{\!\!\!&\!\!\!}
\newcommand\pa{{\mathcal P}_{\rm asy}}

\begin{document}

\title{Box-Ball System with Reflecting End}

\author{A. Kuniba}
\address{Institute of Physics, University of Tokyo, Tokyo 153-8902, Japan}
\email{atsuo@gokutan.c.u-tokyo.ac.jp}

\author{M. Okado}
\address{Department of Mathematical Science,
    Graduate School of Engineering Science,
Osaka University, Osaka 560-8531, Japan}
\email{okado@sigmath.es.osaka-u.ac.jp}

\author{Y. Yamada}
\address{Department of Mathematics, Faculty of Science,
Kobe University, Hyogo 657-8501, Japan}
\email{yamaday@math.kobe-u.ac.jp}

\begin{abstract}
A soliton cellular automaton on a one dimensional semi-infinite lattice 
with a reflecting end is presented.
It extends a box-ball system on an infinite lattice 
associated with the crystal base of $U_q(\widehat{\mathfrak{sl}}_n)$. 
A commuting family of time evolutions are 
obtained by adapting the $K$ matrices and the 
double row construction of transfer matrices 
in solvable lattice models to a crystal setting.
Factorized scattering among the left and the right moving solitons 
and the boundary reflection rule are determined.
\end{abstract}

\maketitle

\section{Introduction}\label{sec:intro}

The box-ball system \cite{TS,T} 
is a soliton cellular automaton on a one dimensional lattice.
It has been studied and generalized {}from 
a variety of aspects; 
ultradiscretization of soliton equations \cite{TTMS},  
solvable vertex models \cite{B} at $q=0$, 
connection to crystal base theory \cite{HHIKTT,FOY,HKT1},
description as particle and anti-particle systems \cite{HKT2,KTT},
factorized scattering \cite{HKOTY,TNS},
Robinson-Schensted-Knuth correspondence \cite{F}, 
geometric crystal and tropical $R$ \cite{KOTY2,KOTY3},
inverse scattering method \cite{KOTY1,Tg},
quantization \cite{HIK,IKO} and so forth.
The automaton is defined on an infinite lattice 
and these studies are concerned with the behaviour without 
a boundary effect.
Solitons undergo factorized scattering during their travel 
{}from the left to the right asymptotic regions where 
they acquire individualities being well separated {}from each other. 

In this paper we formulate a soliton cellular 
automaton on a semi-infinite lattice having a reflecting end.
A new feature here is the reflection at the boundary, which 
essentially doubles the soliton degrees of freedom 
into the right and the left moving ones.
When there are only right moving solitons, 
their behaviour far away from the boundary
is the same as the original box-ball system \cite{T} with 
$n\!-\!2$ kinds of balls.
(The case $n=2$ is trivial and we assume $n \ge 3$ throughout.)

Let $B_l$ and $B^\vee_l$ 
be the crystals of the $l$-fold symmetric tensor 
of the first and the $(n\!-\!1)$ th 
fundamental representations of $U_q(\widehat{\mathfrak{sl}}_n)$, 
respectively.
$B_l$ and $B^\vee_l$ are dual in the sense that 
their crystal graphs are obtained by reversing 
the arrows altogether.
In the crystal formulation, 
the original box-ball system with $n\!-\!1$ kinds of balls 
is a dynamical system on an infinite lattice whose 
state is an element of 
$\cdots \ot B_1 \ot B_1 \ot \cdots$ obeying a certain boundary
condition in the distance.
Our first finding is that the left moving solitons can 
properly be incorporated by replacing this with  
$\cdots \ot (B_1 \ot B^\vee_1)\ot (B_1 \ot B^\vee_1)$.
Then the double row construction of transfer matrices 
in solvable lattice models \cite{S,C,Ku} is adapted 
to the crystal setting  
to generate a commuting family of time evolutions.
The basic ingredients in the construction are the 
three kinds of combinatorial $R$; 
$R: \aff{B_l} \ot \aff{B_k} \simeq \aff{B_k} \ot \aff{B_l}$, 
$R^\vee: \aff{B_l} \ot \aff{B^\vee_k} \simeq 
\aff{B^\vee_k} \ot \aff{B_l}$ and 
$R^{\vee\vee}: \aff{B^\vee_l} \ot \aff{B^\vee_k} 
\simeq \aff{B^\vee_k} \ot \aff{B^\vee_l}$, 
where ${\rm Aff}$ denotes the affinization 
(Section \ref{subsec:B}).
In addition we introduce the {\em combinatorial $K$} 
as the map 
$K: \aff{B_l} \rightarrow \aff{B^\vee_l}$ that 
satisfies the reflection equation (\ref{eq:remigi}).
A computer experiment suggests that those $K$ expressible  
as a permutation of certain coordinates is unique for $n$ odd, while 
three variations are allowed for $n$ even.
See (\ref{eq:rl})--(\ref{eq:12}) and Remark \ref{rem:K}.
It would be an interesting problem to find a quantum $K$ matrix $K$
satisfying
the reflection equation (\ref{eq:remigi}) with the quantum $R$ matrices
$R,R^\vee, R^{\vee\vee}$ that agrees with 
our combinatorial $K$ in $q\rightarrow0$
limit.

By using $R, R^\vee, R^{\vee\vee}$ and $K$, the 
commuting family of time evolutions $\{T_l\}$ as well as 
conserved quantities $\{N_l\}$ counting solitons are 
constructed.

We identify special patterns that behave as solitons and 
investigate their reflection at 
the boundary and the scattering involving the left and the right moving ones.
Solitons bear internal degrees of freedom as well as
phases under collisions and reflection.
They are most naturally described in terms of the scattering data,
an element of a certain affine crystal.
We find that the scattering data 
for a right (left) moving soliton of length $l$ is 
an element in $\aff{B_l}$ ($\aff{B^\vee_l}$) of the algebra
$U_q(\widehat{\mathfrak{sl}}_{n-2})$.
This contrasts with the infinite system \cite{HKOTY},  
where the analogous data is that for the algebra 
$U_q(\widehat{\mathfrak{sl}}_{n-1})$.
Accordingly, the reflection and the scattering rules 
in our automaton are identified essentially 
with the combinatorial $K$ and 
the combinatorial $R$ of $U_q(\widehat{\mathfrak{sl}}_{n-2})$, 
respectively.

Here is a summary of some characteristic features of the 
automata associated with $U_q(\widehat{\mathfrak{sl}}_{n})$ 
on the infinite and the semi-infinite lattices.

\begin{center}
\begin{tabular}{c|c|c}
& $\underset{\phantom 1}{\phantom 2}$ infinite system & 
semi-infinite system\\
\hline 
local state & $\overset{\phantom{1}}{B_1}$ & $B_1 \ot B^\vee_1$ \\
symmetry & $\overset{\phantom{1}}{\mathfrak{sl}_{n-1}}$-invariance & 
asymptotic $\mathfrak{sl}_{n-2}$-invariance \\
soliton & $\overset{\phantom 1}{\aff{B_l}}$ 
of $U_q(\widehat{\mathfrak{sl}}_{n-1})$ &
$\aff{B_l}$, $\aff{B^\vee_l}$ of $U_q(\widehat{\mathfrak{sl}}_{n-2})$ \\
reflection rule & & $\overset{\phantom 1}{K}$ 
of $U_q(\widehat{\mathfrak{sl}}_{n-2})$ \\
two body scattering rule & $\overset{\phantom 1}{R}$ 
of $U_q(\widehat{\mathfrak{sl}}_{n-1})$ &
$R, R^\vee, R^{\vee\vee}$ of $U_q(\widehat{\mathfrak{sl}}_{n-2})$ 
\end{tabular}
\end{center}

\vspace{0.1cm}\noindent
For symmetry, see section 3.1 in \cite{HKOTY} 
and Section \ref{subsec:asy} in this paper.
The reflection and the scattering rules in the semi-infinite 
system are obtained (and more precisely stated) 
in Theorem \ref{th:L} and Theorem \ref{th:S}, respectively.

The paper is arranged as follows.
In Section \ref{sec:ybe}, we recall basic facts on crystals, 
combinatorial $R$ and the Yang-Baxter equation.
A combinatorial version of the reflection equation is 
formulated and the combinatorial $K$ is given.
In Section \ref{sec:automaton}, we present the automaton 
with a reflecting end. 
Conserved quantities $\{E_l\}$ are constructed and the 
asymptotic $\mathfrak{sl}_{n-2}$-invariance is explained.
In Section \ref{sec:soliton}, we identify solitons and determine
their reflection and scattering rules.
We omit most of the proofs but provide several examples.
In Section \ref{sec:finite}, we formulate an automaton 
on a finite lattice surrounded by two reflecting ends and 
having a commuting family of time evolutions.
We illustrate reflecting solitons with a few examples 
leaving a thorough study as a future problem.
Appendix \ref{app:re} contains a proof of the 
reflection equation in a tropical setting.

\section{Yang-Baxter and reflection equations}\label{sec:ybe}

\subsection{Crystals $B_l, \Bc_l$}\label{subsec:B}

Here we fix notations 
concerning the $U_q(\widehat{\mathfrak{sl}}_n)$ crystals used in this paper.
We omit standard facts on crystals such as weight decomposition 
and tensor product rule, etc.,
for which we refer to \cite{K1,K2,KKM,KMN}.

Let $B_1$ and $B^\vee_1$ be the crystals of 
the vector and the $n\!-\!1$ fold antisymmetric tensor representations 
of $U_q(\widehat{\mathfrak{sl}}_n)$, and let $B_l$ and $B^\vee_l$ 
be the ones corresponding to their 
$l$ fold symmetric tensor product representations.
As a set $B_l$ and $B^\vee_l$ are both presented as
\begin{equation*}
\{x=(x_1,\ldots, x_n) \in (\Z_{\ge 0})^n
\mid x_1 + \cdots + x_n = l \}.
\end{equation*}
This parameterization originates in the basis 
labelled with the semi-standard tableaux.
For $B_l$, $x_i$ corresponds to the number of the letter $i$ 
in the length $l$ row shape tableaux.
For $B^\vee_l$, $x_i$ corresponds to the number of 
columns without the letter $i$ in the $n\!-\!1$ by $l$ rectangular
shape tableaux.
A given array $x=(x_1,\ldots, x_n)$ can specify two crystal elements, 
one in $B_l$ and the other one in $B^\vee_l$.
When necessary we distinguish them by writing, for instance, 
$(2,0,1) \in B_3$ as $113$ and $(0,3,1,2) \in B^\vee_6$ as
$\ol{2}\ol{2}\ol{2}\ol{3}\ol{4}\ol{4}$, etc.

Action of Kashiwara operators
$\ft{i}, \et{i} (0 \le i \le n\!-\!1): B_l \rightarrow B_l \sqcup \{0\}$ or 
$B^\vee_l \rightarrow B^\vee_l \sqcup \{0\}$ is defined by
\begin{align*}
&(\ft{i} x)_j = x_j - \delta_{j,i} + \delta_{j,i+1}, \quad 
(\et{i} x)_j = x_j + \delta_{j,i} - \delta_{j,i+1}
\;\;\; \mbox{for } x \in B_l,\\
&(\ft{i} x)_j = x_j + \delta_{j,i} - \delta_{j,i+1}, \quad 
(\et{i} x)_j = x_j - \delta_{j,i} + \delta_{j,i+1}
\;\;\; \mbox{for } x \in B^\vee_l.
\end{align*}
Here and in the remainder of the paper 
all the indices are to be understood in $\Z_n$.
The right hand side is to be interpreted as 0 
if it does not belong to $B_l$ or $B^\vee_l$.
We see that the crystal graphs of $B_l$ and $B^\vee_l$ are 
identical by reversing all the arrows.
$\varphi_i(x) = \max\{k\ge 0 \mid \ft{i}^k x \neq 0\}$ and 
$\varepsilon_i(x) = \max\{k\ge 0 \mid \et{i}^k x \neq 0\}$ are given by
\begin{align*}
\varphi(x) = x_i,\quad \varepsilon_i(x) = x_{i+1} 
\;\;\; \mbox{for } x \in B_l,\\
\varphi(x) = x_{i+1},\quad \varepsilon_i(x) = x_{i} 
\;\;\; \mbox{for } x \in B^\vee_l.
\end{align*}

For crystals $B_l$ and $B^\vee_l$, we define their affinization 
$\aff{B_l} = \{ z^d x \mid x \in B_l, d \in \Z\}$ and 
$\aff{B^\vee_l} = \{ z^d x \mid x \in B^\vee_l, d \in \Z\}$.
$z$ is called the {\em spectral parameter}. 
They also admit the crystal
structure by $\et{i}\cdot z^d x=z^{d+\delta_{i0}}(\et{i}x)$ and 
$\ft{i}\cdot z^d x=z^{d-\delta_{i0}}(\ft{i}x)$.

\subsection{Combinatorial $R$ and Yang-Baxter equation}
\label{subsec:R}

The isomorphism of crystals
$\aff{B}\ot\aff{B'}\stackrel{\sim}{\rightarrow}\aff{B'}\ot\aff{B}$
is called the combinatorial $R$.
It has the following form:
\begin{eqnarray*}
R\;:\;\aff{B}\ot\aff{B'}&\longrightarrow&\aff{B'}\ot\aff{B}\\
z^d x\ot z^{e} y&\longmapsto&
z^{e+H(x\ot y)}{\tilde y}\ot z^{d-H(x\ot y)}{\tilde x},
\end{eqnarray*}
where $x\ot y\mapsto {\tilde y}\ot{\tilde x}$ 
under the isomorphism (classical combinatorial $R$) $B\ot B'
\stackrel{\sim}{\rightarrow}B'\ot B$. 
$H(x\ot y)$ is called the 
energy function and determined up to a global additive constant by
\[
H(\et{i}(x\ot y))=\left\{%
\begin{array}{ll}
H(x\ot y)+1&\mbox{ if }i=0,\ \vphi_0(x)\geq\veps_0(y),\ 
\vphi_0({\tilde y})\geq\veps_0({\tilde x}),\\
H(x\ot y)-1&\mbox{ if }i=0,\ \vphi_0(x)<\veps_0(y),\ 
\vphi_0({\tilde y})<\veps_0({\tilde x}),\\
H(x\ot y)&\mbox{ otherwise}.
\end{array}\right.
\]
In this paper we are concerned with the 
following combinatorial $R$:
\begin{align}
R: \;\;& \aff{B_l} \ot \aff{B_m}  \; \longrightarrow 
\qquad  \aff{B_m} \ot \aff{B_l} \label{eq:R}\\
& \quad \;\; z^{d}x \ot z^{e}y  \;\qquad  \; \longmapsto 
\;\; z^{e-Q_0(x,y)}{\tilde y} \ot z^{d+Q_0(x,y)}{\tilde x}, 
\nonumber \\
&{\tilde x}_i = x_i+Q_i(x,y)-Q_{i-1}(x,y),\quad 
{\tilde y}_i = y_i+Q_{i-1}(x,y)-Q_i(x,y), \nonumber \\
&Q_i(x,y) = \min \{ \sum_{j=1}^{k-1}x_{i+j} + \sum_{j=k+1}^n y_{i+j} 
\mid 1 \le k \le n \}. \nonumber 
\end{align}
\begin{align}
R^\vee: \;\;& \aff{B_l} \ot \aff{B^\vee_m}  \; \longrightarrow 
\qquad  \; \aff{B^\vee_m} \ot \aff{B_l} \label{eq:Rv}\\
& \quad \;\; z^{d}x \ot z^{e}y  \;\qquad  \; \longmapsto 
\;\; z^{e-P_0(x,y)}{\tilde y} \ot z^{d+P_0(x,y)}{\tilde x}, 
\nonumber \\
&{\tilde x}_i = x_i+P_i(x,y)-P_{i-1}(x,y),\quad 
{\tilde y}_i = y_i+P_i(x,y)-P_{i-1}(x,y), \nonumber \\
&P_i(x,y) = P_i(y,x) = \min(x_{i+1}, y_{i+1}). \nonumber 
\end{align}
\begin{align}
{}^\vee\!R: \;\;& \aff{B^\vee_l} \ot \aff{B_m}  \; \longrightarrow 
\qquad  \; \aff{B_m} \ot \aff{B^\vee_l} \label{eq:vR}\\
& \quad \;\; z^{d}x \ot z^{e}y  \;\qquad  \; \longmapsto 
\;\; z^{e-P_{-1}(x,y)}{\tilde y} \ot z^{d+P_{-1}(x,y)}{\tilde x}, 
\nonumber \\
&{\tilde x}_i = x_i+P_{i-2}(x,y)-P_{i-1}(x,y),\quad 
{\tilde y}_i = y_i+P_{i-2}(x,y)-P_{i-1}(x,y). \nonumber
\end{align}
\begin{align}
R^{\vee\vee}: \;\;& \aff{B^\vee_l} \ot \aff{B^\vee_m}  
\; \;\;\longrightarrow 
\qquad \;\;  \aff{B^\vee_m} \ot \aff{B^\vee_l} \quad 
\label{eq:Rvv}\\
& \quad \quad  z^{d}x \ot z^{e}y  \;\qquad  \;\;\longmapsto 
\;\;   z^{e-Q_0(y,x)}{\tilde y} 
\ot z^{d+Q_0(y,x)}{\tilde x}, \nonumber \\
&{\tilde x}_i = x_i+Q_{i-1}(y,x)-Q_i(y,x),\quad 
{\tilde y}_i = y_i+Q_i(y,x)-Q_{i-1}(y,x). \nonumber 
\end{align}
Except (\ref{eq:inversion}) below, 
${}^\vee\!R$ will not be used 
in the rest of the paper.
We have normalized the energy as 
\begin{equation}\label{eq:R-energy}
H(x\ot y) = \begin{cases}
-Q_0(x,y) & \mbox{ for } R,\\
-P_0(x,y) & \mbox{ for } R^\vee,\\
-Q_0(y,x) & \mbox{ for } R^{\vee\vee}
\end{cases}
\end{equation}
so as to simplify the definition of the conserved quantity 
$E_l$ in Section \ref{subsec:El}.

These combinatorial $R$ commute with the Kashiwara operators 
$\et{i}$ and $\ft{i}$.
They satisfy the inversion relations:
\begin{equation}\label{eq:inversion}
\begin{split}
R R &= {\rm id},\\
R^\vee ({}^\vee\!R) = {}^\vee\!RR^\vee &= {\rm id},\\
R^{\vee\vee} R^{\vee\vee} &= {\rm id},
\end{split}
\end{equation}
where we have suppressed the $l, m$-dependence.
For instance, the first line actually means that
$R_{B_m, B_l} R_{B_l, B_m}$ is the identity map on
$\aff{B_l} \ot \aff{B_m}$.
They also satisfy the Yang-Baxter equations:
\begin{equation}\label{eq:ybe}
\begin{split}
(1 \ot R)(R \ot 1)(1 \ot R) 
&= (R \ot 1)(1 \ot R)(R \ot 1),\\
(1 \ot R)(R^\vee \ot 1)(1 \ot R^\vee) 
&= (R^{\vee} \ot 1)(1 \ot R^\vee)(R \ot 1),\\
(1 \ot R^\vee)(R^\vee \ot 1)(1 \ot R^{\vee\vee}) 
&= (R^{\vee\vee} \ot 1)(1 \ot R^\vee)(R^\vee \ot 1),\\
(1 \ot R^{\vee\vee})(R^{\vee\vee} \ot 1)(1 \ot R^{\vee\vee}) 
&= (R^{\vee\vee} \ot 1)(1 \ot R^{\vee\vee})(R^{\vee\vee} \ot 1),
\end{split}
\end{equation}
which correspond to the isomorphisms that reverse the order of 
tensor products 
$\aff{B_k}\ot\aff{B_l}\ot\aff{B_m}, \,
\aff{B_k}\ot\aff{B_l}\ot\aff{B^\vee_m}, \,
\aff{B_k}\ot\aff{B^\vee_l}\ot\aff{B^\vee_m}$ and 
$\aff{B^\vee_k}\ot\aff{B^\vee_l}\ot\aff{B^\vee_m}$.

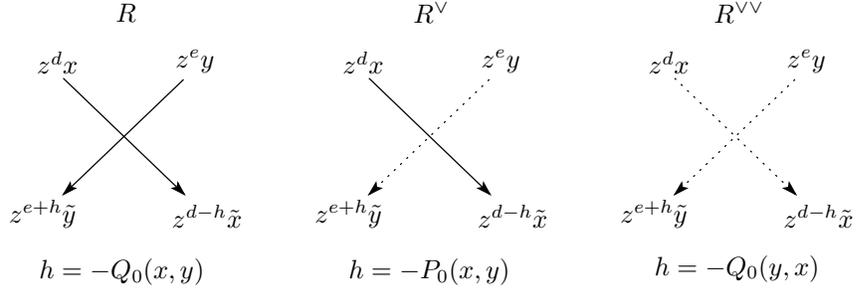
\begin{figure}[h]
\caption{Diagrams for $R, R^\vee$ and $R^{\vee\vee}$}\label{fig:Rvee}
\vspace{0.3cm}

\unitlength 0.1in
\begin{picture}( 41.1200, 13.6000)( 11.2500,-20.7500)
%
\special{pn 8}%
\special{pa 1406 1146}%
\special{pa 2038 1752}%
\special{fp}%
\special{sh 1}%
\special{pa 2038 1752}%
\special{pa 2004 1692}%
\special{pa 2000 1716}%
\special{pa 1976 1720}%
\special{pa 2038 1752}%
\special{fp}%
\put(12.9500,-18.5500){\makebox(0,0){$z^{e+h}{\tilde y}$}}%
\put(21.5500,-18.6500){\makebox(0,0){$z^{d-h}{\tilde x}$}}%
%
\special{pn 8}%
\special{pa 2032 1152}%
\special{pa 1406 1752}%
\special{fp}%
\special{sh 1}%
\special{pa 1406 1752}%
\special{pa 1468 1720}%
\special{pa 1444 1716}%
\special{pa 1440 1692}%
\special{pa 1406 1752}%
\special{fp}%
\put(20.9300,-10.5500){\makebox(0,0){$z^e y$}}%
\put(13.7300,-10.6200){\makebox(0,0){$z^d x$}}%
%
\special{pn 8}%
\special{pa 3006 1146}%
\special{pa 3638 1752}%
\special{fp}%
\special{sh 1}%
\special{pa 3638 1752}%
\special{pa 3604 1692}%
\special{pa 3600 1716}%
\special{pa 3576 1720}%
\special{pa 3638 1752}%
\special{fp}%
\put(29.7300,-10.6200){\makebox(0,0){$z^d x$}}%
\put(36.9300,-10.5500){\makebox(0,0){$z^e y$}}%
%
\special{pn 8}%
\special{pa 3632 1152}%
\special{pa 3006 1752}%
\special{dt 0.045}%
\special{sh 1}%
\special{pa 3006 1752}%
\special{pa 3068 1720}%
\special{pa 3044 1716}%
\special{pa 3040 1692}%
\special{pa 3006 1752}%
\special{fp}%
\put(37.5500,-18.6500){\makebox(0,0){$z^{d-h}{\tilde x}$}}%
\put(28.9500,-18.5500){\makebox(0,0){$z^{e+h}{\tilde y}$}}%
%
\special{pn 8}%
\special{pa 4606 1146}%
\special{pa 5238 1752}%
\special{dt 0.045}%
\special{sh 1}%
\special{pa 5238 1752}%
\special{pa 5204 1692}%
\special{pa 5200 1716}%
\special{pa 5176 1720}%
\special{pa 5238 1752}%
\special{fp}%
\put(44.9500,-18.5500){\makebox(0,0){$z^{e+h}{\tilde y}$}}%
\put(53.5500,-18.6500){\makebox(0,0){$z^{d-h}{\tilde x}$}}%
%
\special{pn 8}%
\special{pa 5232 1152}%
\special{pa 4606 1752}%
\special{dt 0.045}%
\special{sh 1}%
\special{pa 4606 1752}%
\special{pa 4668 1720}%
\special{pa 4644 1716}%
\special{pa 4640 1692}%
\special{pa 4606 1752}%
\special{fp}%
\put(52.9300,-10.5500){\makebox(0,0){$z^e y$}}%
\put(45.7300,-10.6200){\makebox(0,0){$z^d x$}}%
\put(17.3000,-8.0000){\makebox(0,0){$R$}}%
\put(33.3000,-8.0000){\makebox(0,0){$R^\vee$}}%
\put(49.4000,-8.0000){\makebox(0,0){$R^{\vee\vee}$}}%
\put(17.1000,-21.6000){\makebox(0,0){$h=-Q_0(x,y)$}}%
\put(33.2000,-21.6000){\makebox(0,0){$h=-P_0(x,y)$}}%
\put(49.3000,-21.4000){\makebox(0,0){$h=-Q_0(y,x)$}}%
\end{picture}%

\vspace{0.3cm}
\end{figure}

We attach the elements in $B_m$ and $B^\vee_m$ with solid and 
dotted lines, respectively.
The combinatorial $R$ are depicted as in Figure \ref{fig:Rvee}.

For convenience the classical combinatorial $R$
(combinatorial $R$ up to $z^d$ part) will be denoted by 
$\ol{R}: B_l\ot B_m \rightarrow B_m\ot B_l$, 
$\ol{R}^\vee: B_l\ot B^\vee_m \rightarrow B^\vee_m\ot B_l$ and 
$\ol{R}^{\vee\vee}: B^\vee_l\ot B^\vee_m 
\rightarrow B^\vee_m\ot B^\vee_l$.
The relations like 
$R(z^dx\ot z^ey) = z^{\tilde{e}}\tilde{y}\ot z^{\tilde{d}}\tilde{x}$
and $\ol{R}(x\ot y) = \tilde{y}\ot \tilde{x}$ will also be 
denoted by
$z^dx\ot z^ey \simeq z^{\tilde{e}}\tilde{y}\ot z^{\tilde{d}}\tilde{x}$
and $x\ot y \simeq \tilde{y}\ot \tilde{x}$.

There is a simple rule to compute $R$ graphically \cite{NY}.
In particular, $Q_0$ is known as the `unwinding number'.
Although $R$ and $R^{\vee\vee}$ act on different kind of 
crystals, their formulas are intertwined by the 
transposition of the components of the tensor product.
Thus $R^{\vee\vee}$ is also calculated graphically.
Here we illustrate a graphical procedure to compute $R^\vee$ 
along the example 
$R^\vee: z^d(3,1,1,1)\ot z^e(2,2,0,1) \mapsto 
z^{e-2}(1,1,1,2)\ot z^{d+2}(2,0,2,2)$, 
which reads 
$z^d111234 \ot z^e\ol{11224} \simeq 
z^{e-2}\ol{12344}\ot z^{d+2}113344$ 
in the tableau notation.

\vspace{0.2cm}\noindent
(i) Draw pictures corresponding to 
$(3,1,1,1) \ot (2,2,0,1)$.

\noindent
(ii) Connect dots horizontally to make as many pairs as possible.

\noindent
(iii) Shift the pairs upward by one cyclically 
leaving the unpaired dots. 

\noindent
(iv) Exchange the left and the right components.

\vspace{0.2cm}\noindent
The result yields the image $(1,1,1,2) \ot (2,0,2,2)$.
The minus energy $P_0 = 2$ is the number of 
pairs sent from the top to the bottom in step (iii).

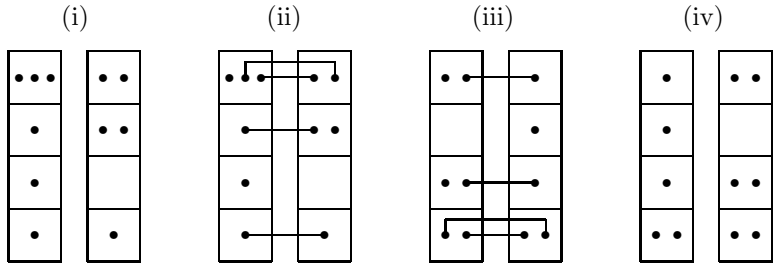
\setlength{\unitlength}{0.7mm}
\begin{picture}(150,58)(-10,-4)

\put(10,45){(i)}
\put(49,45){(ii)}
\put(88,45){(iii)}
\put(128,45){(iv)}

\multiput(0,0)(40,0){4}{
\multiput(0,0)(15,0){2}{
\multiput(0,0)(0,10){5}{
\put(0,0){\line(1,0){10}}}
\put(0,0){\line(0,1){40}}
\put(10,0){\line(0,1){40}}}}

\multiput(0,0)(40,0){2}{
\put(2,35){\circle*{1.3}}
\put(5,35){\circle*{1.3}}
\put(8,35){\circle*{1.3}}
\put(5,25){\circle*{1.3}}
\put(5,15){\circle*{1.3}}
\put(5,5){\circle*{1.3}}
\put(18,35){\circle*{1.3}}
\put(22,35){\circle*{1.3}}
\put(18,25){\circle*{1.3}}
\put(22,25){\circle*{1.3}}
\put(20,5){\circle*{1.3}}}

\put(45,35){\line(0,1){3}}
\put(45,38){\line(1,0){17}}
\put(62,35){\line(0,1){3}}
\put(48,35){\line(1,0){10}}
\put(45,25){\line(1,0){13}}
\put(45,5){\line(1,0){15}}

\put(83,35){\circle*{1.3}}\put(87,35){\circle*{1.3}}
\put(83,15){\circle*{1.3}}\put(87,15){\circle*{1.3}}
\put(83,5){\circle*{1.3}}\put(87,5){\circle*{1.3}}
\put(100,35){\circle*{1.3}}
\put(100,25){\circle*{1.3}}
\put(100,15){\circle*{1.3}}
\put(98,5){\circle*{1.3}}\put(102,5){\circle*{1.3}}

\put(83,5){\line(0,1){3}}
\put(83,8){\line(1,0){19}}
\put(102,5){\line(0,1){3}}
\put(87,5){\line(1,0){11}}
\put(87,15){\line(1,0){13}}
\put(87,35){\line(1,0){13}}

\put(125,35){\circle*{1.3}}
\put(125,25){\circle*{1.3}}
\put(125,15){\circle*{1.3}}
\put(123,5){\circle*{1.3}}
\put(127,5){\circle*{1.3}}
\put(138,35){\circle*{1.3}}
\put(142,35){\circle*{1.3}}
\put(138,15){\circle*{1.3}}
\put(142,15){\circle*{1.3}}
\put(138,5){\circle*{1.3}}
\put(142,5){\circle*{1.3}}
\end{picture}

\begin{example}\label{ex:R}
For $\widehat{\mathfrak{sl}}_3$, one has
\begin{alignat*}{2}
R(z^{d}123 \ot z^{e}12) &= z^{e+H}13 \ot z^{d-H}122, & &\quad H = -1,\\
R^\vee(z^{d}112 \ot z^{e}\,\ol{1123}) &= 
z^{e+H}\,\ol{1333} \ot z^{d-H}133, & &\quad H = -2,\\
R^{\vee\vee}(z^{d}\,\ol{12} \ot z^{e}\,\ol{2233}) &= 
z^{e+H}\,\ol{1222} \ot z^{d-H}\,\ol{33}, & &\quad H = 0.
\end{alignat*}
\end{example}

\subsection{Combinatorial $K$ and Reflection equation}

\begin{figure}[h]
\caption{Diagrams for $K$ and $K^\vee$}\label{fig:K}
\setlength{\unitlength}{1mm}
\begin{picture}(80,30)(0,-5)
\put(30,0){\line(0,1){20}}
\multiput(29.3,9.3)(-0.5,-0.5){14}{$\scriptstyle \cdot$}
\put(23,3){\vector(-1,-1){1}}
\put(22.9,17.1){\line(1,-1){7}}
\put(17,17){$\scriptstyle z^{d}x$}
\put(10,0){$\scriptstyle z^{-d+I(x)}\kappa(x)$}

\put(50,0){\line(0,1){20}}
\put(50,10){\vector(1,-1){7}}
\multiput(50,9.5)(0.5,0.5){14}{$\scriptstyle \cdot$}
\put(58,17){$\scriptstyle z^{d}x$}
\put(55,0){$\scriptstyle z^{-d-I(x)}\kappa(x)$}

\put(23,-7){$K$}\put(54,-7){$K^\vee$}
\end{picture}
\end{figure}
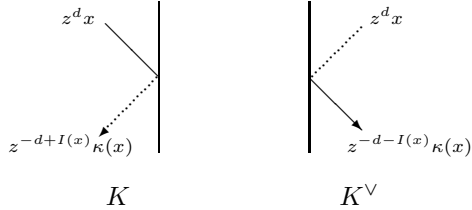

We introduce the maps
\begin{align}
K: \quad \aff{B_l} &\longrightarrow \aff{B^\vee_l} 
\label{eq:Kdef}\\
z^d x &\mapsto z^{-d+I(x)}\kappa(x), \nonumber\\
K^\vee: \quad \aff{B^\vee_l} &\longrightarrow \aff{B_l} 
\label{eq:Kvdef}\\
z^d x &\mapsto z^{-d-I(x)}\kappa(x).\nonumber
\end{align}
Here we have three choices for the pair $(\kappa, I)$.
For $x=(x_1,x_2, \ldots, x_n)$, they are given by
\begin{align}
&\begin{cases}
\kappa(x) = \hbox{Rotateleft}(x) = (x_2,\ldots, x_n,x_1)\\ 
I(x) = -x_1
\end{cases}
\;(\hbox{any } n), \label{eq:rl}\\
&\begin{cases}
\kappa(x) = \hbox{Switch}_{1n}(x) = 
(x_n,x_3,x_2,x_5,x_4,\ldots, x_{n-1},x_{n-2}, x_1)\\
I(x) = x_n-x_1
\end{cases}
\; (\hbox{even } n),\label{eq:1n}\\
&\begin{cases}
\kappa(x) = \hbox{Switch}_{12}(x) = 
(x_2,x_1,x_4,x_3,\ldots, x_{n-1},x_n)\\
I(x) = 0
\end{cases}
\; (\hbox{even } n).\label{eq:12}
\end{align}
Rotateleft is a cyclic permutation, whereas Switch$_{1n}$ and 
Switch$_{12}$ are transposition of adjacent coordinates.
We call $K$ and $K^\vee$ the combinatorial $K$.
To the relations (\ref{eq:Kdef}) and (\ref{eq:Kvdef})
we assign the diagrams in Figure \ref{fig:K}.

In Figure \ref{fig:K}, the vertical line in  
$K$ and $K^\vee$ stands for the right and left
reflecting end, respectively.
The diagonal line segments switch {}from the one
corresponding to $\aff{B_l}$ and $\aff{B^\vee_l}$
upon the contact with the reflecting ends.
Actually $K^\vee$ is irrelevant to our automaton until 
Section \ref{sec:finite}, where its classical part is 
denoted by $\kappa_{\rm left}$.

We let $K_2$ and $K^\vee_1$ denote the operators acting as
\begin{equation*}
K_2(z^dx \ot z^ey) = z^dx \ot K(z^ey),\quad 
K^\vee_1(z^dx \ot z^ey) = K^\vee(z^dx) \ot z^ey,
\end{equation*}
where the untouched tensor component can be either  
$B_m$ or $B^\vee_m$.

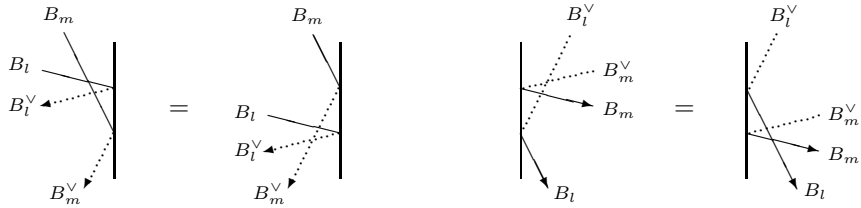
\begin{figure}[h]
\caption{Diagrams for (\ref{eq:remigi}) and (\ref{eq:rehidari})}
\label{fig:RE}
\setlength{\unitlength}{1.2mm}
\begin{picture}(110,23)(7,-1)

\put(25,0){\line(0,1){15}}
\put(25,5){\line(-1,2){5.6}}
\multiput(24.5,4.5)(-0.3,-0.6){10}{$\scriptstyle \cdot$}
\put(21.7,-1){\vector(-1,-2){0.1}}
\put(25,10){\line(-4,1){8}}
\multiput(24.5,9.5)(-0.64,-0.16){13}{$\scriptstyle \cdot$}
\put(16.8,8){\vector(-4,-1){0.1}} 
\put(17,17.5){$\scriptstyle B_m$}
\put(13.2,12){$\scriptstyle B_l$}
\put(13.2,7.5){$\scriptstyle B^\vee_l$}
\put(17.7,-2.3){$\scriptstyle B^\vee_m$}

\put(31,7){$=$}

\put(50,0){\line(0,1){15}}
\put(50,10){\line(-1,2){3}}
\multiput(49.5,9.5)(-0.3,-0.6){18}{$\scriptstyle \cdot$}
\put(44.3,-1){\vector(-1,-2){0.1}} 
\put(50,5){\line(-4,1){8}}
\multiput(49.5,4.5)(-0.64,-0.16){13}{$\scriptstyle \cdot$}
\put(41.5,3){\vector(-4,-1){0.1}} 
\put(44.5,17.4){$\scriptstyle B_m$}
\put(38.2,6.7){$\scriptstyle B_l$}
\put(38.2,2.5){$\scriptstyle B^\vee_l$}
\put(40.3,-2.3){$\scriptstyle B^\vee_m$}

\put(70,0){\line(0,1){15}}
\multiput(70,5)(0.3,0.6){18}{$\scriptstyle \cdot$}
\put(70,5){\vector(1,-2){3}}
\multiput(70,9.5)(0.64,0.16){13}{$\scriptstyle \cdot$}
\put(70,10){\vector(4,-1){8}}
\put(75,17.5){$\scriptstyle B^\vee_l$}
\put(79,11.3){$\scriptstyle B^\vee_m$}
\put(79,7){$\scriptstyle B_m$}
\put(73.6,-2.3){$\scriptstyle B_l$}

\put(87,7){$=$}

\put(95,0){\line(0,1){15}}
\multiput(95,9.5)(0.3,0.6){11}{$\scriptstyle \cdot$}
\put(95,10){\vector(1,-2){5.6}}
\multiput(95,4.5)(0.64,0.16){13}{$\scriptstyle \cdot$}
\put(95,5){\vector(4,-1){8}}
\put(97.5,17.4){$\scriptstyle B^\vee_l$}
\put(104,6.5){$\scriptstyle B^\vee_m$}
\put(104,2){$\scriptstyle B_m$}
\put(101.3,-2){$\scriptstyle B_l$}

\end{picture}
\end{figure}

\begin{proposition}[Combinatorial reflection equation]\label{pr:re}
\begin{align}
&K_2R^\vee K_2R = R^{\vee\vee}K_2R^\vee K_2, \label{eq:remigi}\\
&K^\vee_1R^\vee K^\vee_1R^{\vee \vee} 
= RK^\vee_1R^\vee K^\vee_1.\label{eq:rehidari}
\end{align}
\end{proposition}
This is proved in a tropical setting in Appendix \ref{app:re}.
(\ref{eq:remigi}) is an identity as the maps 
$\aff{B_l} \ot \aff{B_m} \rightarrow 
\aff{B^\vee_l} \ot \aff{B^\vee_m}$, 
and (\ref{eq:rehidari}) is the one for 
$\aff{B^\vee_l} \ot \aff{B^\vee_m} \rightarrow 
\aff{B_l} \ot \aff{B_m}$.
They are depicted in Figure \ref{fig:RE}. 
(`Aff' omitted)

\begin{remark}\label{rem:K}
By a computer experiment we have checked for small $n$ that 
solutions of the combinatorial reflection equation are 
exhausted by those in (\ref{eq:rl})--(\ref{eq:12}) provided that  
they are given as a permutation of coordinates $x_1, \ldots, x_n$.
\end{remark}

\section{Automaton on semi-infinite lattice}\label{sec:automaton}

\subsection{States}\label{subsec:states}

Consider the one dimensional semi-infinite lattice which 
extends towards left from an end.
To the end we assign the coordinate $1$ and successively 
$2, 3, \ldots,$ to the sites located in the left.
To each site
we assign an element of 
\begin{equation*}
B = B_1 \ot B^\vee_1,
\end{equation*}
which we call a local state.
Thus there are $n^2$ local states.
We impose the boundary condition that all the 
local states sufficiently distant from the end 
are $\vac \in B$ defined by
\begin{equation}\label{eq:vac}
\vac = 1 \ot 
\kappa(1) = \begin{cases}
1 \ot \ol{n}  & \mbox{ for } \kappa = \mbox{Rotateleft, Switch$_{1n}$},\\ 
1 \ot \ol{2}  & \mbox{ for } \kappa = \mbox{Switch$_{12}$}.
\end{cases}
\end{equation}
See (\ref{eq:rl}), (\ref{eq:1n}) and (\ref{eq:12}).
An assignment of local states to the semi-infinite lattice satisfying the 
boundary condition is called a state. See Figure \ref{fig:lattice}.

\begin{figure}[h]
\caption{States}
\hspace{8cm}
\label{fig:lattice}

\unitlength 0.1in
\begin{picture}( 24.0300,  2.7500)( 18.0000,-10.5000)
%
\special{pn 8}%
\special{sh 1}%
\special{ar 4200 1000 10 10 0  6.28318530717959E+0000}%
%
\special{pn 8}%
\special{sh 1}%
\special{ar 2200 1000 10 10 0  6.28318530717959E+0000}%
%
\special{pn 8}%
\special{sh 1}%
\special{ar 2600 1000 10 10 0  6.28318530717959E+0000}%
%
\special{pn 8}%
\special{sh 1}%
\special{ar 3400 1000 10 10 0  6.28318530717959E+0000}%
%
\special{pn 8}%
\special{sh 1}%
\special{ar 3800 1000 10 10 0  6.28318530717959E+0000}%
%
\special{pn 8}%
\special{pa 4200 1000}%
\special{pa 3200 1000}%
\special{fp}%
\special{pa 2800 1000}%
\special{pa 2000 1000}%
\special{fp}%
%
\special{pn 8}%
\special{pa 3190 1000}%
\special{pa 2800 1000}%
\special{dt 0.045}%
\special{pa 2000 1000}%
\special{pa 1800 1000}%
\special{dt 0.045}%
\put(37.9000,-8.6000){\makebox(0,0){$B$}}%
\put(41.9000,-8.6000){\makebox(0,0){$B$}}%
\put(33.9000,-8.6000){\makebox(0,0){$B$}}%
\put(25.9000,-8.6000){\makebox(0,0){$B$}}%
\put(21.9000,-8.6000){\makebox(0,0){$B$}}%
\put(42.1500,-11.3500){\makebox(0,0){$p_1$}}%
\put(38.1500,-11.3500){\makebox(0,0){$p_2$}}%
\put(26.1500,-11.3500){\makebox(0,0){vac}}%
\put(22.1500,-11.3500){\makebox(0,0){vac}}%
\put(34.1500,-11.3500){\makebox(0,0){$p_3$}}%
\end{picture}%

\vspace{0.2cm}
\end{figure}

Let ${\mathcal P}$ denote the set of states:
\begin{equation}\label{eq:P}
{\mathcal P} = \{\cdots \ot p_2 \ot p_1 \in 
\cdots \ot B \ot B
\mid p_k = \vac \, \mbox{ for } k \gg 1 \}.
\end{equation}
Our automaton is a dynamical system on ${\mathcal P}$.

\subsection{Time evolution}\label{subsec:Tl}

The dynamics is given by a commuting family of time evolution operators
$T_l: {\mathcal P} \rightarrow {\mathcal P}$
with $l \in \Z_{\ge 1}$.
We set
\begin{align*}
&u_l = (l,0,\ldots,0) = \; \stackrel{l}{\overbrace{11\ldots 1}}
\; \; \in B_l,\\
&u^\vee_l = \kappa(u_l) = 
\begin{cases}
\ol{nn \ldots n}  \; \in B^\vee_l
& \mbox{ for } \kappa = \mbox{Rotateleft, Switch$_{1n}$}, \\
\ol{22 \ldots 2} \; \in B^\vee_l 
& \mbox{ for }  \kappa = \mbox{Switch$_{12}$}.
\end{cases}
\end{align*}
For any $l$ and $m$, the relations
\begin{equation}\label{eq:uiso}
z^du_l \ot z^eu_m \simeq z^eu_m \ot z^du_l,\;
z^du_l \ot z^eu^\vee_m \simeq z^eu^\vee_m \ot z^du_l,\;
z^du^\vee_l \ot z^eu^\vee_m \simeq z^eu^\vee_m \ot z^du^\vee_l
\end{equation}
are valid.
It is easy to verify
\begin{lemma}\label{lem:converge}
By iterating  $B_l\ot B\simeq B\ot B_l$ and 
$B^\vee_l\ot B\simeq B\ot B^\vee_l$, we consider maps
\begin{eqnarray*}
(i)\qquad\;\; B_l\ot B\ot\cd\ot B&\stackrel{\sim}{\longrightarrow}&
B\ot\cd\ot B\ot B_l\\
x \ot b_1 \ot \cd \ot b_N &\mapsto&
\tilde{b}_1\ot\cd\ot\tilde{b}_N\ot\tilde{x},\\
(ii)\qquad B\ot\cd\ot B \ot B^\vee_l &\stackrel{\sim}{\longrightarrow}&
B^\vee_l \ot B\ot\cd\ot B\\
c_N \ot \cd\ot c_1 \ot y &\mapsto&
\tilde{y} \ot \tilde{c}_N\ot\cd\ot\tilde{c}_1,
\end{eqnarray*}
where we assume (i) $b_j = \vac$ ((ii) $c_j = \vac$) for $N-N_0 \le j \le N$
with sufficiently large $N_0$. Then we have (i) $\tilde{x}=u_l$  
((ii) $\tilde{y} = u^\vee_l$).
\end{lemma}

\begin{figure}[h]
\caption{Diagram for $T_l(p)$}
\hspace{-1cm}
\label{fig:tl}

\unitlength 0.1in
\begin{picture}( 30.6000, 13.3000)( 13.4000,-16.9500)
%
\special{pn 8}%
\special{pa 3400 600}%
\special{pa 3400 960}%
\special{fp}%
\put(33.1000,-10.8000){\makebox(0,0){$p^\dag$}}%
%
\special{pn 8}%
\special{sh 1}%
\special{ar 2600 700 10 10 0  6.28318530717959E+0000}%
%
\special{pn 8}%
\special{sh 1}%
\special{ar 2200 700 10 10 0  6.28318530717959E+0000}%
%
\special{pn 8}%
\special{sh 1}%
\special{ar 2400 700 10 10 0  6.28318530717959E+0000}%
%
\special{pn 8}%
\special{sh 1}%
\special{ar 2610 1500 10 10 0  6.28318530717959E+0000}%
%
\special{pn 8}%
\special{sh 1}%
\special{ar 2210 1500 10 10 0  6.28318530717959E+0000}%
%
\special{pn 8}%
\special{sh 1}%
\special{ar 2410 1500 10 10 0  6.28318530717959E+0000}%
\put(33.0000,-4.5000){\makebox(0,0){$p$}}%
\put(32.9000,-17.8000){\makebox(0,0){$T_l(p)$}}%
%
\special{pn 8}%
\special{pa 3800 600}%
\special{pa 3800 960}%
\special{fp}%
%
\special{pn 8}%
\special{pa 3600 600}%
\special{pa 3600 960}%
\special{fp}%
%
\special{pn 8}%
\special{pa 3200 600}%
\special{pa 3200 960}%
\special{fp}%
%
\special{pn 8}%
\special{pa 3000 600}%
\special{pa 3000 960}%
\special{fp}%
%
\special{pn 8}%
\special{pa 2800 600}%
\special{pa 2800 960}%
\special{fp}%
%
\special{pn 8}%
\special{pa 3800 1600}%
\special{pa 3800 1240}%
\special{fp}%
%
\special{pn 8}%
\special{pa 3600 1600}%
\special{pa 3600 1240}%
\special{fp}%
%
\special{pn 8}%
\special{pa 3400 1600}%
\special{pa 3400 1240}%
\special{fp}%
%
\special{pn 8}%
\special{pa 3200 1600}%
\special{pa 3200 1240}%
\special{fp}%
%
\special{pn 8}%
\special{pa 3000 1600}%
\special{pa 3000 1240}%
\special{fp}%
%
\special{pn 8}%
\special{pa 2800 1600}%
\special{pa 2800 1240}%
\special{fp}%
\put(18.0000,-8.0000){\makebox(0,0){$u_l$}}%
\put(17.9000,-14.0000){\makebox(0,0){$u^\vee_l$}}%
\put(41.8000,-8.2000){\makebox(0,0){$v$}}%
\put(42.0000,-13.8000){\makebox(0,0){$\kappa(v)$}}%
%
\special{pn 8}%
\special{pa 4400 600}%
\special{pa 4400 1600}%
\special{fp}%
%
\special{pn 8}%
\special{pa 1980 800}%
\special{pa 3890 800}%
\special{pa 4400 1080}%
\special{pa 4400 1080}%
\special{fp}%
%
\special{pn 8}%
\special{pa 4400 1070}%
\special{pa 3900 1400}%
\special{pa 3900 1400}%
\special{dt 0.045}%
%
\special{pn 8}%
\special{pa 3900 1400}%
\special{pa 1990 1400}%
\special{dt 0.045}%
\special{sh 1}%
\special{pa 1990 1400}%
\special{pa 2058 1420}%
\special{pa 2044 1400}%
\special{pa 2058 1380}%
\special{pa 1990 1400}%
\special{fp}%
\end{picture}%

\vspace{0.3cm}
\end{figure}
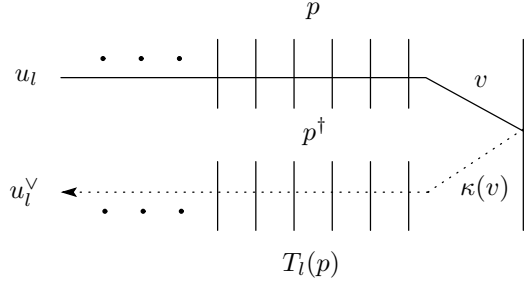

Pick any state $p \in {\mathcal P}$ and $l \in \Z_{\ge 1}$.
By repeated applications of combinatorial $R$,  
one can determine another state $T_l(p) \in {\mathcal P}$ along with 
the subsidiary data $v \in B_l$ and $p^\dag \in {\mathcal P}$ 
as follows:
\begin{eqnarray}
B_l \ot (\cd \ot B \ot B) &\simeq& (\cd \ot B \ot B) \ot B_l \nonumber\\
u_l \ot p &\simeq& p^\dag \ot v, \label{eq:tl1}\\
(\cd \ot B \ot B) \ot B^\vee_l 
&\simeq& B^\vee_l\ot(\cd \ot B \ot B) \nonumber \\
p^\dag \ot \kappa(v) &\simeq& u^\vee_l \ot T_l(p). \label{eq:tl2}
\end{eqnarray}
In (\ref{eq:tl1}), the tensor product $u_l \ot p$ makes sense
due to the boundary condition on ${\mathcal P}$ and 
the property $u_l \ot \vac \simeq \vac \ot u_l$ which follows from 
(\ref{eq:uiso}).
Similarly in (\ref{eq:tl2}), the right hand side takes the specified form 
due to the boundary condition, Lemma \ref{lem:converge} (ii) 
and $u^\vee_l \ot \vac \simeq \vac \ot u^\vee_l$.
The definition of $T_l(p)$ via (\ref{eq:tl1}) and 
(\ref{eq:tl2}) is a crystal theory analogue of the 
well known double row construction of transfer matrices 
in solvable lattice models with a boundary.
Schematically it is shown in Figure \ref{fig:tl}, 
where the vertical solid lines except the rightmost reflecting wall
represent elements of $B$ instead of $B_1$.
(The same convention is used in Figure \ref{fig:commute} and 
Figure \ref{fig:El}.)

The family $\{T_l\}$ possesses a saturation property.
\begin{proposition}\label{pr:saturation}
For any fixed element $p \in {\mathcal P}$, 
there exists an integer $l_0$ such that 
$T_l(p) = T_{l_0}(p)$ for $l \ge l_0$.
\end{proposition}
\begin{proof}
For any $1 \le i \le n$, let
$x^+$ be the array $x=(x_1,\ldots, x_n)$ with only $x_i$ increased by 1.
Suppose $x \ot y \simeq \tilde{y}\ot\tilde{x}$ under 
$\ol{R}, \ol{R}^\vee$ or $\ol{R}^{\vee\vee}$.
Then from the concrete form of the combinatorial $R$, 
it follows that there exist an integer $N$ such that 
$x^+ \ot y \simeq \tilde{y} \ot (\tilde{x})^+$ for $x_i \ge N$.
Similarly there exists an integer $N'$ such that 
$x \ot y^+ \simeq (\tilde{y})^+ \ot \tilde{x}$ for $y_i \ge N'$.
The assertion follows from $i=1,2,n$ cases of this and the 
analogous property of the map $\kappa$.
\end{proof}
\begin{remark}
In the infinite box-ball system without a boundary \cite{T}, 
$T_1$ is the global translation.
This is not the case in the present automaton where 
the translational invariance is absent.
See Example \ref{ex:ex3}.
\end{remark}
\begin{example}\label{ex:ex1}
$\widehat{\mathfrak{sl}}_5$, $\kappa = $Rotateleft.
Time evolution $T^t_3$ of the state in the first line 
is presented for $0 \le t \le 4$.
The symbol $\ot$ is omitted and 
$..$ stands for $\vac = 1\ot \ol{5}$ and 
$4\ol{5} = 4 \ot \ol{5} \in B$, etc.
There supposed to be an infinite tail 
consisting of $..$ only in the left on each line.
$$
\begin{array}{ccccccccccccccccccccccccccc}
0:\quad ..\s4\ol{5}\s2\ol{5}\s2\ol{5}\s..\s..\s..\s..\\
1:\quad ..\s..\s..\s..\s4\ol{5}\s2\ol{5}\s2\ol{5}\s..\\
2:\quad ..\s..\s..\s..\s..\s..\s1\ol{1}\s4\ol{1}\\
3:\quad ..\s..\s..\s..\s..\s1\ol{3}\s1\ol{4}\s1\ol{4}\\
4:\quad ..\s..\s1\ol{3}\s1\ol{4}\s1\ol{4}\s..\s..\s..
\end{array}
$$
In this case $T_l$ with any $l \ge 3$ 
yields the same evolution pattern.
The action of $T_3$ on the $t=1$ state is calculated 
according to the rule (\ref{eq:tl1}) and (\ref{eq:tl2}) as in 
Figure \ref{fig:ex1}.

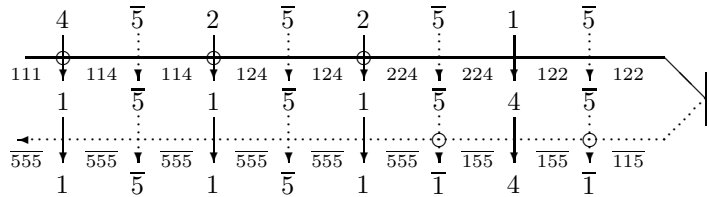
\begin{figure}[h]
\setlength{\unitlength}{1mm}
\caption{Calculation of $T_3$ on the $t=1$ state}
\label{fig:ex1}
\begin{picture}(100,30)(0,-4)

\multiput(89.5,2.5)(-1,0){85}{$\scriptstyle \cdot$}
\put(5,3){\vector(-1,0){1}}
\put(90,14){\line(-1,0){85}}

\multiput(89.5,2.5)(0.68,0.68){9}{$\scriptstyle \cdot$}
\put(90,14){\line(1,-1){5.5}}
\put(95.5,5){\line(0,1){7}}

\multiput(0,0)(20,0){4}{
\put(10,6){\vector(0,-1){6}}
\put(10,17){\vector(0,-1){6}}
\multiput(19.6,5.5)(0,-1){6}{$\scriptstyle \cdot$}
\put(20,1){\vector(0,-1){1}}
\multiput(19.6,16.5)(0,-1){6}{$\scriptstyle \cdot$}
\put(20,12){\vector(0,-1){1}}
}

\put(9,18){4}\put(19,18){$\ol{5}$}
\put(29,18){2}\put(39,18){$\ol{5}$}
\put(49,18){2}\put(59,18){$\ol{5}$}
\put(69,18){1}\put(79,18){$\ol{5}$}

\put(3,11){$\scriptstyle 111$}
\put(13,11){$\scriptstyle 114$}
\put(23,11){$\scriptstyle 114$}
\put(33,11){$\scriptstyle 124$}
\put(43,11){$\scriptstyle 124$}
\put(53,11){$\scriptstyle 224$}
\put(63,11){$\scriptstyle 224$}
\put(73,11){$\scriptstyle 122$}
\put(83,11){$\scriptstyle 122$}

\put(9.1,7){1}\put(19.1,7){$\ol{5}$}
\put(29.1,7){1}\put(39.1,7){$\ol{5}$}
\put(49.1,7){1}\put(59.1,7){$\ol{5}$}
\put(69.1,7){4}\put(79.1,7){$\ol{5}$}

\put(3,-0.5){$\scriptstyle \ol{555}$}
\put(13,-0.5){$\scriptstyle \ol{555}$}
\put(23,-0.5){$\scriptstyle \ol{555}$}
\put(33,-0.5){$\scriptstyle \ol{555}$}
\put(43,-0.5){$\scriptstyle \ol{555}$}
\put(53,-0.5){$\scriptstyle \ol{555}$}
\put(63,-0.5){$\scriptstyle \ol{155}$}
\put(73,-0.5){$\scriptstyle \ol{155}$}
\put(83,-0.5){$\scriptstyle \ol{115}$}

\put(9,-4){1}\put(19,-4.1){$\ol{5}$}
\put(29,-4){1}\put(39,-4.1){$\ol{5}$}
\put(49,-4){1}\put(59,-4.1){$\ol{1}$}
\put(69,-4){4}\put(79,-4.1){$\ol{1}$}

\put(10,14){\circle{1.6}}
\put(30,14){\circle{1.6}}
\put(50,14){\circle{1.6}}
\put(60,3){\circle{1.6}}
\put(80,3){\circle{1.6}}

\end{picture}
\end{figure}

In Figure \ref{fig:ex1}, the 
combinatorial $\ol{R}^\vee$ is acting just as a transposition 
of the components everywhere. 
The symbol $\circ$ is put for convenience in Example \ref{ex:El}.
\end{example}

\begin{example}\label{ex:ex2}
$\widehat{\mathfrak{sl}}_4$, $\kappa = $Switch$_{14}$.
Time evolution $T^t_2$ of the state in the first line 
is presented for $0 \le t \le 7$.
The symbol $..$ stands for $\vac = 1\ot \ol{4}$ and 
$3\ol{4} = 3 \ot \ol{4} \in B$, etc.
$$
\begin{array}{cccccccccccccccccccccccc}
0:\quad ..\s..\s..\s..\s3\ol{4}\s3\ol{4}\s2\ol{4}\s..\s..\s..\s..\s1\ol{3}\\
1:\quad ..\s..\s..\s..\s..\s..\s3\ol{4}\s3\ol{4}\s2\ol{4}\s..\s1\ol{3}\s..\\
2:\quad ..\s..\s..\s..\s..\s..\s..\s..\s3\ol{4}\s3\ol{3}\s2\ol{4}\s..\\
3:\quad ..\s..\s..\s..\s..\s..\s..\s..\s1\ol{2}\s..\s3\ol{4}\s2\ol{3}\\
4:\quad ..\s..\s..\s..\s..\s..\s..\s1\ol{2}\s..\s1\ol{2}\s1\ol{3}\s1\ol{3}\\
5:\quad ..\s..\s..\s..\s..\s1\ol{2}\s1\ol{2}\s..\s1\ol{3}\s1\ol{3}\s..\s..\\
6:\quad ..\s..\s..\s1\ol{2}\s1\ol{2}\s1\ol{3}\s..\s1\ol{3}\s..\s..\s..\s..\\
7:\quad ..\s1\ol{2}\s1\ol{2}\s1\ol{3}\s..\s..\s1\ol{3}\s..\s..\s..\s..\s..
\end{array}
$$
These states have the same evolution pattern under $T_l$ with any $l \ge 3$.

The action of $T_2$ on the $t=2$ state is calculated 
as in Figure \ref{fig:ex2}.

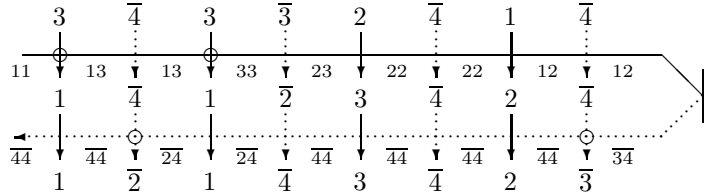
\begin{figure}[h]
\setlength{\unitlength}{1mm}
\caption{Calculation of $T_2$ on the $t=2$ state}
\label{fig:ex2}
\begin{picture}(100,30)(0,-4)

\multiput(89.5,2.5)(-1,0){85}{$\scriptstyle \cdot$}
\put(5,3){\vector(-1,0){1}}
\put(90,14){\line(-1,0){85}}

\multiput(89.5,2.5)(0.68,0.68){9}{$\scriptstyle \cdot$}
\put(90,14){\line(1,-1){5.5}}
\put(95.5,5){\line(0,1){7}}

\multiput(0,0)(20,0){4}{
\put(10,6){\vector(0,-1){6}}
\put(10,17){\vector(0,-1){6}}
\multiput(19.6,5.5)(0,-1){6}{$\scriptstyle \cdot$}
\put(20,1){\vector(0,-1){1}}
\multiput(19.6,16.5)(0,-1){6}{$\scriptstyle \cdot$}
\put(20,12){\vector(0,-1){1}}
}

\put(9,18){3}\put(19,18){$\ol{4}$}
\put(29,18){3}\put(39,18){$\ol{3}$}
\put(49,18){2}\put(59,18){$\ol{4}$}
\put(69,18){1}\put(79,18){$\ol{4}$}

\put(3.4,11){$\scriptstyle 11$}
\put(13.4,11){$\scriptstyle 13$}
\put(23.4,11){$\scriptstyle 13$}
\put(33.4,11){$\scriptstyle 33$}
\put(43.4,11){$\scriptstyle 23$}
\put(53.4,11){$\scriptstyle 22$}
\put(63.4,11){$\scriptstyle 22$}
\put(73.4,11){$\scriptstyle 12$}
\put(83.4,11){$\scriptstyle 12$}

\put(9.1,7){1}\put(19.1,7){$\ol{4}$}
\put(29.1,7){1}\put(39.1,7){$\ol{2}$}
\put(49.1,7){3}\put(59.1,7){$\ol{4}$}
\put(69.1,7){2}\put(79.1,7){$\ol{4}$}

\put(3.4,-0.5){$\scriptstyle \ol{44}$}
\put(13.4,-0.5){$\scriptstyle \ol{44}$}
\put(23.4,-0.5){$\scriptstyle \ol{24}$}
\put(33.4,-0.5){$\scriptstyle \ol{24}$}
\put(43.4,-0.5){$\scriptstyle \ol{44}$}
\put(53.4,-0.5){$\scriptstyle \ol{44}$}
\put(63.4,-0.5){$\scriptstyle \ol{44}$}
\put(73.4,-0.5){$\scriptstyle \ol{44}$}
\put(83.4,-0.5){$\scriptstyle \ol{34}$}

\put(9,-4){1}\put(19,-4.1){$\ol{2}$}
\put(29,-4){1}\put(39,-4.1){$\ol{4}$}
\put(49,-4){3}\put(59,-4.1){$\ol{4}$}
\put(69,-4){2}\put(79,-4.1){$\ol{3}$}

\put(10,14){\circle{1.6}}
\put(30,14){\circle{1.6}}
\put(20,3){\circle{1.6}}
\put(80,3){\circle{1.6}}

\end{picture}
\end{figure}

Again the symbol $\circ$ is put for convenience in Example \ref{ex:El}.
\end{example}

\begin{example}\label{ex:ex3}
$\widehat{\mathfrak{sl}}_4$, $\kappa = $Switch$_{12}$.
Time evolution $T^t_1$ of the state in the first line 
is presented for $0 \le t \le 4$.
The symbol $..$ stands for $\vac = 1\ot \ol{2}$ and 
$2\ol{3} = 2 \ot \ol{3} \in B$, etc.
$$
\begin{array}{cccccccccccccccccccccccccccccccccccc}
0:\quad ..\sa ..\sa ..\sa..\sa..\sa..\sa..\sa..\sa..\sa2\ol{3}\sa..\sa..\sa..\sa..\sa4\ol{1}\sa..\sa..\sa..\sa..\sa..\sa..\sa..\sa..\sa..\sa2\ol{1}\sa..\sa..\sa..\\
1:\quad ..\sa..\sa..\sa..\sa..\sa..\sa..\sa..\sa1\ol{3}\sa1\ol{1}\sa..\sa..\sa..\sa1\ol{4}\sa4\ol{2}\sa4\ol{2}\sa..\sa..\sa..\sa..\sa..\sa..\sa..\sa1\ol{4}\sa4\ol{1}\sa..\sa..\sa..\\
2:\quad ..\sa..\sa..\sa..\sa..\sa..\sa..\sa1\ol{3}\sa1\ol{4}\sa..\sa4\ol{2}\sa..\sa1\ol{4}\sa..\sa..\sa4\ol{2}\sa4\ol{2}\sa..\sa..\sa..\sa..\sa..\sa1\ol{4}\sa1\ol{4}\sa4\ol{2}\sa4\ol{2}\sa..\sa..\\
3:\quad ..\sa..\sa..\sa..\sa..\sa..\sa1\ol{3}\sa1\ol{4}\sa..\sa..\sa..\sa4\ol{4}\sa..\sa..\sa..\sa..\sa4\ol{2}\sa4\ol{2}\sa..\sa..\sa..\sa1\ol{4}\sa1\ol{4}\sa..\sa..\sa4\ol{2}\sa4\ol{2}\sa..\\
4:\quad ..\s\,..\s\,..\s\,..\s\,..\s1 \ol{3}\sa1\ol{4}\sa..\sa..\sa..\sa1\ol{3}\sa..\sa3\ol{2}\sa..\sa..\sa..\sa..\sa4\ol{2}\sa 4\ol{2}\s..\s 1\ol{4}\sa1\ol{4}\sa..\sa..\sa..\sa..\sa4\ol{2}\sa4\ol{2}
\end{array}
$$

On the other hand, time evolution of the initial state under 
$T^t_l$ with any $l \ge 2$ are the same, 
and given as follows:
$$
\begin{array}{ccccccccccccccccccccccccccccccccccccccccccccccc}
0:\quad ..\sa..\sa..\sa..\sa..\sa..\sa..\sa..\sa..\sa2\ol{3}\sa..\sa..\sa..\sa..\sa4\ol{1}
\sa..\sa..\sa..\sa..\sa..\sa..\sa..\sa..\sa..\sa2\ol{1}\sa..\sa..\sa..\\
1:\quad ..\sa..\sa..\sa..\sa..\sa..\sa..\sa1\ol{3}\sa1\ol{4}\sa4\ol{2}\sa..\sa..\sa..\sa
1\ol{4}\sa..\sa4\ol{2}\sa4\ol{2}\sa..\sa..\sa..\sa..\sa..\sa1\ol{4}\sa1\ol{4}\sa4\ol{2}\sa4\ol{2}\sa..\sa..\\
2:\quad ..\sa..\sa..\sa..\sa..\sa1\ol{3}\sa1\ol{4}\sa..\sa..\sa..\sa4\ol{2}\sa..\sa1\ol{4}
\sa..\sa..\sa..\sa..\sa4\ol{2}\sa4\ol{2}\sa..\sa1\ol{4}\sa1\ol{4}\sa..\sa..\sa..\sa..\sa
4\ol{2}\sa4\ol{2}\\
3:\quad ..\sa..\sa..\sa1\ol{3}\sa1\ol{4}\sa..\sa..\sa..\sa..\sa..\sa..\sa4\ol{4}\sa..\sa..\sa..\sa..\sa..\sa..\sa1\ol{3}\sa4\ol{4}\sa3\ol{2}\sa..\sa..\sa..\sa..\sa..\sa1\ol{3}\sa1\ol{3}\\
4:\quad ..\s1\ol{3}\sa1\ol{4}\sa..\sa..\sa..\sa..\sa..\sa..\sa..\sa1\ol{3}\sa..\sa3\ol{2}
\sa..\sa..\sa..\sa1\ol{3}\s1\ol{3}\sa..\sa..\sa..\sa3\ol{2}\sa3\ol{2}\sa..\sa1\ol{3}
\sa1\ol{3}\sa..\sa..
\end{array}
$$

\end{example}

\begin{proposition}\label{pr:commute}
$\{ T_l \}$ forms a commuting family, i.e., 
$T_lT_m = T_mT_l$.
\end{proposition}
\begin{proof}
This is shown by a standard argument 
based on the successive transformations 
in Figure \ref{fig:commute}.
\begin{figure}[h]
\caption{Explanation of $T_mT_l(p) = T_lT_m(p)$}
\vspace{1cm}
\hspace{-1cm}
\label{fig:commute}

\unitlength 0.1in
\begin{picture}( 52.0000, 38.2300)(  9.000,-44.2300)
\put(34.8000,-24.6000){\makebox(0,0){$=$}}%
\put(34.6000,-38.6000){\makebox(0,0){$=$}}%
\put(34.8000,-10.5000){\makebox(0,0){$=$}}%
%
\special{pn 8}%
\special{pa 3156 650}%
\special{pa 3156 1400}%
\special{fp}%
%
\special{pn 8}%
\special{pa 2410 530}%
\special{pa 2410 1530}%
\special{fp}%
\special{pa 1910 530}%
\special{pa 1910 1530}%
\special{fp}%
\put(11.8000,-7.1000){\makebox(0,0)[rb]{$u_l$}}%
\put(12.0000,-9.6000){\makebox(0,0)[rb]{$u^\vee_l$}}%
\put(12.0000,-14.7000){\makebox(0,0)[rb]{$u^\vee_m$}}%
\put(12.0000,-12.2000){\makebox(0,0)[rb]{$u_m$}}%
\put(21.5000,-5.1000){\makebox(0,0){$p$}}%
\put(21.6000,-10.4000){\makebox(0,0){$T_l(p)$}}%
\put(21.6000,-16.5000){\makebox(0,0){$T_mT_l(p)$}}%
\put(8.1000,-38.6000){\makebox(0,0){$=$}}%
\put(21.6000,-19.0000){\makebox(0,0){$p$}}%
\put(12.1000,-23.8000){\makebox(0,0)[rb]{$u_m$}}%
\put(12.1000,-26.2000){\makebox(0,0)[rb]{$u^\vee_m$}}%
\put(12.1000,-28.6000){\makebox(0,0)[rb]{$u^\vee_l$}}%
\put(11.8000,-21.1000){\makebox(0,0)[rb]{$u_l$}}%
%
\special{pn 8}%
\special{pa 1910 1920}%
\special{pa 1910 2920}%
\special{fp}%
%
\special{pn 8}%
\special{pa 2410 1930}%
\special{pa 2410 2930}%
\special{fp}%
%
\special{pn 8}%
\special{pa 3156 2050}%
\special{pa 3156 2800}%
\special{fp}%
\put(21.7000,-30.5000){\makebox(0,0){$T_mT_l(p)$}}%
%
\special{pn 8}%
\special{pa 4630 1920}%
\special{pa 4630 2920}%
\special{fp}%
%
\special{pn 8}%
\special{pa 5876 2040}%
\special{pa 5876 2790}%
\special{fp}%
%
\special{pn 8}%
\special{pa 5130 1930}%
\special{pa 5130 2930}%
\special{fp}%
\put(39.0000,-21.0000){\makebox(0,0)[rb]{$u_l$}}%
\put(39.2000,-28.5000){\makebox(0,0)[rb]{$u^\vee_l$}}%
\put(39.2000,-26.1000){\makebox(0,0)[rb]{$u^\vee_m$}}%
\put(39.3000,-23.6000){\makebox(0,0)[rb]{$u_m$}}%
\put(48.8000,-18.9000){\makebox(0,0){$p$}}%
\put(48.7000,-30.5000){\makebox(0,0){$T_mT_l(p)$}}%
\put(8.1000,-24.5000){\makebox(0,0){$=$}}%
\put(21.7000,-44.5000){\makebox(0,0){$T_mT_l(p)$}}%
\put(21.8000,-33.0000){\makebox(0,0){$p$}}%
\put(12.4000,-37.7000){\makebox(0,0)[rb]{$u_m$}}%
\put(12.3000,-40.1000){\makebox(0,0)[rb]{$u^\vee_m$}}%
\put(12.3000,-42.7000){\makebox(0,0)[rb]{$u^\vee_l$}}%
\put(12.0000,-35.2000){\makebox(0,0)[rb]{$u_l$}}%
%
\special{pn 8}%
\special{pa 3176 3450}%
\special{pa 3176 4200}%
\special{fp}%
%
\special{pn 8}%
\special{pa 2426 3326}%
\special{pa 2426 4326}%
\special{fp}%
%
\special{pn 8}%
\special{pa 1926 3326}%
\special{pa 1926 4326}%
\special{fp}%
\put(21.7000,-38.3000){\makebox(0,0){$T_m(p)$}}%
%
\special{pn 8}%
\special{pa 5884 3446}%
\special{pa 5884 4196}%
\special{fp}%
%
\special{pn 8}%
\special{pa 5138 3326}%
\special{pa 5138 4326}%
\special{fp}%
\special{pa 4638 3326}%
\special{pa 4638 4326}%
\special{fp}%
\put(38.9800,-40.2500){\makebox(0,0)[rb]{$u_l$}}%
\put(39.2800,-42.5500){\makebox(0,0)[rb]{$u^\vee_l$}}%
\put(39.2800,-37.7500){\makebox(0,0)[rb]{$u^\vee_m$}}%
\put(39.3800,-35.1500){\makebox(0,0)[rb]{$u_m$}}%
\put(48.7800,-32.9500){\makebox(0,0){$p$}}%
\put(46.4800,-43.7500){\makebox(0,0)[lt]{$T_mT_l(p) = T_lT_m(p)$}}%
\put(48.6800,-38.2500){\makebox(0,0){$T_m(p)$}}%
%
\special{pn 4}%
\special{sh 1}%
\special{ar 4538 4406 10 10 0  6.28318530717959E+0000}%
%
\special{pn 4}%
\special{sh 1}%
\special{ar 4588 4470 10 10 0  6.28318530717959E+0000}%
%
\special{pn 4}%
\special{sh 1}%
\special{ar 4490 4470 10 10 0  6.28318530717959E+0000}%
%
\special{pn 8}%
\special{pa 5876 650}%
\special{pa 5876 1400}%
\special{fp}%
%
\special{pn 8}%
\special{pa 5130 530}%
\special{pa 5130 1530}%
\special{fp}%
\special{pa 4630 530}%
\special{pa 4630 1530}%
\special{fp}%
\put(39.2000,-14.8000){\makebox(0,0)[rb]{$u^\vee_l$}}%
\put(39.0000,-7.1000){\makebox(0,0)[rb]{$u_l$}}%
\put(39.2000,-12.1000){\makebox(0,0)[rb]{$u^\vee_m$}}%
\put(39.3000,-9.5000){\makebox(0,0)[rb]{$u_m$}}%
\put(48.7000,-5.1000){\makebox(0,0){$p$}}%
\put(48.7000,-10.4000){\makebox(0,0){$T_l(p)$}}%
\put(48.7000,-16.5000){\makebox(0,0){$T_mT_l(p)$}}%
%
\special{pn 8}%
\special{pa 3980 2800}%
\special{pa 5380 2800}%
\special{pa 5880 2670}%
\special{pa 5880 2670}%
\special{pa 5880 2670}%
\special{dt 0.045}%
%
\special{pn 8}%
\special{pa 5870 2670}%
\special{pa 5380 2080}%
\special{pa 3980 2080}%
\special{pa 3980 2080}%
\special{pa 3980 2080}%
\special{fp}%
%
\special{pn 8}%
\special{pa 3970 2800}%
\special{pa 3946 2800}%
\special{fp}%
\special{sh 1}%
\special{pa 3946 2800}%
\special{pa 4012 2820}%
\special{pa 3998 2800}%
\special{pa 4012 2780}%
\special{pa 3946 2800}%
\special{fp}%
%
\special{pn 8}%
\special{pa 1260 650}%
\special{pa 2660 650}%
\special{pa 3160 780}%
\special{pa 3160 780}%
\special{pa 3160 780}%
\special{fp}%
%
\special{pn 8}%
\special{pa 1260 900}%
\special{pa 1236 900}%
\special{fp}%
\special{sh 1}%
\special{pa 1236 900}%
\special{pa 1302 920}%
\special{pa 1288 900}%
\special{pa 1302 880}%
\special{pa 1236 900}%
\special{fp}%
%
\special{pn 8}%
\special{pa 1260 900}%
\special{pa 2660 900}%
\special{pa 3160 770}%
\special{pa 3160 770}%
\special{pa 3160 770}%
\special{dt 0.045}%
%
\special{pn 8}%
\special{pa 1260 1150}%
\special{pa 2660 1150}%
\special{pa 3160 1280}%
\special{pa 3160 1280}%
\special{pa 3160 1280}%
\special{fp}%
%
\special{pn 8}%
\special{pa 1260 1400}%
\special{pa 1236 1400}%
\special{fp}%
\special{sh 1}%
\special{pa 1236 1400}%
\special{pa 1302 1420}%
\special{pa 1288 1400}%
\special{pa 1302 1380}%
\special{pa 1236 1400}%
\special{fp}%
%
\special{pn 8}%
\special{pa 1260 1400}%
\special{pa 2660 1400}%
\special{pa 3160 1270}%
\special{pa 3160 1270}%
\special{pa 3160 1270}%
\special{dt 0.045}%
%
\special{pn 8}%
\special{pa 3980 3450}%
\special{pa 5380 3450}%
\special{pa 5880 3580}%
\special{pa 5880 3580}%
\special{pa 5880 3580}%
\special{fp}%
%
\special{pn 8}%
\special{pa 3980 3700}%
\special{pa 3956 3700}%
\special{fp}%
\special{sh 1}%
\special{pa 3956 3700}%
\special{pa 4022 3720}%
\special{pa 4008 3700}%
\special{pa 4022 3680}%
\special{pa 3956 3700}%
\special{fp}%
%
\special{pn 8}%
\special{pa 3980 3700}%
\special{pa 5380 3700}%
\special{pa 5880 3570}%
\special{pa 5880 3570}%
\special{pa 5880 3570}%
\special{dt 0.045}%
%
\special{pn 8}%
\special{pa 3980 3950}%
\special{pa 5380 3950}%
\special{pa 5880 4080}%
\special{pa 5880 4080}%
\special{pa 5880 4080}%
\special{fp}%
%
\special{pn 8}%
\special{pa 3980 4200}%
\special{pa 3956 4200}%
\special{fp}%
\special{sh 1}%
\special{pa 3956 4200}%
\special{pa 4022 4220}%
\special{pa 4008 4200}%
\special{pa 4022 4180}%
\special{pa 3956 4200}%
\special{fp}%
%
\special{pn 8}%
\special{pa 3980 4200}%
\special{pa 5380 4200}%
\special{pa 5880 4070}%
\special{pa 5880 4070}%
\special{pa 5880 4070}%
\special{dt 0.045}%
\put(48.6300,-21.7500){\makebox(0,0){$T_m(p)$}}%
%
\special{pn 8}%
\special{pa 3976 2300}%
\special{pa 5376 2300}%
\special{pa 5876 2430}%
\special{pa 5876 2430}%
\special{pa 5876 2430}%
\special{fp}%
%
\special{pn 8}%
\special{pa 3976 2550}%
\special{pa 3950 2550}%
\special{fp}%
\special{sh 1}%
\special{pa 3950 2550}%
\special{pa 4018 2570}%
\special{pa 4004 2550}%
\special{pa 4018 2530}%
\special{pa 3950 2550}%
\special{fp}%
%
\special{pn 8}%
\special{pa 3976 2550}%
\special{pa 5376 2550}%
\special{pa 5876 2420}%
\special{pa 5876 2420}%
\special{pa 5876 2420}%
\special{dt 0.045}%
%
\special{pn 8}%
\special{pa 1260 2300}%
\special{pa 2660 2300}%
\special{pa 3160 2430}%
\special{pa 3160 2430}%
\special{pa 3160 2430}%
\special{fp}%
%
\special{pn 8}%
\special{pa 1260 2550}%
\special{pa 1236 2550}%
\special{fp}%
\special{sh 1}%
\special{pa 1236 2550}%
\special{pa 1302 2570}%
\special{pa 1288 2550}%
\special{pa 1302 2530}%
\special{pa 1236 2550}%
\special{fp}%
%
\special{pn 8}%
\special{pa 1260 2550}%
\special{pa 2660 2550}%
\special{pa 3160 2420}%
\special{pa 3160 2420}%
\special{pa 3160 2420}%
\special{dt 0.045}%
%
\special{pn 8}%
\special{pa 3990 1390}%
\special{pa 3976 1410}%
\special{fp}%
\special{sh 1}%
\special{pa 3976 1410}%
\special{pa 4032 1370}%
\special{pa 4008 1368}%
\special{pa 4000 1346}%
\special{pa 3976 1410}%
\special{fp}%
%
\special{pn 8}%
\special{pa 5860 780}%
\special{pa 5360 900}%
\special{pa 4360 900}%
\special{pa 3980 1400}%
\special{pa 3980 1400}%
\special{dt 0.045}%
%
\special{pn 8}%
\special{pa 3980 660}%
\special{pa 5380 660}%
\special{pa 5880 790}%
\special{pa 5880 790}%
\special{pa 5880 790}%
\special{fp}%
%
\special{pn 8}%
\special{pa 1280 3450}%
\special{pa 1770 3950}%
\special{pa 2690 3950}%
\special{pa 3170 4050}%
\special{pa 3170 4050}%
\special{pa 3170 4050}%
\special{fp}%
%
\special{pn 8}%
\special{pa 1280 4180}%
\special{pa 2680 4180}%
\special{pa 3180 4050}%
\special{pa 3180 4050}%
\special{pa 3180 4050}%
\special{dt 0.045}%
%
\special{pn 8}%
\special{pa 1280 4180}%
\special{pa 1256 4180}%
\special{fp}%
\special{sh 1}%
\special{pa 1256 4180}%
\special{pa 1322 4200}%
\special{pa 1308 4180}%
\special{pa 1322 4160}%
\special{pa 1256 4180}%
\special{fp}%
%
\special{pn 8}%
\special{pa 1290 2790}%
\special{pa 1266 2790}%
\special{fp}%
\special{sh 1}%
\special{pa 1266 2790}%
\special{pa 1332 2810}%
\special{pa 1318 2790}%
\special{pa 1332 2770}%
\special{pa 1266 2790}%
\special{fp}%
%
\special{pn 8}%
\special{pa 3150 2200}%
\special{pa 2660 2790}%
\special{pa 1260 2790}%
\special{pa 1260 2790}%
\special{pa 1260 2790}%
\special{dt 0.045}%
%
\special{pn 8}%
\special{pa 1250 2070}%
\special{pa 2650 2070}%
\special{pa 3150 2200}%
\special{pa 3150 2200}%
\special{pa 3150 2200}%
\special{fp}%
%
\special{pn 8}%
\special{pa 1270 3950}%
\special{pa 1250 3964}%
\special{fp}%
\special{sh 1}%
\special{pa 1250 3964}%
\special{pa 1316 3942}%
\special{pa 1294 3934}%
\special{pa 1294 3910}%
\special{pa 1250 3964}%
\special{fp}%
%
\special{pn 8}%
\special{pa 1270 3690}%
\special{pa 1650 3450}%
\special{pa 2670 3450}%
\special{pa 3170 3560}%
\special{pa 3170 3560}%
\special{pa 3170 3560}%
\special{fp}%
%
\special{pn 8}%
\special{pa 1260 3950}%
\special{pa 1660 3690}%
\special{pa 2670 3690}%
\special{pa 3160 3570}%
\special{pa 3160 3570}%
\special{dt 0.045}%
%
\special{pn 8}%
\special{pa 3970 1150}%
\special{pa 3950 1136}%
\special{fp}%
\special{sh 1}%
\special{pa 3950 1136}%
\special{pa 3994 1192}%
\special{pa 3994 1168}%
\special{pa 4016 1158}%
\special{pa 3950 1136}%
\special{fp}%
%
\special{pn 8}%
\special{pa 3980 1160}%
\special{pa 4360 1400}%
\special{pa 5380 1400}%
\special{pa 5880 1290}%
\special{pa 5880 1290}%
\special{pa 5880 1290}%
\special{dt 0.045}%
%
\special{pn 8}%
\special{pa 3970 900}%
\special{pa 4370 1160}%
\special{pa 5380 1160}%
\special{pa 5870 1280}%
\special{pa 5870 1280}%
\special{fp}%
\end{picture}%

\vspace{0.5cm}
\end{figure}
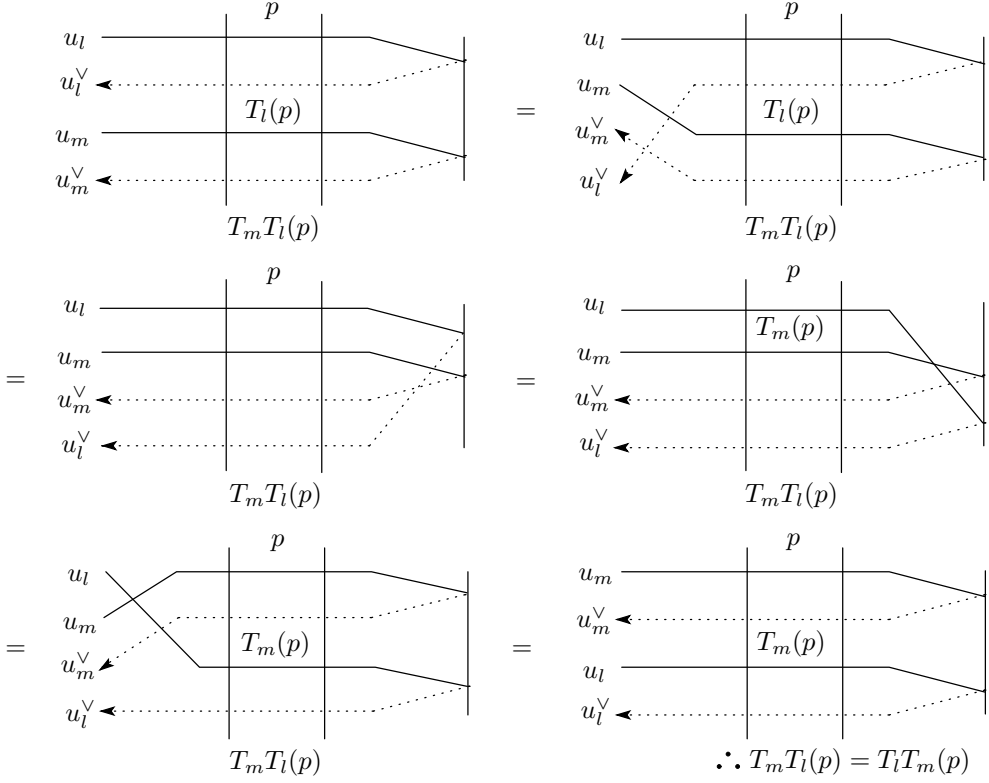
In the figure, each line segment is assigned with 
an element of a crystal and each cross represents a relation 
$x \ot y \simeq \tilde{y} \ot \tilde{x}$ under
$\ol{R}, \ol{R}^\vee$ or $\ol{R}^{\vee\vee}$.
A touch with the end stands for an application of $\kappa$.
The first and the last equalities are due to (\ref{eq:uiso}).
The second and the fourth ones are results of repeated 
applications of the Yang-Baxter equation.
The third one is the reflection equation.
\end{proof}

\subsection{Conserved quantity $E_l$}\label{subsec:El}

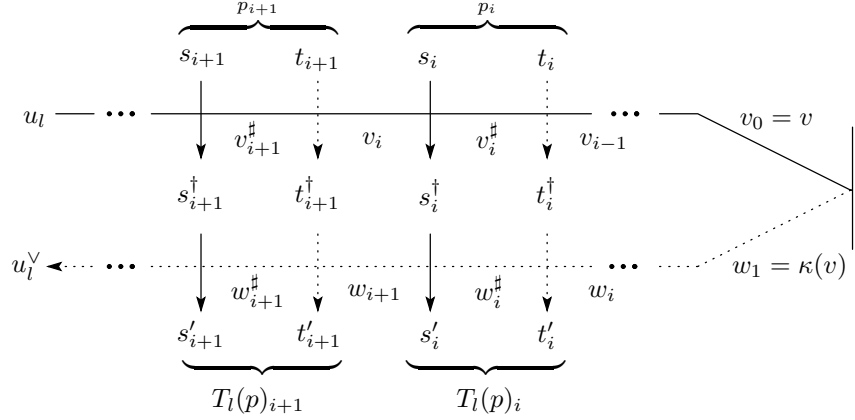
\begin{figure}[h]
\caption{Another diagram for $T_l(p)$}
\vspace{0.1cm}
\hspace{-3.0cm}
\label{fig:detail}

\unitlength 0.1in
\begin{picture}( 54.5000, 20.5200)( 10.3000,-26.1500)
%
\special{pn 8}%
\special{pa 2400 1030}%
\special{pa 2400 1430}%
\special{fp}%
\special{sh 1}%
\special{pa 2400 1430}%
\special{pa 2420 1364}%
\special{pa 2400 1378}%
\special{pa 2380 1364}%
\special{pa 2400 1430}%
\special{fp}%
\special{pa 2400 1830}%
\special{pa 2400 2230}%
\special{fp}%
\special{sh 1}%
\special{pa 2400 2230}%
\special{pa 2420 2164}%
\special{pa 2400 2178}%
\special{pa 2380 2164}%
\special{pa 2400 2230}%
\special{fp}%
\special{pa 1404 564}%
\special{pa 1404 564}%
\special{fp}%
\put(39.1000,-27.0000){\makebox(0,0){$T_l(p)_i$}}%
\put(27.0000,-27.0000){\makebox(0,0){$T_l(p)_{i+1}$}}%
\put(27.2000,-25.3000){\makebox(0,0){$\underbrace{\qquad\qquad\qquad}$}}%
\put(38.9000,-25.3000){\makebox(0,0){$\underbrace{\qquad\qquad\qquad}$}}%
\put(42.1000,-23.6000){\makebox(0,0){$t'_i$}}%
\put(36.0000,-23.6000){\makebox(0,0){$s'_i$}}%
\put(30.2000,-23.6000){\makebox(0,0){$t'_{i+1}$}}%
\put(24.0000,-23.6000){\makebox(0,0){$s'_{i+1}$}}%
\put(39.0000,-7.0000){\makebox(0,0){$\stackrel{p_i}{\overbrace{\qquad\qquad\qquad}}$}}%
\put(27.0000,-7.0000){\makebox(0,0){$\stackrel{p_{i+1}}{\overbrace{\qquad\qquad\qquad}}$}}%
\put(42.1000,-9.2000){\makebox(0,0){$t_i$}}%
\put(35.9000,-9.2000){\makebox(0,0){$s_i$}}%
\put(30.1000,-9.2000){\makebox(0,0){$t_{i+1}$}}%
%
\special{pn 8}%
\special{pa 5810 1270}%
\special{pa 5810 1910}%
\special{fp}%
\put(51.8000,-19.3000){\makebox(0,0)[lt]{$w_1 = \kappa(v)$}}%
\put(52.2000,-12.8000){\makebox(0,0)[lb]{$v_0 = v$}}%
\put(24.1000,-9.1000){\makebox(0,0){$s_{i+1}$}}%
\put(45.0000,-21.4000){\makebox(0,0){$w_i$}}%
\put(39.1000,-21.3000){\makebox(0,0){$w^\sharp_i$}}%
\put(33.1000,-21.4000){\makebox(0,0){$w_{i+1}$}}%
\put(27.0000,-21.3000){\makebox(0,0){$w^\sharp_{i+1}$}}%
\put(42.1000,-16.2000){\makebox(0,0){$t^\dag_i$}}%
\put(36.0000,-16.3000){\makebox(0,0){$s^\dag_i$}}%
\put(30.1000,-16.2000){\makebox(0,0){$t^\dag_{i+1}$}}%
\put(45.1000,-13.5000){\makebox(0,0){$v_{i-1}$}}%
\put(39.0000,-13.3000){\makebox(0,0){$v^\sharp_i$}}%
\put(33.0000,-13.4000){\makebox(0,0){$v_i$}}%
\put(15.9000,-12.7000){\makebox(0,0)[rb]{$u_l$}}%
\put(15.7000,-20.7000){\makebox(0,0)[rb]{$u^\vee_l$}}%
%
\special{pn 8}%
\special{pa 1830 2000}%
\special{pa 1600 2000}%
\special{dt 0.045}%
\special{sh 1}%
\special{pa 1600 2000}%
\special{pa 1668 2020}%
\special{pa 1654 2000}%
\special{pa 1668 1980}%
\special{pa 1600 2000}%
\special{fp}%
%
\special{pn 8}%
\special{pa 1640 1200}%
\special{pa 1840 1200}%
\special{fp}%
\put(27.0000,-13.3000){\makebox(0,0){$v^\sharp_{i+1}$}}%
\put(24.0000,-16.2000){\makebox(0,0){$s^\dag_{i+1}$}}%
%
\special{pn 8}%
\special{pa 3012 1030}%
\special{pa 3012 1430}%
\special{dt 0.045}%
\special{sh 1}%
\special{pa 3012 1430}%
\special{pa 3032 1364}%
\special{pa 3012 1378}%
\special{pa 2992 1364}%
\special{pa 3012 1430}%
\special{fp}%
\special{pa 3012 1830}%
\special{pa 3012 2230}%
\special{dt 0.045}%
\special{sh 1}%
\special{pa 3012 2230}%
\special{pa 3032 2164}%
\special{pa 3012 2178}%
\special{pa 2992 2164}%
\special{pa 3012 2230}%
\special{fp}%
\special{pa 2016 564}%
\special{pa 2016 564}%
\special{dt 0.045}%
%
\special{pn 8}%
\special{pa 3600 1030}%
\special{pa 3600 1430}%
\special{fp}%
\special{sh 1}%
\special{pa 3600 1430}%
\special{pa 3620 1364}%
\special{pa 3600 1378}%
\special{pa 3580 1364}%
\special{pa 3600 1430}%
\special{fp}%
\special{pa 3600 1830}%
\special{pa 3600 2230}%
\special{fp}%
\special{sh 1}%
\special{pa 3600 2230}%
\special{pa 3620 2164}%
\special{pa 3600 2178}%
\special{pa 3580 2164}%
\special{pa 3600 2230}%
\special{fp}%
\special{pa 2604 564}%
\special{pa 2604 564}%
\special{fp}%
%
\special{pn 8}%
\special{pa 4212 1030}%
\special{pa 4212 1430}%
\special{dt 0.045}%
\special{sh 1}%
\special{pa 4212 1430}%
\special{pa 4232 1364}%
\special{pa 4212 1378}%
\special{pa 4192 1364}%
\special{pa 4212 1430}%
\special{fp}%
\special{pa 4212 1830}%
\special{pa 4212 2230}%
\special{dt 0.045}%
\special{sh 1}%
\special{pa 4212 2230}%
\special{pa 4232 2164}%
\special{pa 4212 2178}%
\special{pa 4192 2164}%
\special{pa 4212 2230}%
\special{fp}%
\special{pa 3216 564}%
\special{pa 3216 564}%
\special{dt 0.045}%
%
\special{pn 8}%
\special{pa 2140 1200}%
\special{pa 4444 1200}%
\special{fp}%
%
\special{pn 8}%
\special{pa 2140 2000}%
\special{pa 4444 2000}%
\special{dt 0.045}%
%
\special{pn 8}%
\special{pa 4800 1200}%
\special{pa 5000 1200}%
\special{pa 5800 1600}%
\special{pa 5800 1600}%
\special{pa 5800 1600}%
\special{fp}%
%
\special{pn 8}%
\special{pa 4800 2000}%
\special{pa 5000 2000}%
\special{pa 5800 1600}%
\special{pa 5800 1600}%
\special{pa 5800 1600}%
\special{dt 0.045}%
%
\special{pn 8}%
\special{sh 1}%
\special{ar 1920 2000 10 10 0  6.28318530717959E+0000}%
\special{sh 1}%
\special{ar 1980 2000 10 10 0  6.28318530717959E+0000}%
\special{sh 1}%
\special{ar 2040 2000 10 10 0  6.28318530717959E+0000}%
\special{sh 1}%
\special{ar 2040 2000 10 10 0  6.28318530717959E+0000}%
%
\special{pn 8}%
\special{sh 1}%
\special{ar 4560 1200 10 10 0  6.28318530717959E+0000}%
\special{sh 1}%
\special{ar 4620 1200 10 10 0  6.28318530717959E+0000}%
\special{sh 1}%
\special{ar 4680 1200 10 10 0  6.28318530717959E+0000}%
\special{sh 1}%
\special{ar 4680 1200 10 10 0  6.28318530717959E+0000}%
%
\special{pn 8}%
\special{sh 1}%
\special{ar 1920 1200 10 10 0  6.28318530717959E+0000}%
\special{sh 1}%
\special{ar 1980 1200 10 10 0  6.28318530717959E+0000}%
\special{sh 1}%
\special{ar 2040 1200 10 10 0  6.28318530717959E+0000}%
\special{sh 1}%
\special{ar 2040 1200 10 10 0  6.28318530717959E+0000}%
%
\special{pn 8}%
\special{sh 1}%
\special{ar 4550 2000 10 10 0  6.28318530717959E+0000}%
\special{sh 1}%
\special{ar 4610 2000 10 10 0  6.28318530717959E+0000}%
\special{sh 1}%
\special{ar 4670 2000 10 10 0  6.28318530717959E+0000}%
\special{sh 1}%
\special{ar 4670 2000 10 10 0  6.28318530717959E+0000}%
\end{picture}%

\vspace{0.1cm}
\end{figure}

To each time evolution $T_l$, a conserved quantity 
$E_l: {\mathcal P} \rightarrow \Z_{\ge 0}$ is associated.
It is a consequence of the affinization 
of the construction (\ref{eq:tl1}), (\ref{eq:tl2}) and 
Proposition \ref{pr:commute}.
To see this we set $p_i = s_i \ot t_i$,  
$p^\dag_i = s^\dag_i \ot t^\dag_i$ and 
$T_l(p)_i = s'_i \ot t'_i \in B = B_1\ot B^\vee_1$
in (\ref{eq:tl1}) and (\ref{eq:tl2}).
Then Figure \ref{fig:tl} is detailed as Figure \ref{fig:detail}.

In Figure \ref{fig:detail}, we have introduced 
$v_i, v^\sharp_i \in B_l$ such that 
$u_l\ot(\cd\ot p_{i+1}) \simeq (\cd\ot p^\dag_{i+1})\ot v_i$
and $v_i\ot s_i\simeq s^\dag_i\ot v^\sharp_i$.
Similarly, $w_i, w^\sharp_i \in B^\vee_l$ are determined by
$(p^\dag_{i-1}\ot\cd\ot p^\dag_1)\ot\kappa(v) 
\simeq w_i\ot (T_l(p)_{i-1}\ot\cd\ot T_l(p)_1)$ and 
$t^\dag_i\ot w_i \simeq w^\sharp_i\ot t'_i$.

\begin{figure}[h]
\caption{Definition of $E_l(p)$}
\vspace{3cm}
\hspace{6.0cm}
\label{fig:El}

\unitlength 0.1in
\begin{picture}(  0.0000,  0.0000)( 37.0000,-14.0000)
\put(48.0000,-14.0000){\makebox(0,0)[lt]{$z^{I(v)-d}\kappa(v)$}}%
\put(36.5000,-17.3000){\makebox(0,0){$\underbrace{\qquad\qquad\qquad\qquad\quad}$}}%
\put(36.5000,-4.4000){\makebox(0,0){$\overbrace{\qquad\qquad\qquad\qquad\quad}$}}%
\put(38.3000,-11.4000){\makebox(0,0){$p^\dag_2$}}%
\put(42.3000,-11.5000){\makebox(0,0){$p^\dag_1$}}%
\put(38.3000,-5.7000){\makebox(0,0){$p_2$}}%
\put(42.3000,-5.7000){\makebox(0,0){$p_1$}}%
%
\special{pn 8}%
\special{sh 1}%
\special{ar 3546 1570 10 10 0  6.28318530717959E+0000}%
%
\special{pn 8}%
\special{sh 1}%
\special{ar 3136 1570 10 10 0  6.28318530717959E+0000}%
%
\special{pn 8}%
\special{sh 1}%
\special{ar 3340 1570 10 10 0  6.28318530717959E+0000}%
%
\special{pn 8}%
\special{sh 1}%
\special{ar 3546 770 10 10 0  6.28318530717959E+0000}%
%
\special{pn 8}%
\special{sh 1}%
\special{ar 3136 770 10 10 0  6.28318530717959E+0000}%
%
\special{pn 8}%
\special{sh 1}%
\special{ar 3340 770 10 10 0  6.28318530717959E+0000}%
%
\special{pn 8}%
\special{pa 3830 1270}%
\special{pa 3830 1630}%
\special{fp}%
%
\special{pn 8}%
\special{pa 4230 1270}%
\special{pa 4230 1630}%
\special{fp}%
%
\special{pn 8}%
\special{pa 3830 670}%
\special{pa 3830 1030}%
\special{fp}%
%
\special{pn 8}%
\special{pa 4230 670}%
\special{pa 4230 1030}%
\special{fp}%
%
\special{pn 8}%
\special{pa 5630 870}%
\special{pa 5630 1470}%
\special{fp}%
\put(27.9000,-15.4000){\makebox(0,0)[rb]{$z^{-E_l(p)+I(u_l)}u^\vee_l$}}%
\put(36.6000,-2.9000){\makebox(0,0){$p$}}%
\put(36.6000,-19.2000){\makebox(0,0){$T_l(p)$}}%
\put(27.9000,-9.3000){\makebox(0,0)[rb]{$z^0u_l$}}%
\put(49.4000,-9.6000){\makebox(0,0)[lb]{$z^dv$}}%
%
\special{pn 8}%
\special{pa 2820 870}%
\special{pa 4420 870}%
\special{pa 5620 1170}%
\special{pa 5620 1170}%
\special{pa 5620 1170}%
\special{fp}%
%
\special{pn 8}%
\special{pa 2820 1470}%
\special{pa 4420 1470}%
\special{pa 5620 1170}%
\special{pa 5620 1170}%
\special{pa 5620 1170}%
\special{dt 0.045}%
%
\special{pn 8}%
\special{pa 2860 1470}%
\special{pa 2820 1470}%
\special{dt 0.045}%
\special{sh 1}%
\special{pa 2820 1470}%
\special{pa 2888 1490}%
\special{pa 2874 1470}%
\special{pa 2888 1450}%
\special{pa 2820 1470}%
\special{fp}%
\end{picture}%

\vspace{1cm}
\end{figure}
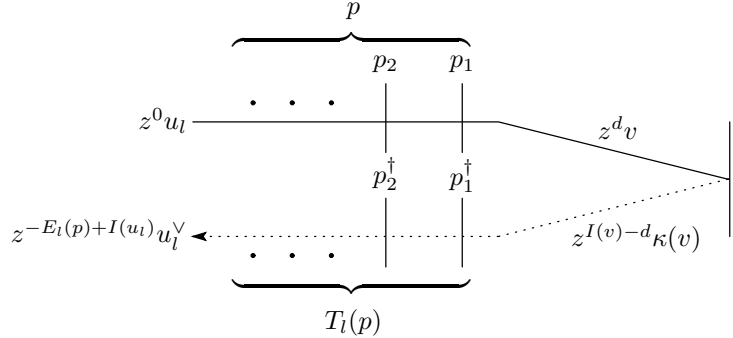

Now consider the affinization of Figure \ref{fig:detail}.
We replace $u_l \in B_l$ by attaching the 
spectral parameter as $z^0u_l \in \aff{B_l}$ 
and keep track of it along the arrow, namely the affinization of
$\ldots, v_i, v_{i-1},\ldots, v_0, w_1, w_2, \ldots w_i, \ldots$.
Due to the boundary condition and the property (\ref{eq:uiso}), 
as the classical part of 
$w_i$ converges to $u^\vee_l$ for $i$ large, 
the spectral parameter also tends to a finite value.
We define $E_l(p) \in \Z$ by saying that 
$z^0u_l$ comes back as 
$z^{-E_l(p)+I(u_l)}u^\vee_l \in \aff{B^\vee_l}$ as in 
Figure \ref{fig:El}.

Figure \ref{fig:detail} is useful to write down a formula
for $E_l(p)$.
{}From (\ref{eq:R}) and (\ref{eq:Rv}) we have
$d = \sum_{i \ge 1}(Q_0(v_i,s_i)+ P_0(v^\sharp_i,t_i))$ 
in Figure \ref{fig:El}.
Combining this with a similar calculation 
in the bottom horizontal arrow we obtain
\begin{equation}\label{eq:El}
E_l(p) = \sum_{i\ge 1}(Q_0(v_i,s_i)+ P_0(v^\sharp_i,t_i))
+ \sum_{i\ge 1}(Q_0(w_i,t^\dag_i)+ P_0(w^\sharp_i,s^\dag_i))
- I(v) + I(u_l).
\end{equation}
Here $I(u_l) = -l$ (Rotateleft, Switch$_{1n}$), 
$= 0$ (Switch$_{12}$) is a normalization constant 
for a later convenience.

\begin{proposition}\label{pr:El}
$E_l$ is invariant under time evolutions, i.e.,
$E_l(T_m(p)) = E_l(p)$ for any $l, m \in \Z_{\ge 1}$ 
and $p \in {\mathcal P}$.
\end{proposition}
\begin{proof}
We consider the affinization of the series of identities 
depicted in Figure \ref{fig:commute}, which makes sense 
since the Yang-Baxter and the reflection equations are both valid 
in the affine setting. 
In the first diagram, let the affinization of 
$u_l$ and $u_m$ be $z^0u_l$ and $z^0u_m$, respectively.
Then by definition $u^\vee_l$ and $u^\vee_m$ are 
replaced by $z^{-E_l(p)+I(u_l)}u^\vee_l$ and 
$z^{-E_m(T_l(p))+I(u_m)}u^\vee_m$, respectively.
In the diagrams in Figure \ref{fig:commute}, 
these four elements on the left change their positions only
because of (\ref{eq:uiso}).
Therefore in the last diagram they are aligned as 
$z^0u_m, \, z^{-E_m(T_l(p))+I(u_m)}u^\vee_m,\, z^0u_l$
and $z^{-E_l(p)+I(u_l)}u^\vee_l$ {}from the top to the bottom.
On the other hand, the last diagram itself tells that
if the first and the third ones
are chosen as $z^0u_m$ and $z^0u_l$,
the second and the fourth ones should be 
$z^{-E_m(p)+I(u_m)}u^\vee_m$ and 
$z^{-E_l(T_m(p))+I(u_l)}u^\vee_l$, respectively.
Therefore we conclude $E_l(T_m(p)) = E_l(p)$ 
(and $E_m(T_l(p)) = E_m(p)$ as well).
\end{proof}

\begin{remark}\label{rem:El}
Given a state $p$, $E_l(p) = E_{l_0}(p)$ holds for any $l \ge l_0$  
with the same $l_0$ as in Proposition \ref{pr:saturation}.
\end{remark}

\begin{example}\label{ex:El}
$E_l$ in the previous examples reads
$$
\begin{array}{ccccccc}
& E_1 & E_2 & E_3 & E_4 & \ldots \\
\mbox{Example } \ref{ex:ex1}: & 1 & 2 & 3 & 3 & \ldots\\
\mbox{Example } \ref{ex:ex2}: & 2 & 3 & 4 & 4 & \ldots\\
\mbox{Example } \ref{ex:ex3}: & 6 & 10 & 10 & 10 & \ldots
\end{array}
$$
In Example \ref{ex:ex1}, $E_3=3$ can be read off Figure \ref{fig:ex1}.
Only those vertices marked with $\circ$ 
contribute to the $i-$sums in (\ref{eq:El}) each by 1, adding up to 5. 
Together with $I(v) = I(122)= -1$ and $I(u_3) = -3$, one finds
$E_3 = 5+1-3=3$.
Similarly in Example \ref{ex:ex2}, the $i-$sums 
for $E_2$ (\ref{eq:El}) equal to 4 
from the marked vertices in Figure \ref{fig:ex2}.
Therefore $E_2 = 4-I(12)+I(u_2) = 4-(-1)+(-2) = 3$.
\end{example}

\subsection{Asymptotic $\mathfrak{sl}_{n-2}$-invariance}
\label{subsec:asy}

Suppose a Kashiwara operator $\ft{i}$ act on 
the defining relations (\ref{eq:tl1}) and (\ref{eq:tl2}) of $T_l$ as
\begin{equation}\label{eq:scheme}
\begin{array}{rclrcl}
\ft{i}(u_l\ot p) &\simeq & \ft{i}(p^\dag\ot v),
\qquad \quad 
\ft{i}(p^\dag\ot\kappa(v)) &\simeq &\ft{i}(u^\vee_l\ot T_l(p)),
\\
|| \;\;\quad &&\quad \quad || \;\;\quad\qquad\qquad\qquad||&& ||\\
u_l\ot\ft{i}p &\simeq &\;\;\ft{i}p^\dag\ot v,
\qquad \quad \;\;
\ft{i}p^\dag\ot\kappa(v) &\simeq &u^\vee_l\ot\ft{i}T_l(p).
\end{array}
\end{equation}
Then by definition the bottom line implies the commutativity
$T_l(\ft{i}p) = \ft{i}T_l(p)$.
The same argument holds also for $\et{i}$.
The maximal set of Kashiwara operators $\{\et{i}, \ft{i}\}$
commuting with the time evolutions is the fundamental data
to characterize the quantum group symmetry of the system.
In the scheme (\ref{eq:scheme}), the leftmost and the rightmost equalities 
(for $\et{i}$ as well) are automatically satisfied if 
\begin{equation}\label{eq:set}
\begin{split}
&i \in \{ 2, 3, \ldots, n-2 \} \quad \mbox{ for } 
\kappa = \mbox{Rotateleft}, \; \mbox{Switch}_{1n},\\
&i \in \{ 3, 4, \ldots, n-1 \} \quad \mbox{ for } 
\kappa = \mbox{Switch}_{12}.
\end{split}
\end{equation}
On the other hand the middle two equalities in (\ref{eq:scheme}) 
are essentially dependent on $p^\dag$ and $v$, and are not 
guaranteed by the simple condition (\ref{eq:set}).
To remedy this, we give up coping with the full set of states 
${\mathcal P}$ but confine ourselves to the subset 
\begin{equation}\label{eq:pasy}
\pa = \{p \in {\mathcal P} \mid 
u_l \ot p \simeq p^\dag \ot u_l \mbox{ for any }l
\mbox{ with some ($l$-dependent) } p^\dag \in {\mathcal P}\}.
\end{equation}
We call an element of $\pa$ an {\em asymptotic} state.
In view of Lemma \ref{lem:converge} (i), the states whose 
deviation from $\cd\ot \vac \ot \vac$ is 
sufficiently far from the right end of the lattice are 
asymptotic states.
In (\ref{eq:pasy}) there are actually finitely many conditions on $p$,  
because for sufficiently large $l$ they become equivalent owing to 
the saturation property explained in the proof of 
Proposition \ref{pr:saturation}.
The set $\pa$ is not stable under the time evolutions
$\{T_l\}$.
For $p \in \pa$, one has $v=u_l$ and $\kappa(v) = u^\vee_l$ in 
(\ref{eq:scheme}), hence all the equalities therein are valid
under the choice (\ref{eq:set}).
To summarize we have shown
\begin{proposition}[asymptotic ${\mathfrak{sl}}_{n-2}$-invariance]
\label{pr:inv}
For any asymptotic state $p \in \pa$ and 
$i$ specified in (\ref{eq:set}), the commutativity
$T_l(\et{i}p) = \et{i}T_l(p),\,T_l(\ft{i}p) = \ft{i}T_l(p)$
is valid.
\end{proposition}

Note that the choice of the $\widehat{\mathfrak{sl}}_{n-2}$ subalgebra 
is dependent on $\kappa$.
For a comparison, the box-ball system without a boundary
associated with $U_q(\widehat{\mathfrak{sl}}_{n})$ \cite{T} 
possesses $\mathfrak{sl}_{n-1}$-invariance. 
See \cite{HHIKTT,FOY,HKOTY}.

\begin{remark}
The identity $\varepsilon_i(\kappa(v)) = \varepsilon_i(v) (= v_{i+1})$ 
is valid for $\kappa = $Rotateleft (any $i$), Switch$_{1n}$ (even $i$) and 
Switch$_{12}$ (odd $i$).
Therefore for these $i$ further satisfying (\ref{eq:set}), 
the commutativity 
$T_l(\et{i}p) = \et{i}T_l(p)$ (resp. $T_l(\ft{i}p) = \ft{i}T_l(p)$) holds
under the condition $\varphi_i(p^\dag) \ge v_{i+1}$ 
(resp. $\varphi_i(p^\dag) > v_{i+1}$) in (\ref{eq:scheme}).
There are such $p$'s not belonging to $\pa$.
\end{remark}

\begin{example}
A commutative diagram exemplifying 
the asymptotic $\mathfrak{sl}_2$-invariance in the 
$\widehat{\mathfrak{sl}}_4$ automaton with 
$\kappa = $Switch$_{14}$ ($.. = 1\ol{4}$).
\begin{align*}
..\;\;..\;\; 1\ol{4}\;\; 2\ol{1}\;\; 3\ol{2} \;\; 1\ol{4}\;\; 1\ol{4} 
\quad\; &\stackrel{\et{2}}{\longmapsto}\quad 
..\;\;..\;\; 1\ol{4} \;\; 2\ol{1}\;\; 3\ol{3}\;\; 1\ol{4}\;\; 1\ol{4} 
\vspace{0.15cm}\\
\vspace{0.15cm}
T_2 \;\downarrow \qquad\quad & \qquad\qquad\qquad\downarrow\;  T_2\\
..\;\;..\;\; 1\ol{4}\;\; 4\ol{3}\;\; 1\ol{4}\;\; 3\ol{4}\;\; 3\ol{4} 
\quad\; &\stackrel{\et{2}}{\longmapsto}\quad 
..\;\;..\;\; 1\ol{4}\;\; 4\ol{3}\;\; 1\ol{4}\;\; 3\ol{4}\;\; 2\ol{4} 
\end{align*}
The states on the top line are asymptotic states.
They exhibit the same evolution pattern 
under $T_l$ with any $l \ge 2$.
\end{example}

\section{Solitons}\label{sec:soliton}

By a soliton we mean some localized pattern of local states.
Under time evolutions it remains stable 
as long as it is surrounded by sufficiently many 
$\vac$ states.
Here we formulate solitons in our automaton and 
study their scattering and reflection rule.
We demonstrate the results mainly with examples and include only 
sketchy proofs, for they are straightforward or quite similar to 
those for the infinite systems \cite{HKT1,FOY,HKOTY}.

In this section we let $B'_l$ and  $B'^\vee_l$ 
denote the crystals $B_l$ and $B^\vee_l$ for 
$U_q(\widehat{\mathfrak{sl}}_{n-2})$.

\subsection{Basic feature of solitons}

Example \ref{ex:ex1} illustrates a reflection of a soliton at the end.  
There are two kinds of solitons, one moving to the right and 
the other one moving to the left.
In order to describe them we introduce an injection
$$
\imath_l: B'_l \sqcup B'^\vee_l \rightarrow B^{\ot l}
$$
for each $l \in \Z_{\ge 1}$.
In terms of the tableau notation 
$x = i_1i_2 \ldots i_{l} \in B'_l$  
or
$x = \ol{i_1}\ol{i_2} \ldots \ol{i_{l}} \in B'^\vee_l$  
with $1 \le i_1 \le \cd \le i_l \le n-2$, 
it is defined by
\begin{align*}
&\kappa = \mbox{Rotateleft, Switch}_{1n}:\\
&\quad \imath_l(x) = \begin{cases}
\bigl((i_l+1)\ot\ol{n}\bigr) \ot\cd\ot 
\bigl((i_2+1)\ot\ol{n}\bigr) \ot \bigl((i_1+1)\ot\ol{n}\bigr)
&\mbox{for } x \in B'_l,\\
\bigl(1\ot\ol{i_1+1}\bigr)\ot
\bigl(1\ot\ol{i_2+1}\bigr) \ot\cd\ot 
\bigl(1\ot\ol{i_l+1}\bigr)
&\mbox{for }  x \in B'^\vee_l.
\end{cases}\\
&\kappa = \mbox{Switch}_{12}:\\
&\quad \imath_l(x) = \begin{cases}
\bigl((i_l+2)\ot\ol{2}\bigr) \ot\cd\ot 
\bigl((i_2+2)\ot\ol{2}\bigr) \ot \bigl((i_1+2)\ot\ol{2}\bigr)
&\mbox{for } x \in B'_l,\\
\bigl(1\ot\ol{i_1+2}\bigr)\ot
\bigl(1\ot\ol{i_2+2}\bigr) \ot\cd\ot 
\bigl(1\ot\ol{i_l+2}\bigr)
&\mbox{for }  x \in B'^\vee_l.
\end{cases}
\end{align*}

A state of the form 
\begin{equation}\label{eq:m-soliton}
......[l_1]........[l_2]..... \cdots .....[l_m]....... 
\end{equation}
is an example of $m$-soliton states of
length $l_1,l_2,\ldots,l_m$ defined generally in the sequel.
Here $..[l]..$ denotes a local configuration such as
\begin{equation}\label{eq:1-soliton}
\cd\ot \vac \ot \vac \ot \imath_l(x) \ot \vac \ot \vac \ot\cd
\quad \mbox{ for some } x \in B'_l \sqcup B'^\vee_l
\end{equation}
surrounded by sufficiently many $\vac$'s.
It is called a right (resp. left) soliton if $x \in B'_l$ 
(resp. $x \in B'^\vee_l$).
We call $x$ the {\em label} of a soliton.
(For (\ref{eq:m-soliton}) to be an $m$-soliton state, 
it is sufficient (but not necessary in general) 
to assume there are at least 
$\vac^{\ot l_i}$ between $[l_i]$ and $[l_{i+1}]$ 
if they are both right solitons, 
$\vac^{\ot l_{i+1}}$ if the both are left solitons, and 
$\vac^{\ot l_i+l_{i+1}}$ if $[l_i]$ is a right 
and $[l_{i+1}]$ is a left soliton.)

According to these terminologies, Example \ref{ex:ex1} is 
the reflection of a length 3 soliton 
with label $113$ into $\ol{233}$.

By a direct calculation one can check
\begin{lemma} \label{lem:1-soliton}
Let $p$ be the length $l$ one-soliton state (\ref{eq:1-soliton}).

\begin{itemize}
\item[(1)]
The $k$-th conserved quantity of $p$ is given by $E_k(p)=\min(k,l)$.
\item[(2)]
Under the evolution $p \rightarrow  T_k(p)$, 
the right (resp. left) soliton moves to the right (resp. left)
by $\min(k,l)$ lattice steps.
(For a right soliton we assume there are at least 
$\vac^{\ot \min(k,l)}$ 
in its right until the end.)
\end{itemize}
\end{lemma}

For any state $p$, define the numbers $N_l=N_l(p)$ ($l=1,2,\ldots$) by 
\begin{equation}
E_k(p)=\sum_{l \geq 1} \min(k,l) N_l
\end{equation}
By definition $\{N_l\}$ is a set of conserved quantities  
such that $N_l = 0$ for sufficiently large $l$.  
See also Remark \ref{rem:El}.

Motivated by the property of individual solitons 
in Lemma \ref{lem:1-soliton}
and apparent independence of sufficiently separated solitons, 
we define the number of length $l$ solitons in $p$ 
to be $N_l$ for $l=1,2, \ldots$.
The total number of solitons is $N_1 + N_2 + \cdots = E_1$.
The general definition of $m$-soliton states is 
those $p$ satisfying $E_1(p) = m$, 
and the state (\ref{eq:m-soliton}) is such an example.
In fact it is not difficult to show
\begin{lemma}\label{lem:E1=1}
Suppose a state of the form 
$p = \cd\ot\vac\ot\vac \ot b \ot 
\stackrel{L}{\overbrace{\vac \ot \cd \ot \vac}}$ with 
$b \in B^{\ot l}$ and $L \gg l$ satisfies the 
one-soliton condition $E_1(p) = 1$.
Then $b = \imath_l(x)$ for some $x \in B'_l \sqcup B'^\vee_l$.
\end{lemma}
 
\begin{example}\label{ex:Nl}
Example \ref{ex:El} is translated into $\{N_l\}$ as
$$
\begin{array}{ccccccl}
& N_1 & N_2 & N_3 & N_4 & \ldots & \quad \mbox{label of solitons}\\
\mbox{Example } \ref{ex:ex1}: & 0 & 0 & 1 & 0 & \ldots & \quad 
\ol{233}\; (t\!=\!4)\\
\mbox{Example } \ref{ex:ex2}: & 1 & 0 & 1 & 0 & \ldots & \quad 
\ol{112},\, \ol{2}\; (t\!=\!7)\\
\mbox{Example } \ref{ex:ex3}: & 2 & 4 & 0 & 0 & \ldots & \quad 
\ol{12},\, \ol{1},\, 1,\, 22,\, \ol{22},\, 22\; (T_1^4 \mbox{ case})
\end{array}
$$
\end{example}

Labels of solitons are not conserved quantities.
It is our subject in the following subsections 
to determine the transformation rule of labels combined with 
another important data, phase, 
under collisions and reflection.

\subsection{Scattering data}\label{subsec:sd}

Let us introduce the {\em position} and the {\em phase} of a soliton.
These are the data defined only for those states where 
solitons are enough separated as in (\ref{eq:m-soliton}).
Consider a length $l$ label $x$ soliton 
\begin{equation*}
\cd \ot \vac \ot \vac \ot 
\stackrel{\imath_l(x)}
{\overbrace{p_{i+l} \ot \cd \ot p_{i+2} \ot p_{i+1}}}
\ot \vac \ot \vac \ot \cd
\end{equation*}
appearing as a portion of the state $p = \cd \ot p_2 \ot p_1$.
See Figure \ref{fig:lattice}.
We define its position to be $i+1$ if it is a right soliton and 
$1-i-l$ for a left soliton.
By the definition, the position of the length $l$ soliton, 
either it is a right or left one, always decreases by $\min(k,l)$ 
under $T_k$ from Lemma \ref{lem:1-soliton} (2) as long as it remains 
isolated. 
Therefore within such a time interval, 
the position $\gamma(t)$ of the soliton at time $t$ should be 
written as $\gamma(t) = -\min(k,l)t + d$ for some $d$.
This constant $d$ is called the phase of the soliton 
during the time interval in question.
The arbitrariness of $d$ due to the freedom $t \rightarrow t + \mbox{const}$ 
does not matter since we will actually be concerned only with 
the phase shift from one time interval to another.

To the one-soliton state 
$...[l]...$ containing a length $l$ soliton 
with label $x \in B'_l \sqcup B'^\vee_l$ 
and phase $d$, we assign the {\em scattering data}:
\begin{align*}
z^dx &\in \aff{B'_l} \;\;\;\; \mbox{ for a right soliton},\\
z^{-d}x &\in \aff{B'^\vee_l} \;\; \mbox{ for a left soliton}.
\end{align*}
Similarly the scattering data for the $m$-soliton state 
(\ref{eq:m-soliton}) is defined by
\begin{equation}\label{eq:sd}
z^{\pm d_1}x_1 \ot z^{\pm d_2}x_2 \ot\cd\ot z^{\pm d_m}x_m
\in \aff{B'^{\pm}_{l_1}}\ot
\aff{B'^{\pm}_{l_2}}\ot\cd\ot\aff{B'^{\pm}_{l_m}},
\end{equation}
where $B'^+_l = B'_l, B'^-_l = B'^\vee_l$ and the 
$\pm$ symbols are to be taken $+\,(-)$ for right (left) solitons.

\begin{example}
$\widehat{\mathfrak{sl}}_5$, $\kappa = $Rotateleft. 
The symbol $..$ stands for $1\ol{5}$.
Time evolution $T^t_2(p)$ for $t=0,1$ is given. ($p=$ the middle line.)
The top line shows the coordinates of the lattice.
$$
\begin{array}{cccccccccccccccccccccccccccccc}
\qquad\qquad 17 & 16 & 15 & 14 & 13 & 12 & 11 & 
10 & 9 & 8 & 7 & 6 & 5 & 4 & 3 & 2 &1 \\
\hline\vspace{-0.2cm}\\ 
t\!=\!0: \quad..\s 4\ol{5}\s 3\ol{5}\s 2\ol{5}\s ..\s ..\s 
4\ol{5}\s 2\ol{5}\s ..\s ..\s ..\s ..\s 1\ol{3}\s ..\s ..\s 1\ol{2}\s 1\ol{4}\\
t\!=\!1: \quad..\s ..\s ..\s 4\ol{5}\s 3\ol{5}\s 2\ol{5}\s 
..\s ..\s 4\ol{5}\s 2\ol{5}\s ..\s 1\ol{3}\s ..\s 1\ol{2}\s 1\ol{4}\s ..\s ..\\
\end{array}
$$
The number of solitons are $N_1 = 1,\, N_2 = 2$ and $N_3 = 1$.
Under $T_2$, phases $d$ are determined by fitting the positions to the formula 
$-t+d$ for the length 1 soliton and $-2t+d$ for the other solitons.
The resulting scattering data is 
$$
z^{14}123 \ot z^{10}13 \ot z^4\, \ol{2} \ot z^1\, \ol{13}
$$ 
for the both $t=0$ and $t=1$ states.
\end{example}

\subsection{Reflection rule}\label{subsec:rr}

Let $d_1$ and $d_2$ be the phases of a soliton 
before and after a reflection, respectively.
We define the phase shift of the reflection to be $d_2-d_1$.

\begin{example}\label{ex:phase}
In Example \ref{ex:ex1}, the position $-3t+d$ and the phase $d$ of the 
length 3 soliton are given as follows:
$$
\begin{array}{cccccccccccccc}
t & \mbox{position} & \mbox{phase} \\
0 & 5 & 5 \\
1 & 2 & 5 \\
2 &   & \\
3 &  -2 & 7 \\
4 &  -5 & 7
\end{array}
$$
The phase shift of this reflection is $7-5=2$.
At $t=2$, neither data are defined nor needed to determine the 
phase shift.
\end{example}

Suppose a length $l$ soliton with label $x \in B'_l$ and 
phase $d$ is transformed into 
the one with label $y \in B'^\vee_l$ and phase $d+\delta$ 
by the reflection.
We introduce the following map on the 
scattering data that integrates the reflection rule:
\begin{equation}\label{eq:L}
\begin{split}
L: \;\; \aff{B'_l} \; &\longrightarrow \aff{B'^\vee_l}\\
z^d x \;\;&\longmapsto \; z^{-\delta-d}y.
\end{split}
\end{equation}
We call $L$ the {\em reflection operator}.
(The letter $L$ is chosen as the next one to $K$.)
\begin{example}\label{ex:Lrl}
In view of Example \ref{ex:phase},
the reflection in Example \ref{ex:ex1} 
($\widehat{\mathfrak{sl}}_{5}$, $\kappa=$ Rotateleft) is expressed as
$L(z^5 113) = z^{-7}\, \ol{233}$.
\end{example}

\begin{theorem}\label{th:L}
Let $K'_{\rm Rotateleft}$ and $K'_{{\rm Switch}_{12}}$ 
denote the $K$ defined by (\ref{eq:rl}) and (\ref{eq:12}) for  
$\widehat{\mathfrak{sl}}_{n-2}$, i.e., those with 
$n$ replaced by $n\!-\!2$, 
respectively.
Then the reflection operator is given by
\begin{equation*}
L = \begin{cases}
K'_{\rm Rotateleft}  & \mbox{ for } \kappa = \mbox{Rotateleft},\\
K'_{{\rm Switch}_{12}} & \mbox{ for } 
\kappa = \mbox{Switch}_{1n}, \, \mbox{Switch}_{12}.
\end{cases}
\end{equation*}
In particular, the phase shift in (\ref{eq:L}) 
is given by $\delta = -I(x)$.
\end{theorem}
\begin{proof}
We give the proof only for $\kappa=$Rotateleft.
By Proposition \ref{pr:commute} and Lemma \ref{lem:1-soliton} (2) one can
assume that at $t=0$
our right soliton is given by
\[
p=\vac^{\ot L}\ot i_l\ol{n} \ot \cd \ot i_2\ol{n} \ot i_1\ol{n},
\]
where $1<i_1\le i_2\le\cd\le i_l<n$ and $L$ is sufficiently large.
Set $a=\sharp\{s\mid i_s=2\}$. A direct calculation shows
\[
T_l(p)=\left\{
\begin{array}{l}
\vac^{\ot(L+a)}\ot 1\ol{i_{a+1}-1}\ot\cd\ot 1\ol{i_{l-a}-1}
\ot n\ol{i_{l-a+1}-1}\ot\cd\ot n\ol{i_l-1}\\
\hspace{7.5cm}\mbox{if }l-2a\ge0,\\
\vac^{\ot(L+\frac{l}2)}\ot(n\ol{1})^{\ot\frac{2a-l}2}
\ot n\ol{i_{a+1}-1}\ot\cd\ot n\ol{i_l-1}\\
\hspace{7.5cm}\mbox{if }l-2a<0,\;l\mbox{ is even},\\
\vac^{\ot(L+\frac{l-1}2)}\ot 1\ol{1}\ot(n\ol{1})^{\ot\frac{2a-l-1}2}
\ot n\ol{i_{a+1}-1}\ot\cd\ot n\ol{i_l-1}\\
\hspace{7.5cm}\mbox{if }l-2a<0,\;l\mbox{ is odd},
\end{array}
\right.
\]
\[
T_l^2(p)=
\vac^{\ot(L-l+a)}\ot 1\ol{i_{a+1}-1}\ot\cd\ot 1\ol{i_l-1}
\ot(1\ol{n-1})^{\ot a}\ot\vac^{\ot(l-a)}.
\]
Under our notation this reflection is written as
\[
L:\quad z^1\cdot 1^a(i_{a+1}-1)\cd(i_l-1)\longmapsto
z^{-(1+a)}\cdot (i_{a+1}-2)\cd(i_l-2)(n-2)^a.
\]
This agrees with $K'_{\rm Rotateleft}$ as desired.
\end{proof}

In Example \ref{ex:Lrl}, $L=K'_{\rm Rotateleft}$ indeed holds 
as $I(113) = -2$.
Since $I=0$ in (\ref{eq:12}), no phase shift takes place 
for $\kappa =$ Switch$_{1n}$ and Switch$_{12}$.
The following two are such examples.

\begin{example}\label{ex:L1n}
$\widehat{\mathfrak{sl}}_6$, $\kappa =$Switch$_{16}$.
Time evolution under $T^t_3$ for $0 \le t \le 4$.
The symbol $..$ stands for $1\ol{6}$.
$$
\begin{array}{cccccccccccccccccccccccc}
0:\quad ..\s..\s5\ol{6}\s4\ol{6}\s4\ol{6}\s2\ol{6}\s2\ol{6}\s..\s..\s..\\
1:\quad ..\s..\s..\s..\s..\s5\ol{6}\s4\ol{6}\s4\ol{6}\s2\ol{6}\s2\ol{6}\\
2:\quad ..\s..\s..\s..\s..\s..\s..\s1\ol{3}\s5\ol{3}\s4\ol{5}\\
3:\quad ..\s..\s..\s..\s1\ol{3}\s1\ol{3}\s1\ol{4}\s1\ol{5}\s1\ol{5}\s..\\
4:\quad ..\s1\ol{3}\s1\ol{3}\s1\ol{4}\s1\ol{5}\s1\ol{5}\s..\s..\s..\s..
\end{array}
$$  
Fitting the positions of the length $5$ soliton to $-3t+d$, 
we find the phase remains $d=4$ before and after the reflection.
Hence $L(z^4 11334) = z^{-4}\, \ol{22344}$, which agrees with 
$K'_{{\rm Switch}_{12}}(z^4 11334)$.
\end{example}

\begin{example}\label{ex:L12}
$\widehat{\mathfrak{sl}}_6$, $\kappa =$Switch$_{12}$.
Time evolution under $T^t_4$ for $0 \le t \le 4$.
The symbol $..$ stands for $1\ol{2}$.
$$
\begin{array}{cccccccccccccccccccccccc}
0:\quad ..\s6\ol{2}\s5\ol{2}\s5\ol{2}\s3\ol{2}\s3\ol{2}\s3\ol{2}
\s..\s..\s..\s..\s..\s..\\
1:\quad ..\s..\s..\s..\s..\s6\ol{2}\s5\ol{2}\s5\ol{2}\s3\ol{2}
\s3\ol{2}\s3\ol{2}\s..\s..\\
2:\quad ..\s..\s..\s..\s..\s..\s..\s..\s..\s6\ol{2}\s5\ol{2}
\s5\ol{4}\s3\ol{4}\\
3:\quad ..\s..\s..\s..\s..\s..\s..\s1\ol{4}\s1\ol{4}\s1\ol{4}
\s1\ol{5}\s1\ol{6}\s1\ol{6}\\
4:\quad ..\s..\s..\s1\ol{4}\s1\ol{4}\s1\ol{4}\s1\ol{5}\s1\ol{6}
\s1\ol{6}\s..\s..\s..\s..
\end{array}
$$
Fitting the positions of the length $6$ soliton to $-4t+d$, 
we find the phase remains $d=7$ before and after the reflection.
Hence $L(z^7 111334) = z^{-7}\, \ol{222344}$, which agrees with 
$K'_{{\rm Switch}_{12}}(z^7 111334)$.
\end{example}

\subsection{Scattering rule}\label{subsec:sr}

Here we consider the scattering among solitons which takes place 
far from the end without a boundary effect.
The result is the same for all the choices 
$\kappa =$Rotateleft, Switch$_{1n}$ and Switch$_{12}$ 
when expressed in terms of the scattering data (\ref{eq:sd}).
(All the examples in this subsection are taken from the first two.)
In particular the scattering involving only right solitons 
are essentially the same as the infinite system 
studied in \cite{HHIKTT,FOY,TNS}.

Let us observe scattering of two solitons, 
which is the most fundamental case. 

\begin{example}\label{ex:rr}
$\widehat{\mathfrak{sl}}_5$. Scattering of length 3 and 2 right 
solitons under $T_l$ with any $l \ge 3$.
$$
\begin{array}{ccccccccccccccccccccccccccccc}
\cd \;  ..\s4\ol{5}\s3\ol{5}\s2\ol{5}\s..\s..\s3\ol{5}\s2\ol{5}
\s..\s..\s..\s..\s..\s..\s..\s..\s..\; \cd\\
\cd \;  ..\s..\s..\s..\s4\ol{5}\s3\ol{5}\s2\ol{5}\s..\s3\ol{5}
\s2\ol{5}\s..\s..\s..\s..\s..\s..\s..\; \cd\\
\cd \;  ..\s..\s..\s..\s..\s..\s..\s4\ol{5}\s2\ol{5}\s..\s3\ol{5}
\s3\ol{5}\s2\ol{5}\s..\s..\s..\s..\; \cd\\
\cd \;  ..\s..\s..\s..\s..\s..\s..\s..\s..\s4\ol{5}\s2\ol{5}
\s..\s..\s3\ol{5}\s3\ol{5}\s2\ol{5}\s..\; \cd
\end{array}
$$
The phases, say $d_1$ and $d_2$, of the larger and smaller 
solitons before the collision
have been changed into $d_1-3$ and $d_2+3$, respectively.
In terms of the scattering data the event is expressed as 
$$
z^{d_1}123 \ot z^{d_2}12 \; \mapsto z^{d_2+3}13 \ot z^{d_1-3}122.
$$
\end{example}

\begin{example}\label{ex:rl}
$\widehat{\mathfrak{sl}}_5$. Scattering of a length 3 right and 
a length 4 left soliton under $T_l$ with any $l \ge 4$.
$$
\begin{array}{ccccccccccccccccccccccccccccc}
\cd \;  ..\s..\s..\s..\s3\ol{5}\s2\ol{5}\s2\ol{5}
\s..\s..\s..\s..\s1\ol{2}\s1\ol{2}\s1\ol{3}\s1\ol{4}\s..\; \cd \\
\cd \;  ..\s..\s..\s..\s..\s..\s..\s3\ol{1}\s2\ol{2}\s1\ol{3}
\s1\ol{4}\s..\s..\s..\s..\s..\; \cd \\
\cd \;  ..\s..\s..\s..\s..\s1\ol{2}\s1\ol{4}\s1\ol{4}\s5\ol{5}
\s4\ol{5}\s2\ol{5}\s..\s..\s..\s..\s..\; \cd \\
\cd \;  ..\s1\ol{2}\s1\ol{4}\s1\ol{4}\s1\ol{4}\s..\s..\s..\s..
\s..\s..\s4\ol{5}\s4\ol{5}\s2\ol{5}\s..\s..\; \cd
\end{array}
$$
The phases, say $d_1$ and $d_2$, of the right and left 
solitons before the collision
have been changed into $d_1+2$ and $d_2+2$, respectively.
In terms of the scattering data the event is expressed as 
$$
z^{d_1}112 \ot z^{-d_2}\,\ol{1123} \;
\mapsto z^{-d_2-2}\,\ol{1333} \ot z^{d_1+2}133.
$$
The same result is valid under $T_l$ with any $l \ge 1$.
\end{example}

\begin{example}\label{ex:ll}
$\widehat{\mathfrak{sl}}_5$. Scattering of a length 2 and 4 
left solitons under $T_l$ with any $l \ge 4$.
$$
\begin{array}{ccccccccccccccccccccccccccccc}
\cd \;  ..\s..\s..\s..\s..\s..\s..\s..\s..\s..\s1
\ol{2}\s1\ol{3}\s..\s..\s..\s..\s..\s1\ol{3}\s1
\ol{3}\s1\ol{4}\s1\ol{4}\s..\; \cd \\
\cd \;  ..\s..\s..\s..\s..\s..\s..\s..\s1\ol{2}\s1
\ol{3}\s..\s..\s..\s1\ol{3}\s1\ol{3}\s1\ol{4}\s1
\ol{4}\s..\s..\s..\s..\s..\; \cd \\
\cd \;  ..\s..\s..\s..\s..\s1\ol{2}\s1\ol{3}\s1
\ol{3}\s..\s..\s1\ol{3}\s1\ol{4}\s1\ol{4}
\s..\s..\s..\s..\s..\s..\s..\s..\s..\; \cd \\
\cd \;  ..\s1\ol{2}\s1\ol{3}\s1\ol{3}\s1\ol{3}
\s..\s..\s..\s1\ol{4}\s1\ol{4}\s..\s..\s..\s..
\s..\s..\s..\s..\s..\s..\s..\s..\; \cd
\end{array}
$$
The phases, say $d_1$ and $d_2$, of the smaller and larger 
solitons before the collision
have been changed into $d_1+4$ and $d_2-4$, respectively.
In terms of the scattering data the event is expressed as 
$$
z^{-d_1}\,\ol{12} \ot z^{-d_2}\,\ol{2233}\; 
\mapsto z^{-d_2+4}\,\ol{1222} \ot z^{-d_1-4}\,\ol{33}.
$$
The same result is valid also under $T_3$.
\end{example}

In general suppose that 
the initially enough separated two solitons with 
length $l, m$, labels $x, y$ and phases $d_1, d_2$ 
are scattered into those with labels $\tilde{x}, \tilde{y}$ 
and phases $\tilde{d}_1, \tilde{d}_2$, respectively.
\begin{equation*}
 ......\,\imath_l(x)_{d_1}.......\,\imath_m(y)_{d_2}......... \quad 
\longrightarrow \quad 
......\,\imath_m(\tilde{y})_{\tilde{d}_2}...
......\,\imath_l(\tilde{x})_{\tilde{d}_1}.......,
\end{equation*}
where the phases are attached as further indices and $.$ denotes $\vac$.
We define the three kinds of {\em scattering operators} by
\begin{eqnarray}
S: \;\; \aff{B'_l} \ot \aff{B'_m} & \longrightarrow & 
\aff{B'_m} \ot \aff{B'_l} \;\;(l \neq m)\label{eq:S}\\
z^{d_1}x \ot z^{d_2}y & \longmapsto & 
z^{\tilde{d}_2}\tilde{y}\ot z^{\tilde{d}_1}\tilde{x},\nonumber \\
S^\vee: \;\; \aff{B'_l} \ot \aff{B'^\vee_m} & \longrightarrow & 
\aff{B'^\vee_m} \ot \aff{B'_l} \label{eq:Sv}\\
z^{d_1}x \ot z^{-d_2}y & \longmapsto & 
z^{-\tilde{d}_2}\tilde{y}\ot z^{\tilde{d}_1}\tilde{x}, \nonumber \\
S^{\vee\vee}: \;\; \aff{B'^\vee_l} \ot \aff{B'^\vee_m} & \longrightarrow & 
\aff{B'^\vee_m} \ot \aff{B'^\vee_l} \;\;(l \neq m)\label{eq:Svv}\\
z^{-d_1}x \ot z^{-d_2}y & \longmapsto & 
z^{-\tilde{d}_2}\tilde{y}\ot z^{-\tilde{d}_1}\tilde{x}.\nonumber 
\end{eqnarray}
These operators are the fundamental objects that
integrates the two body scattering rule, which is common
under $T_k$ with $k > \min(l,m)$ for $S$ and $S^{\vee\vee}$, and 
any $k \ge 1$ for $S^\vee$.
The action of $S, S^\vee$ and $S^{\vee\vee}$ is illustrated 
in Examples \ref{ex:rr}, \ref{ex:rl} and \ref{ex:ll}, respectively.
The operators $S$ and $S^{\vee\vee}$ with $l=m$ are not determined 
by (\ref{eq:S}) and (\ref{eq:Svv}) 
because solitons of equal length and direction do not collide.
Here we formally define them by declaring that the following 
theorem can be extrapolated to the $l=m$ case.
\begin{theorem}\label{th:S}
Let $R', R'^{\vee}$ and $R'^{\vee\vee}$ be the combinatorial $R$ 
defined by (\ref{eq:R}), (\ref{eq:Rv}) and (\ref{eq:Rvv}) for 
$\widehat{\mathfrak{sl}}_{n-2}$ with the energy $H(x\ot y)$ 
(\ref{eq:R-energy}) replaced with
\begin{equation}\label{eq:Delta}
\Delta(x \ot y) := \begin{cases}
2\min(l,m)-Q_0(x,y) & \mbox{ for } R',\\
-P_0(x,y) & \mbox{ for } R'^\vee, \\
2\min(l,m)-Q_0(y,x) & \mbox{ for } R'^{\vee\vee}.
\end{cases}
\end{equation}
Then scattering operators are given by
\begin{equation*}
S = R',\quad 
S^\vee = R'^{\vee}, \quad 
S^{\vee\vee} = R'^{\vee\vee}.
\end{equation*}
\end{theorem}
\begin{proof}
Let
\[
p=\vac^{\ot L}\ot\imath_l(x)\ot\vac^{\ot d}\ot\imath_m(y)\ot\vac^{\ot M},
\]
be a truncated state in the semi-infinite system. One may observe the
scattering after applying
the following isomorphism of crystals.
\[
\begin{array}{rccc}
\Phi\;:&\quad (B_1\ot\Bc_1)^{\ot\tilde{L}}&\longrightarrow
&B_1^{\ot\tilde{L}}\ot(\Bc_1)^{\ot\tilde{L}}\\
&p&\longmapsto&p^{(1)}\ot p^{(2)},
\end{array}
\]
since one can show $\Phi$ commutes with $T_l$ by the inversion relation
(\ref{eq:inversion}) and
the Yang-Baxter equation (\ref{eq:ybe}). Here $\tilde{L}=L+l+d+m+M$ and we
assume $L,M$ are
sufficiently large.

First suppose both solitons are right ones. Then one finds
$p^{(2)}=\ol{n}^{\ot\tilde{L}}$ and
notices that the transition $p^{(1)}\mapsto(\mbox{the left component
of }\Phi(T_l(p)))$ is
nothing but the time evolution of the original box-ball system. Therefore it
is natural to see
our scattering rule agrees with that in \cite{HHIKTT,FOY}. If both solitons
are left ones, the
proof reduces to the above by taking the dual.

The proof is left when $\imath_l(x)$ is a right soliton and $\imath_m(y)$ is a
left one. By the
asymptotic $\mathfrak{sl}_{n-2}$-invariance (Proposition \ref{pr:inv}) it
suffices to show when
$p$ is an $\mathfrak{sl}_{n-2}$ highest weight element, i.e.,
\[
p=\vac^{\ot L}\ot(2\ol{n})^{\ot l}\ot\vac^{\ot d}\ot
(1\ol{2})^{\ot a}\ot(1\ol{n-1})^{\ot(m-a)}\ot\vac^{\ot M}.
\]
Here $a\le l$ and one can assume $l+m>d$. Then one has
\[
\Phi(p)=1^L\ot 2^l\ot
1^{d+m+M}\ot\ol{n}^{L+l+d}\ot\ol{2}^a\ot\ol{n-1}^{m-a}\ot\ol{n}^M.
\]
(We omitted $\ot$ in the superscript.) A direct calculation shows
\[
T_l^j(\Phi(p))=1^{L+jl}\ot 2^l\ot 1^{d+m+M-jl}\ot
\ol{n}^{L+l+d-jm}\ot\ol{2}^a\ot\ol{n-1}^{m-a}\ot\ol{n}^{jm+M}.
\]
for $j\ge1$. Applying $\Phi^{-1}$ when $j=3$ (scattering is finished), we
get
\[
\begin{split}
T_l^3(p)&=\vac^{\ot (L+l+d-3m+a)}\ot(1\ol{n-1})^{\ot m}\ot\vac^{\ot
(2l+2m-d-2a)}\\
&\hspace{2cm}\ot(n-1\ol{n})^{\ot a}\ot(2\ol{n})^{\ot(l-a)}\ot\vac^{\ot
(d+m+M-3l+a)}.
\end{split}
\]
In our notation, this scattering reads as
\[
\begin{split}
z^{M+m+d+1}\cdot 1^l&\ot z^{-(1-M-m)}\cdot\ol{1}^a\ol{n-2}^{m-a}\\
&\longmapsto z^{-(1-M-m)-a}\cdot\ol{n-2}^m\ot z^{M+m+d+1+a}\cdot
1^{l-a}(n-2)^a.
\end{split}
\]
This shows $S^\vee=R'^\vee$.
\end{proof}

\begin{example}
The transformations of the scattering data in 
Examples \ref{ex:rr}, \ref{ex:rl} and \ref{ex:ll} should coincide 
with the $\widehat{\mathfrak{sl}}_3$ combinatorial $R$
with the modified energy (\ref{eq:Delta}).
In fact, they agree with the three formulas 
in Example \ref{ex:R} with $H$ replaced by 
$\Delta = 3\, (\mbox{for } R ), -2\, (\mbox{for } R^\vee)$ 
and  $4\, (\mbox{for } R^{\vee\vee})$.
\end{example}

\begin{remark}
The phase $(d_1, d_2)$ of the colliding two solitons are changed into 
$(d_1-\Delta, d_2+\Delta)$ under $S$, 
$(d_1-\Delta, d_2-\Delta)$ under $S^\vee$ and 
$(d_1+\Delta, d_2-\Delta)$ under $S^{\vee\vee}$.
Thus the modified energy $\Delta$ plays the role of the phase shift.
For $S$ and $S^{\vee\vee}$, $\min(l,m) \le \Delta \le 2 \min(l,m)$,  
whereas $-\min(l,m) \le \Delta \le 0$ for $S^\vee$.
\end{remark}

\subsection{Factorized scattering and reflection}
\label{subsec:multi}

Here we briefly discuss the scattering and reflection 
in multi-soliton state.
Theorem \ref{th:S} implies that the scattering operators 
$S, S^\vee$ and $S^{\vee\vee}$ satisfy the Yang-Baxter equations
(\ref{eq:ybe}).
Moreover from Theorem \ref{th:L} and Proposition \ref{pr:re}, 
the reflection operator $L$ and the scattering operators 
satisfy the reflection equation:
\begin{equation}\label{eq:LS}
L_2S^\vee L_2S = S^{\vee\vee}L_2S^\vee L_2.
\end{equation}
These properties lead to the factorized scattering and reflection.
Namely, multi-soliton scattering and reflection 
is expressed as a composition of two body scattering 
and single body reflection 
whose order does not alter the final state.
This statement is made most transparent in terms of the scattering data.
Let $p$ be an $N$-soliton state 
with length $l_1 \ge l_2 \ge \cd \ge l_N$ and 
consider the time evolution $T^t_{r}(p)$ with $r \ge l_1$.
The scattering data of $T^t_{r}(p)$ is convergent in the limit 
$t \rightarrow -\infty$ (resp. $t \rightarrow +\infty$), 
where there are only right (resp. left) solitons.
Thus the time evolution specifies a map 
\begin{equation}\label{eq:rearrange}
\aff{B'_{l_1}}\ot \cd \ot \aff{B'_{l_N}} 
\rightarrow 
\aff{B'^\vee_{l_1}}\ot \cd \ot \aff{B'^\vee_{l_N}}
\end{equation}
between the scattering data.
By factorized scattering and reflection it is meant that this map 
coincides with any product of  
\begin{equation*}
\begin{split}
1 \ot \cd \ot 1 \ot S \ot 1 \ot \cd \ot 1, \quad & \quad 
1 \ot \cd \ot 1 \ot S^\vee \ot 1 \ot \cd \ot 1,\\
1 \ot \cd \ot 1 \ot S^{\vee\vee}  \ot 1 \ot \cd \ot 1,\; &\;\; \qquad \quad 
1 \ot 1 \ot \cd \ot 1 \ot L
\end{split}
\end{equation*}
that achieves the rearrangement (\ref{eq:rearrange}). 

\begin{example} 
$\widehat{\mathfrak{sl}}_6, \kappa = $Switch$_{16}$. 
Time evolution under $T_5$. 
$$
\begin{array}{cccccccccccccccccccccccccccccccccc}
..\s5\ol{6}\s4\ol{6}\s3\ol{6}\s3\ol{6}\s2\ol{6}
\s..\s..\s5\ol{6}\s4\ol{6}\s2\ol{6}
\s..\s..\s..\s..\s..\s..\s..\s..\s..\s..\s..\s..\\
..\s..\s..\s..\s..\s..\s5\ol{6}\s4\ol{6}\s3\ol{6}
\s3\ol{6}\s..\s5\ol{6}\s4\ol{6}\s2\ol{6}\s2\ol{6}
\s..\s..\s..\s..\s..\s..\s..\s..\\
..\s..\s..\s..\s..\s..\s..\s..\s..\s..\s5\ol{6}
\s4\ol{6}\s3\ol{6}\s..\s..\s5\ol{6}\s4\ol{6}\s3\ol{6}
\s2\ol{6}\s2\ol{6}\s..\s..\s..\\
..\s..\s..\s..\s..\s..\s..\s..\s..\s..\s..\s..\s..\s5\ol{6}
\s4\ol{6}\s3\ol{6}\s..\s..\s..\s..\s5\ol{6}\s4\ol{2}\s2\ol{3}\\
..\s..\s..\s..\s..\s..\s..\s..\s..\s..\s..\s..\s..
\s..\s..\s..\s5\ol{2}\s4\ol{2}\s2\ol{3}\s1\ol{4}\s1\ol{5}\s..\s..\\
..\s..\s..\s..\s..\s..\s..\s..\s..\s..\s..\s1\ol{2}
\s1\ol{2}\s1\ol{3}\s1\ol{3}\s1\ol{4}\s..\s..\s..\s4\ol{6}
\s3\ol{6}\s2\ol{6}\s..\\
..\s..\s..\s..\s..\s..\s1\ol{2}\s1\ol{2}\s1\ol{3}
\s1\ol{3}\s1\ol{4}\s..\s..\s..\s..\s..\s..\s..\s..
\s..\s..\s1\ol{2}\s4\ol{3}\\
..\s1\ol{2}\s1\ol{2}\s1\ol{3}\s1\ol{3}\s1\ol{4}
\s..\s..\s..\s..\s..\s..\s..\s..\s..\s..\s..\s..\s1\ol{2}
\s1\ol{3}\s1\ol{5}\s..\s..
\end{array}
$$
Here scattering and reflection are occurring 
along the order in the left hand side of 
the reflection equation (\ref{eq:LS}).
$$
\begin{array}{cccccccccccccccccccccccccccccccccccc}
..\s5\ol{6}\s4\ol{6}\s3\ol{6}\s3\ol{6}\s2\ol{6}
\s..\s..\s..\s..\s..\s..\s..\s..\s..\s..\s..\s5\ol{6}
\s4\ol{6}\s2\ol{6}\s..\s..\s..\\
..\s..\s..\s..\s..\s..\s5\ol{6}\s4\ol{6}\s3\ol{6}
\s3\ol{6}\s2\ol{6}\s..\s..\s..\s..\s..\s..\s..\s..\s..\s5\ol{6}
\s4\ol{6}\s2\ol{6}\\
..\s..\s..\s..\s..\s..\s..\s..\s..\s..\s..\s5\ol{6}
\s4\ol{6}\s3\ol{6}\s3\ol{6}\s2\ol{6}
\s..\s..\s..\s..\s1\ol{3}\s1\ol{4}\s1\ol{5}\\
..\s..\s..\s..\s..\s..\s..\s..\s..\s..\s..\s..
\s..\s..\s..\s..\s5\ol{6}\s4\ol{2}\s3\ol{4}\s2\ol{5}\s2\ol{6}\s..\s..\\
..\s..\s..\s..\s..\s..\s..\s..\s..\s..\s..\s..
\s..\s..\s1\ol{2}\s1\ol{3}\s1\ol{4}\s..\s..\s..\s1\ol{2}\s4\ol{2}\s2\ol{3}\\
..\s..\s..\s..\s..\s..\s..\s..\s..\s..\s..\s1\ol{2}
\s1\ol{3}\s1\ol{4}\s..\s1\ol{2}\s1\ol{2}\s1\ol{3}
\s1\ol{3}\s1\ol{5}\s..\s..\s..\\
..\s..\s..\s..\s..\s..\s1\ol{2}\s1\ol{2}\s1\ol{3}
\s1\ol{3}\s1\ol{4}\s..\s1\ol{2}\s1\ol{3}\s1\ol{5}
\s..\s..\s..\s..\s..\s..\s..\s..\\
..\s1\ol{2}\s1\ol{2}\s1\ol{3}\s1\ol{3}\s1\ol{4}
\s..\s..\s..\s1\ol{2}\s1\ol{3}\s1\ol{5}
\s..\s..\s..\s..\s..\s..\s..\s..\s..\s..\s..
\end{array}
$$
Here scattering and reflection are occurring 
along the order in the right hand side of 
the reflection equation (\ref{eq:LS}).
In terms of the scattering data the both cases 
are expressed as the same transformation (\ref{eq:rearrange}):
\begin{equation*}
z^{d_1}12234 \ot z^{d_2}134 
\mapsto 
z^{-d_1+4}\,\ol{11223}\ot z^{-d_2-4}\,\ol{124}
\end{equation*}
for some $d_1, d_2$ and $d_3$.
\end{example}

For a scattering without a boundary effect, the factorization 
is due to the Yang-Baxter equation of the scattering operators only.
We finish with such an example involving both right and left solitons.

\begin{example}
$\widehat{\mathfrak{sl}}_5$. Scattering of two right and one left solitons 
under $T_3$. The two patterns correspond
 to the two sides of the second Yang-Baxter equation 
in (\ref{eq:ybe}):
$$(1 \ot S)(S^\vee \ot 1)(1 \ot S^\vee) 
= (S^{\vee} \ot 1)(1 \ot S^\vee)(S \ot 1).$$
$$
\begin{array}{ccccccccccccccccccccccccccccccccccccc}
 ..\s4\ol{5}\s3\ol{5}\s2\ol{5}\s..\s..\s..\s..\s..\s..
\s..\s..\s..\s..\s..\s2\ol{5}\s..\s..\s..\s..\s..\s1\ol{2}
\s1\ol{2}\s1\ol{3}\s..\s..\\
 ..\s..\s..\s..\s4\ol{5}\s3\ol{5}\s2\ol{5}
\s..\s..\s..\s..\s..\s..\s..\s..\s..\s2\ol{5}
\s..\s1\ol{2}\s1\ol{2}\s1\ol{3}\s..\s..\s..\s..\s..\\
 ..\s..\s..\s..\s..\s..\s..\s4\ol{5}\s3\ol{5}\s2\ol{5}
\s..\s..\s..\s..\s..\s..\s5\ol{2}\s1\ol{3}
\s..\s..\s..\s..\s..\s..\s..\s..\\
 ..\s..\s..\s..\s..\s..\s..\s..\s..\s..\s4\ol{5}
\s3\ol{5}\s2\ol{5}\s1\ol{2}\s1\ol{3}\s1\ol{4}
\s..\s4\ol{5}\s..\s..\s..\s..\s..\s..\s..\s..\\
 ..\s..\s..\s..\s..\s..\s..\s..\s..\s..\s..\s1\ol{2}
\s5\ol{3}\s3\ol{5}\s2\ol{5}\s..\s..\s..\s4\ol{5}
\s..\s..\s..\s..\s..\s..\s..\\
 ..\s..\s..\s..\s..\s..\s..\s..\s1\ol{2}\s1\ol{3}
\s1\ol{4}\s..\s..\s..\s..\s4\ol{5}\s3\ol{5}\s2\ol{5}
\s..\s4\ol{5}\s..\s..\s..\s..\s..\s..\\
 ..\s..\s..\s..\s..\s1\ol{2}\s1\ol{3}\s1\ol{4}
\s..\s..\s..\s..\s..\s..\s..\s..\s..\s..\s4\ol{5}
\s3\ol{5}\s4\ol{5}\s2\ol{5}\s..\s..\s..\s..\\
 ..\s..\s1\ol{2}\s1\ol{3}\s1\ol{4}\s..\s..\s..\s..
\s..\s..\s..\s..\s..\s..\s..\s..\s..\s..\s..\s3\ol{5}
\s..\s4\ol{5}\s4\ol{5}\s2\ol{5}\s..
\end{array}
$$

\vspace{0.3cm}

$$
\begin{array}{ccccccccccccccccccccccccccccccccccccc}
 ..\s4\ol{5}\s3\ol{5}\s2\ol{5}\s..\s..\s..\s2\ol{5}
\s..\s..\s..\s..\s..\s..\s..\s..\s..\s..\s..\s..
\s..\s1\ol{2}\s1\ol{2}\s1\ol{3}\s..\s..\\
 ..\s..\s..\s..\s4\ol{5}\s3\ol{5}\s2\ol{5}\s..\s2\ol{5}
\s..\s..\s..\s..\s..\s..\s..\s..\s..\s1\ol{2}\s1\ol{2}
\s1\ol{3}\s..\s..\s..\s..\s..\\
 ..\s..\s..\s..\s..\s..\s..\s4\ol{5}\s..\s3\ol{5}
\s2\ol{5}\s2\ol{5}\s..\s..\s..\s1\ol{2}\s1\ol{2}
\s1\ol{3}\s..\s..\s..\s..\s..\s..\s..\s..\\
 ..\s..\s..\s..\s..\s..\s..\s..\s4\ol{5}
\s..\s..\s..\s3\ol{1}\s2\ol{2}\s1\ol{3}
\s..\s..\s..\s..\s..\s..\s..\s..\s..\s..\s..\\
 ..\s..\s..\s..\s..\s..\s..\s..\s..\s4\ol{5}
\s..\s1\ol{2}\s1\ol{4}\s5\ol{5}\s4\ol{5}\s2\ol{5}
\s..\s..\s..\s..\s..\s..\s..\s..\s..\s..\\
 ..\s..\s..\s..\s..\s..\s..\s..\s1\ol{2}\s1\ol{3}
\s3\ol{4}\s..\s..\s..\s..\s..\s4\ol{5}\s4\ol{5}\s2\ol{5}
\s..\s..\s..\s..\s..\s..\s..\\
 ..\s..\s..\s..\s..\s1\ol{2}\s1\ol{3}\s1\ol{4}
\s..\s..\s..\s3\ol{5}\s..\s..\s..\s..\s..\s..\s..\s4\ol{5}
\s4\ol{5}\s2\ol{5}\s..\s..\s..\s..\\
 ..\s..\s1\ol{2}\s1\ol{3}\s1\ol{4}
\s..\s..\s..\s..\s..\s..\s..\s3\ol{5}
\s..\s..\s..\s..\s..\s..\s..\s..\s..\s4\ol{5}
\s4\ol{5}\s2\ol{5}\s..
\end{array}
$$
In terms of the scattering data the both patterns are expressed as
\begin{equation*}
z^{d_1}123 \ot z^{d_2} 1 \ot z^{-d_3}\, \ol{112} 
\mapsto
z^{-d_3-2}\,\ol{123} \ot z^{d_2+2} 2 \ot z^{d_1}133
\end{equation*}
for some $d_1, d_2$ and $d_3$.
\end{example}

\section{Automaton with two reflecting ends}\label{sec:finite}

Here we formulate an automaton 
on a finite lattice surrounded by two reflecting ends.
We will only display its soliton behaviour 
leaving a thorough study as a future problem.

\subsection{General case}\label{subsec:general}

\begin{figure}[h]
\caption{Diagram for $T^{(i)}$}
\label{fig:T}
\vspace{0.5cm}
\hspace{-3cm}

\unitlength 0.1in
\begin{picture}( 58.0000, 11.9500)(  12.5000,-15.7000)
%
\special{pn 8}%
\special{pa 1726 816}%
\special{pa 1726 1536}%
\special{pa 1726 1536}%
\special{fp}%
\put(63.2000,-12.8000){\makebox(0,0){$\kappa_{\rm right}$}}%
\put(35.0500,-10.5500){\makebox(0,0){$\kappa_{\rm right}$}}%
\put(11.6500,-13.0500){\makebox(0,0){$\kappa_{\rm left}$}}%
\put(40.4000,-10.5000){\makebox(0,0){$\kappa_{\rm left}$}}%
\put(59.5000,-10.9500){\makebox(0,0){$b''_i$}}%
\put(14.9500,-11.1500){\makebox(0,0){$b''_i$}}%
\put(43.4000,-8.5000){\makebox(0,0){$b'_i$}}%
\put(31.8500,-9.1500){\makebox(0,0){$b'_i$}}%
%
\special{pn 8}%
\special{pa 6110 1386}%
\special{pa 6110 1146}%
\special{fp}%
%
\special{pn 8}%
\special{pa 4190 1150}%
\special{pa 4190 910}%
\special{fp}%
\put(49.3000,-16.3500){\makebox(0,0){${\tilde b}_{i-1}$}}%
\put(57.7000,-16.2500){\makebox(0,0){${\tilde b}_L$}}%
\put(54.1000,-16.2500){\makebox(0,0){${\tilde b}_{i+1}$}}%
\put(51.7000,-16.2500){\makebox(0,0){${\tilde b}_i$}}%
\put(45.7000,-16.2500){\makebox(0,0){${\tilde b}_1$}}%
\put(53.9000,-6.9500){\makebox(0,0){$b_{i+1}$}}%
\put(49.1000,-6.8500){\makebox(0,0){$b_{i-1}$}}%
\put(45.5000,-6.8500){\makebox(0,0){$b_1$}}%
\put(57.5000,-6.8500){\makebox(0,0){$b_L$}}%
\put(51.6000,-6.9500){\makebox(0,0){$b_i$}}%
%
\special{pn 8}%
\special{pa 4910 1466}%
\special{pa 4910 1506}%
\special{fp}%
\special{sh 1}%
\special{pa 4910 1506}%
\special{pa 4930 1438}%
\special{pa 4910 1452}%
\special{pa 4890 1438}%
\special{pa 4910 1506}%
\special{fp}%
%
\special{pn 8}%
\special{pa 4910 1506}%
\special{pa 4910 786}%
\special{pa 4910 786}%
\special{fp}%
%
\special{pn 8}%
\special{pa 5390 1466}%
\special{pa 5390 1506}%
\special{fp}%
\special{sh 1}%
\special{pa 5390 1506}%
\special{pa 5410 1438}%
\special{pa 5390 1452}%
\special{pa 5370 1438}%
\special{pa 5390 1506}%
\special{fp}%
%
\special{pn 8}%
\special{pa 5390 1506}%
\special{pa 5390 786}%
\special{pa 5390 786}%
\special{fp}%
%
\special{pn 8}%
\special{pa 5750 1466}%
\special{pa 5750 1506}%
\special{fp}%
\special{sh 1}%
\special{pa 5750 1506}%
\special{pa 5770 1438}%
\special{pa 5750 1452}%
\special{pa 5730 1438}%
\special{pa 5750 1506}%
\special{fp}%
%
\special{pn 8}%
\special{pa 5750 1506}%
\special{pa 5750 786}%
\special{pa 5750 786}%
\special{fp}%
%
\special{pn 8}%
\special{pa 5150 1466}%
\special{pa 5150 1506}%
\special{dt 0.045}%
\special{sh 1}%
\special{pa 5150 1506}%
\special{pa 5170 1438}%
\special{pa 5150 1452}%
\special{pa 5130 1438}%
\special{pa 5150 1506}%
\special{fp}%
%
\special{pn 8}%
\special{pa 5150 1506}%
\special{pa 5150 1386}%
\special{pa 5870 1386}%
\special{pa 6110 1266}%
\special{pa 6110 1266}%
\special{dt 0.045}%
%
\special{pn 8}%
\special{pa 4190 1026}%
\special{pa 4430 906}%
\special{pa 5150 906}%
\special{pa 5150 786}%
\special{pa 5150 786}%
\special{pa 5150 786}%
\special{dt 0.045}%
%
\special{pn 8}%
\special{pa 6110 1266}%
\special{pa 5870 1146}%
\special{pa 4430 1146}%
\special{pa 4190 1026}%
\special{pa 4190 1026}%
\special{fp}%
%
\special{pn 8}%
\special{sh 1}%
\special{ar 5504 906 10 10 0  6.28318530717959E+0000}%
\special{sh 1}%
\special{ar 5576 906 10 10 0  6.28318530717959E+0000}%
\special{sh 1}%
\special{ar 5648 906 10 10 0  6.28318530717959E+0000}%
\special{sh 1}%
\special{ar 5648 906 10 10 0  6.28318530717959E+0000}%
%
\special{pn 8}%
\special{sh 1}%
\special{ar 4652 1386 10 10 0  6.28318530717959E+0000}%
\special{sh 1}%
\special{ar 4724 1386 10 10 0  6.28318530717959E+0000}%
\special{sh 1}%
\special{ar 4796 1386 10 10 0  6.28318530717959E+0000}%
\special{sh 1}%
\special{ar 4796 1386 10 10 0  6.28318530717959E+0000}%
%
\special{pn 8}%
\special{pa 4550 1466}%
\special{pa 4550 1506}%
\special{fp}%
\special{sh 1}%
\special{pa 4550 1506}%
\special{pa 4570 1438}%
\special{pa 4550 1452}%
\special{pa 4530 1438}%
\special{pa 4550 1506}%
\special{fp}%
%
\special{pn 8}%
\special{pa 4550 1506}%
\special{pa 4550 786}%
\special{pa 4550 786}%
\special{fp}%
\put(21.0500,-16.5500){\makebox(0,0){${\tilde b}_{i-1}$}}%
\put(29.4500,-16.4500){\makebox(0,0){${\tilde b}_L$}}%
\put(25.8500,-16.4500){\makebox(0,0){${\tilde b}_{i+1}$}}%
\put(23.4500,-16.4500){\makebox(0,0){${\tilde b}_i$}}%
\put(17.4500,-16.4500){\makebox(0,0){${\tilde b}_1$}}%
\put(23.2500,-7.1500){\makebox(0,0){$b_i$}}%
\put(29.2500,-7.0500){\makebox(0,0){$b_L$}}%
\put(25.7500,-7.1500){\makebox(0,0){$b_{i+1}$}}%
\put(20.8500,-7.1500){\makebox(0,0){$b_{i-1}$}}%
\put(17.2500,-7.0500){\makebox(0,0){$b_1$}}%
%
\special{pn 8}%
\special{sh 1}%
\special{ar 1828 936 10 10 0  6.28318530717959E+0000}%
\special{sh 1}%
\special{ar 1900 936 10 10 0  6.28318530717959E+0000}%
\special{sh 1}%
\special{ar 1972 936 10 10 0  6.28318530717959E+0000}%
\special{sh 1}%
\special{ar 1972 936 10 10 0  6.28318530717959E+0000}%
%
\special{pn 8}%
\special{sh 1}%
\special{ar 2680 1416 10 10 0  6.28318530717959E+0000}%
\special{sh 1}%
\special{ar 2752 1416 10 10 0  6.28318530717959E+0000}%
\special{sh 1}%
\special{ar 2824 1416 10 10 0  6.28318530717959E+0000}%
\special{sh 1}%
\special{ar 2824 1416 10 10 0  6.28318530717959E+0000}%
%
\special{pn 8}%
\special{pa 3286 1056}%
\special{pa 3046 1176}%
\special{pa 1606 1176}%
\special{pa 1366 1296}%
\special{pa 1366 1296}%
\special{dt 0.045}%
%
\special{pn 8}%
\special{pa 2326 1512}%
\special{pa 2326 1536}%
\special{fp}%
\special{sh 1}%
\special{pa 2326 1536}%
\special{pa 2346 1468}%
\special{pa 2326 1482}%
\special{pa 2306 1468}%
\special{pa 2326 1536}%
\special{fp}%
%
\special{pn 8}%
\special{pa 1366 1296}%
\special{pa 1606 1416}%
\special{pa 2326 1416}%
\special{pa 2326 1536}%
\special{pa 2326 1536}%
\special{pa 2326 1536}%
\special{fp}%
%
\special{pn 8}%
\special{pa 2326 816}%
\special{pa 2326 936}%
\special{pa 3046 936}%
\special{pa 3286 1056}%
\special{pa 3286 1056}%
\special{fp}%
%
\special{pn 8}%
\special{pa 1366 1176}%
\special{pa 1366 1416}%
\special{fp}%
%
\special{pn 8}%
\special{pa 3300 960}%
\special{pa 3300 1200}%
\special{fp}%
%
\special{pn 8}%
\special{pa 2926 1512}%
\special{pa 2926 1536}%
\special{fp}%
\special{sh 1}%
\special{pa 2926 1536}%
\special{pa 2946 1468}%
\special{pa 2926 1482}%
\special{pa 2906 1468}%
\special{pa 2926 1536}%
\special{fp}%
%
\special{pn 8}%
\special{pa 2926 816}%
\special{pa 2926 1536}%
\special{pa 2926 1536}%
\special{fp}%
%
\special{pn 8}%
\special{pa 2566 1512}%
\special{pa 2566 1536}%
\special{fp}%
\special{sh 1}%
\special{pa 2566 1536}%
\special{pa 2586 1468}%
\special{pa 2566 1482}%
\special{pa 2546 1468}%
\special{pa 2566 1536}%
\special{fp}%
%
\special{pn 8}%
\special{pa 2566 816}%
\special{pa 2566 1536}%
\special{pa 2566 1536}%
\special{fp}%
%
\special{pn 8}%
\special{pa 2086 1512}%
\special{pa 2086 1536}%
\special{fp}%
\special{sh 1}%
\special{pa 2086 1536}%
\special{pa 2106 1468}%
\special{pa 2086 1482}%
\special{pa 2066 1468}%
\special{pa 2086 1536}%
\special{fp}%
%
\special{pn 8}%
\special{pa 2086 816}%
\special{pa 2086 1536}%
\special{pa 2086 1536}%
\special{fp}%
%
\special{pn 8}%
\special{pa 1726 1512}%
\special{pa 1726 1536}%
\special{fp}%
\special{sh 1}%
\special{pa 1726 1536}%
\special{pa 1746 1468}%
\special{pa 1726 1482}%
\special{pa 1706 1468}%
\special{pa 1726 1536}%
\special{fp}%
\put(23.2500,-4.6500){\makebox(0,0){$b_i \in B_{m_i}$ case}}%
\put(51.5000,-4.6000){\makebox(0,0){$b_i \in B^\vee_{m_i}$ case}}%
\end{picture}%

\end{figure}
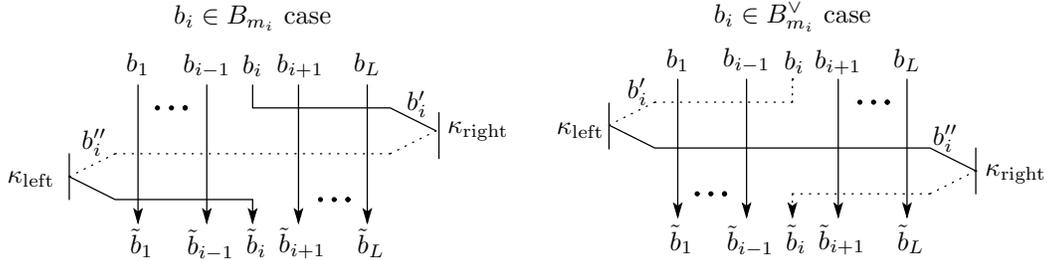

First we consider a rather general setting 
where the set of states of the automaton is taken as
\begin{equation}\label{eq:gP}
{\mathcal P} = B^{\epsilon_1}_{m_1} \ot B^{\epsilon_2}_{m_2}
\ot \cdots \ot B^{\epsilon_L}_{m_L},
\end{equation}
where $\epsilon_i = \pm 1$ and 
$B^+_m = B_m,\, B^-_m = B^\vee_m$.
$m_i$ and $L$ are arbitrary positive integers.
This is an inhomogeneous system in that 
the local states belong to different crystals 
according to the data $\{(\epsilon_i,m_i)\}$.
We denote by ${\mathcal P}^{(i)}$ 
the $L\!-\!1$ fold tensor product without the 
$i$ th component $B^{\epsilon_i}_{m_i}$ in (\ref{eq:gP}).
Let $\kappa_{\rm right}: B_l \rightarrow B^\vee_l$ 
and $\kappa_{\rm left}: B^\vee_l \rightarrow B_l$ 
be any one of the maps (\ref{eq:rl}), (\ref{eq:1n}) and (\ref{eq:12}).
Given $b_1 \ot \cdots \ot b_L \in {\mathcal P}$, 
its time evolution 
$T^{(i)}(b_1 \ot \cdots \ot b_L) 
= {\tilde b}_1 \ot \cdots \ot {\tilde b}_L \in {\mathcal P}$
$(1\! \le \! i \! \le \! L)$ is defined as follows:
\begin{equation}\label{eq:ti1}
\begin{split}
\hbox{If } b_i \in B_{m_i},\quad
b_1 \ot \cdots \ot b_L &\simeq {\bold p} \ot b'_i 
\in {\mathcal P}^{(i)}\ot B_{m_i},\\
{\bold p} \ot \kappa_{\rm right}(b'_i) &\simeq b''_i \ot {\bold p}'
\in B^\vee_{m_i} \ot {\mathcal P}^{(i)},\\
\kappa_{\rm left}(b''_i) \ot {\bold p}' &\simeq 
{\tilde b}_1 \ot \cdots \ot {\tilde b}_L.
\end{split}
\end{equation}
\begin{equation}\label{eq:ti2}
\begin{split}
\hbox{If } b_i \in B^\vee_{m_i},\quad
b_1 \ot \cdots \ot b_L &\simeq b'_i \ot {\bold p}
\in B^\vee_{m_i} \ot {\mathcal P}^{(i)},\\
\kappa_{\rm left}(b'_i) \ot {\bold p} &\simeq 
{\bold p}' \ot b''_i
\in {\mathcal P}^{(i)}\ot B_{m_i},\\
{\bold p}' \ot \kappa_{\rm right}(b''_i) &\simeq
{\tilde b}_1 \ot \cdots \ot {\tilde b}_L.
\end{split}
\end{equation}
Here $\simeq$ is obtained by 
repeated applications of the combinatorial $R$.
The definitions (\ref{eq:ti1}) and 
(\ref{eq:ti2}) are depicted in Figure \ref{fig:T},
where we have assumed 
$\epsilon_1 = \epsilon_{i-1} = \epsilon_{i+1} = \epsilon_L = 1$.
(Otherwise, the corresponding vertical lines therein 
should be dotted ones.)
Unlike the semi-infinite system, 
we impose no boundary 
condition as in (\ref{eq:P}).

By using the Yang-Baxter equation (\ref{eq:ybe}) and 
the reflection equation (\ref{eq:remigi}) and (\ref{eq:rehidari}),
one can show
\begin{proposition}\label{pr:kakan}
$\{T^{(1)}, \ldots, T^{(L)}\}$ forms a commuting family, i.e., 
$T^{(i)}T^{(j)}=T^{(j)}T^{(i)}$.
\end{proposition}

\subsection{Soliton behaviour}\label{subsec:soliton}

Next we report the soliton behaviour in the automaton 
corresponding to the almost homogeneous 
specialization of (\ref{eq:gP}):
\begin{equation}\label{eq:pl}
{\mathcal P} = B_l \ot B^{\ot L},
\end{equation}
where $B = B_1 \ot B^\vee_1$ as before and 
$l$ and $L$ are arbitrary positive integers.
According to Proposition \ref{pr:kakan}, there are 
commuting operators 
$T^{(1)}, T^{(2)}, \ldots, T^{(2L+1)}$.
However they are not independent under the choice (\ref{eq:pl}).
In fact, 
$T^{(2)} = T^{(4)} = \cdots = T^{(2L)}$ and 
$T^{(3)} = T^{(5)} = \cdots = T^{(2L+1)}$ can be proved 
by means of the inversion relation (\ref{eq:inversion}).
In what follows we shall concentrate on 
$T := T^{(1)}$, which is an analogue of 
$T_l$ in the semi-infinite system.
($T^{(2)}$ is an analogue of $T_1$.)
For $u \in B_l$ and $p \in B^{\ot L}$, 
the time evolution $T(u \ot p) \in  {\mathcal P}$ 
defined in Section \ref{subsec:general} is 
rephrased as follows:
\begin{equation}\label{eq:T}
\begin{split}
u \ot p &\simeq p^\dag \ot v \in B^{\ot L} \ot B_l,\\
p^\dag \ot \kappa_{\rm right}(v) &\simeq 
w \ot p' \in B^\vee_l \ot B^{\ot L},\\
T(u \ot p) &= \kappa_{\rm left}(w) \ot p'.
\end{split}
\end{equation}
In place of a boundary condition, 
we postulate that there exist an element
$a \in B_1$ such that 
\begin{equation}\label{eq:ab}
\kappa_{\rm left}(\kappa_{\rm right}(a)) = a.
\end{equation}
Under this condition one can let the local state 
$a\ot\kappa_{\rm right}(a)\in B$ play a role analogous to 
$\vac$ (\ref{eq:vac}).
The condition (\ref{eq:ab}) is satisfied only for even $n$ 
and the following choices:
\begin{center}
\begin{tabular}{c|c|c}
$\kappa_{\rm left}$ & $\kappa_{\rm right}$ & $a$ \\
\hline 
Switch$_{1n}$ & Rotateleft &  odd \\
Switch$_{12}$ & Rotateleft &  even \\
Rotateleft & Switch$_{1n}$ & even \\
Switch$_{1n}$ & Switch$_{1n}$ & arbitrary \\
Rotateleft & Switch$_{12}$ & odd \\
Switch$_{12}$ & Switch$_{12}$ & arbitrary 
\end{tabular}
\end{center}
Here the freedom of $a$ can be absorbed into 
a relabelling and is not essential.
By regarding the local state $a\ot\kappa_{\rm right}(a)$ 
as the vacuum one, the present automaton contains 
the semi-infinite case (with time evolution $T_l$) 
as a natural limit $L \rightarrow \infty$.
By computer experiments
we have observed that in each case in the list, 
the system behaves as a soliton cellular automaton 
on finite lattice with two reflecting ends.
(Unless (\ref{eq:ab}) is satisfied, we observed a chaotic behaviour.)
We include two examples with $a=1$.

\begin{example}\label{ex:finite1}
$\widehat{\mathfrak{sl}}_4,\; 
\kappa_{\rm right} = \hbox{Rotateleft},\; 
\kappa_{\rm left} = \hbox{Switch}_{14}$.
${\mathcal P} = B_3 \ot B^{\ot 13}$. 
The symbol $..$ stands for $1\ol{4}$.
The leftmost components $111, 113$ etc. are the tableau representation of the
elements in $B_3$.
Successive collisions and reflection of solitons of length 2 and 3. 
$$
\begin{array}{cccccccccccccccccccccccccccccccc}
 111\s..\s..\s..\s1\ol{2}\s1\ol{3}\s..\s..\s..\s..\s1\ol{2}\s1\ol{2}\s1\ol{3}\s..\\ 111\s..\s1\ol{2}\s1\ol{3}\s..\s..\s..\s1\ol{2}\s1\ol{2}\s1\ol{3}\s..\s..\s..\s..\\
   113\s1\ol{3}\s..\s..\s1\ol{2}\s1\ol{2}\s1\ol{3}\s..\s..\s..\s..\s..\s..\s..\\ 113\s2\ol{2}\s1\ol{2}\s1\ol{3}\s..\s..\s..\s..\s..\s..\s..\s..\s..\s..\\
   133\s4\ol{4}\s2\ol{4}\s..\s..\s..\s..\s..\s..\s..\s..\s..\s..\s..\\ 113\s2\ol{4}\s..\s3\ol{4}\s3\ol{4}\s2\ol{4}\s..\s..\s..\s..\s..\s..\s..\s..\\
   111\s..\s3\ol{4}\s2\ol{4}\s..\s..\s3\ol{4}\s3\ol{4}\s2\ol{4}\s..\s..\s..\s..\s..\\ 111\s..\s..\s..\s3\ol{4}\s2\ol{4}\s..\s..\s..\s3\ol{4}\s3\ol{4}\s2\ol{4}\s..\s..\\
   111\s..\s..\s..\s..\s..\s3\ol{4}\s2\ol{4}\s..\s..\s..\s..\s3\ol{4}\s3\ol{1}\\ 111\s..\s..\s..\s..\s..\s..\s..\s3\ol{4}\s2\ol{4}\s..\s1\ol{2}\s1\ol{2}\s1\ol{3}\\
   111\s..\s..\s..\s..\s..\s..\s..\s..\s4\ol{2}\s3\ol{3}\s..\s..\s..\\ 111\s..\s..\s..\s..\s..\s1\ol{2}\s1\ol{2}\s1\ol{3}\s..\s..\s3\ol{4}\s2\ol{4}\s..\\
   111\s..\s..\s1\ol{2}\s1\ol{2}\s1\ol{3}\s..\s..\s..\s..\s..\s..\s..\s3\ol{1}\\ 113\s1\ol{2}\s1\ol{3}\s..\s..\s..\s..\s..\s..\s..\s..\s..\s1\ol{2}\s1\ol{3}\\
   133\s2\ol{4}\s..\s..\s..\s..\s..\s..\s..\s..\s1\ol{2}\s1\ol{3}\s..\s..\\ 111\s..\s3\ol{4}\s3\ol{4}\s2\ol{4}\s..\s..\s..\s1\ol{2}\s1\ol{3}\s..\s..\s..\s..\\
   111\s..\s..\s..\s..\s3\ol{4}\s3\ol{1}\s1\ol{3}\s..\s..\s..\s..\s..\s..\\ 111\s..\s..\s..\s..\s1\ol{2}\s1\ol{3}\s3\ol{4}\s3\ol{4}\s2\ol{4}\s..\s..\s..\s..\\
   111\s..\s..\s1\ol{2}\s1\ol{3}\s..\s..\s..\s..\s..\s3\ol{4}\s3\ol{4}\s2\ol{4}\s..\\ 111\s1\ol{2}\s1\ol{3}\s..\s..\s..\s..\s..\s..\s..\s..\s..\s1\ol{1}\s3\ol{2}\\
   123\s..\s..\s..\s..\s..\s..\s..\s..\s..\s1\ol{2}\s1\ol{2}\s1\ol{3}\s..\\ 111\s3\ol{4}\s2\ol{4}\s..\s..\s..\s..\s1\ol{2}\s1\ol{2}\s1\ol{3}\s..\s..\s..\s..\\
   111\s..\s..\s3\ol{4}\s2\ol{2}\s1\ol{2}\s1\ol{3}\s..\s..\s..\s..\s..\s..\s..\\ 111\s..\s1\ol{2}\s1\ol{2}\s4\ol{4}\s2\ol{4}\s..\s..\s..\s..\s..\s..\s..\s..\\
   133\s1\ol{3}\s..\s..\s..\s..\s3\ol{4}\s2\ol{4}\s..\s..\s..\s..\s..\s..\\ 113\s3\ol{4}\s2\ol{4}\s..\s..\s..\s..\s..\s3\ol{4}\s2\ol{4}\s..\s..\s..\s..\\
   111\s..\s..\s3\ol{4}\s3\ol{4}\s2\ol{4}\s..\s..\s..\s..\s3\ol{4}\s2\ol{4}\s..\s..
\end{array}
$$
\end{example}

\begin{example}\label{ex:finite2}
$\widehat{\mathfrak{sl}}_6,\; 
\kappa_{\rm right} = \kappa_{\rm left} = \hbox{Switch}_{12}$.
${\mathcal P} = B_4 \ot B^{\ot 21}$. 
The symbol $..$ stands for $1\ol{2}$. 
Successive collisions and reflection of solitons of length 1, 1, 2 and 3. 
$$
\begin{array}{ccccccccccccccccccccccccccccccccccccccc}
  1111\s..\s..\s..\s1\ol{6}\s1\ol{6}\s1\ol{6}\s..\s..\s..\s..\s..\s1\ol{3}\s1\ol{4}\s..\s..\s..\s1\ol{3}\s..\s..\s..\s1\ol{4}\\  1111\s1\ol{6}\s1\ol{6}\s1\ol{6}\s..\s..\s..\s..\s..\s..\s1\ol{3}\s1\ol{4}\s..\s..\s..\s..\s1\ol{3}\s..\s..\s..\s1\ol{4}\s..\\  1555\s..\s..\s..\s..\s..\s..\s..\s1\ol{3}\s1\ol{4}\s..\s..\s..\s..\s..\s1\ol{3}\s..\s..\s..\s1\ol{4}\s..\s..\\
  1111\s5\ol{2}\s5\ol{2}\s5\ol{2}\s..\s..\s1\ol{3}\s1\ol{4}\s..\s..\s..\s..\s..\s..\s1\ol{3}\s..\s..\s..\s1\ol{4}\s..\s..\s..\\  1111\s..\s..\s..\s5\ol{3}\s5\ol{4}\s5\ol{2}\s..\s..\s..\s..\s..\s..\s1\ol{3}\s..\s..\s..\s1\ol{4}\s..\s..\s..\s..\\
  1111\s..\s1\ol{3}\s1\ol{4}\s..\s..\s..\s5\ol{2}\s5\ol{2}\s5\ol{2}\s..\s..\s1\ol{3}\s..\s..\s..\s1\ol{4}\s..\s..\s..\s..\s..\\ 1114\s1\ol{4}\s..\s..\s..\s..\s..\s..\s..\s..\s5\ol{2}\s5\ol{3}\s5\ol{2}\s..\s..\s1\ol{4}\s..\s..\s..\s..\s..\s..\\
  1114\s3\ol{2}\s..\s..\s..\s..\s..\s..\s..\s..\s1\ol{3}\s..\s..\s5\ol{2}\s5\ol{4}\s5\ol{2}\s..\s..\s..\s..\s..\s..\\
  1111\s..\s4\ol{2}\s3\ol{2}\s..\s..\s..\s..\s..\s1\ol{3}\s..\s..\s..\s1\ol{4}\s..\s..\s5\ol{2}\s5\ol{2}\s5\ol{2}\s..\s..\s..\\  1111\s..\s..\s..\s4\ol{2}\s3\ol{2}\s..\s..\s1\ol{3}\s..\s..\s..\s1\ol{4}\s..\s..\s..\s..\s..\s..\s5\ol{2}\s5\ol{2}\s5\ol{2}\\   1111\s..\s..\s..\s..\s..\s4\ol{2}\s3\ol{3}\s..\s..\s..\s1\ol{4}\s..\s..\s..\s..\s..\s..\s..\s1\ol{6}\s1\ol{6}\s1\ol{6}\\  1111\s..\s..\s..\s..\s..\s..\s1\ol{1}\s4\ol{2}\s..\s1\ol{4}\s..\s..\s..\s..\s..\s1\ol{6}\s1\ol{6}\s1\ol{6}\s..\s..\s..\\
   1111\s..\s..\s..\s..\s..\s1\ol{6}\s..\s..\s6\ol{3}\s3\ol{2}\s..\s..\s1\ol{6}\s1\ol{6}\s1\ol{6}\s..\s..\s..\s..\s..\s..\\   1111\s..\s..\s..\s..\s1\ol{6}\s..\s..\s1\ol{3}\s..\s1\ol{5}\s5\ol{6}\s3\ol{6}\s..\s..\s..\s..\s..\s..\s..\s..\s..\\
  1111\s..\s..\s..\s1\ol{6}\s1\ol{3}\s1\ol{5}\s1\ol{6}\s..\s1\ol{6}\s..\s..\s..\s5\ol{2}\s3\ol{2}\s..\s..\s..\s..\s..\s..\s..\\  1111\s1\ol{3}\s1\ol{6}\s1\ol{6}\s1\ol{5}\s..\s..\s..\s1\ol{6}\s..\s..\s..\s..\s..\s..\s5\ol{2}\s3\ol{2}\s..\s..\s..\s..\s..\\   1455\s..\s..\s1\ol{5}\s..\s..\s..\s1\ol{6}\s..\s..\s..\s..\s..\s..\s..\s..\s..\s5\ol{2}\s3\ol{2}\s..\s..\s..\\
  1111\s5\ol{2}\s5\ol{5}\s4\ol{2}\s..\s..\s1\ol{6}\s..\s..\s..\s..\s..\s..\s..\s..\s..\s..\s..\s..\s5\ol{2}\s3\ol{2}\s..\\
  1111\s1\ol{4}\s..\s..\s5\ol{2}\s4\ol{6}\s4\ol{2}\s..\s..\s..\s..\s..\s..\s..\s..\s..\s..\s..\s..\s..\s..\s5\ol{4}\\
  1113\s..\s..\s..\s1\ol{6}\s..\s..\s5\ol{2}\s4\ol{2}\s4\ol{2}\s..\s..\s..\s..\s..\s..\s..\s..\s..\s1\ol{4}\s1\ol{6}\s..\\
  1111\s3\ol{2}\s..\s1\ol{6}\s..\s..\s..\s..\s..\s..\s5\ol{2}\s4\ol{2}\s4\ol{2}\s..\s..\s..\s..\s1\ol{4}\s1\ol{6}\s..\s..\s..\\   1111\s..\s3\ol{6}\s..\s..\s..\s..\s..\s..\s..\s..\s..\s..\s5\ol{2}\s4\ol{2}\s4\ol{4}\s1\ol{6}\s..\s..\s..\s..\s..\\
 1111\s1\ol{6}\s..\s3\ol{2}\s..\s..\s..\s..\s..\s..\s..\s..\s..\s1\ol{3}\s1\ol{6}\s..\s5\ol{2}\s4\ol{2}\s3\ol{2}\s..\s..\s..\\
 1115\s..\s..\s..\s3\ol{2}\s..\s..\s..\s..\s..\s..\s1\ol{3}\s1\ol{6}\s..\s..\s..\s..\s..\s..\s5\ol{2}\s4\ol{2}\s3\ol{2}\\
 1111\s5\ol{2}\s..\s..\s..\s3\ol{2}\s..\s..\s..\s1\ol{3}\s1\ol{6}\s..\s..\s..\s..\s..\s..\s..\s..\s1\ol{3}\s1\ol{4}\s1\ol{6}\\
 1111\s..\s5\ol{2}\s..\s..\s..\s3\ol{2}\s1\ol{3}\s1\ol{6}\s..\s..\s..\s..\s..\s..\s..\s1\ol{3}\s1\ol{4}\s1\ol{6}\s..\s..\s..
\end{array}
$$
\end{example}

\appendix
\section{Proof of Proposition \ref{pr:re}}\label{app:re}

\subsection{Equivalence of 
(\ref{eq:rehidari}) and (\ref{eq:remigi})}\label{appsub:equiv}

Let $P(z^dx \ot z^ey) = z^ey \ot z^dx$ be the transposition and 
let $\iota$ be the map $\aff{B_l} \rightarrow \aff{B^\vee_l}$ 
defined by $\iota(z^dx) = z^{-d}x$.
\begin{lemma}\label{lem:iota}
\begin{align}
&(\iota^{-1}\ot\iota^{-1})R^{\vee\vee}(\iota\ot\iota) = 
PRP, \label{eq:iota1}\\
&(\iota\ot\iota)K^\vee_1 R^\vee K^\vee_1(\iota\ot\iota) = 
PK_2R^\vee K_2P. \label{eq:iota2}
\end{align}
\end{lemma}
(\ref{eq:iota1}) is also written as 
$(\iota\ot\iota)R(\iota^{-1}\ot\iota^{-1}) = PR^{\vee\vee}P$.
Thus one has $(\iota\ot\iota)(\ref{eq:rehidari})(\iota\ot\iota) 
= P(\ref{eq:remigi})P$, hence the two relations are equivalent.
\begin{proof}
We apply the formulas (\ref{eq:R}), (\ref{eq:Rv}) and (\ref{eq:Rvv}).
Pick any $z^dx \ot z^ey \in \aff{B_l}\ot \aff{B_m}$ and 
set $R(z^ey \ot z^dx) 
= z^{d-Q_0(y,x)}{\tilde x} \ot z^{e+Q_0(y,x)}{\tilde y}$.
Then the both sides of 
(\ref{eq:iota1}) applied to $z^dx \ot z^ey$ become 
$z^{e+Q_0(y,x)}{\tilde y} \ot z^{d-Q_0(y,x)}{\tilde x}$.
Similarly, the both sides of (\ref{eq:iota2}) send $z^dx \ot z^ey$ to 
$z^{-e-p+I({\tilde y})}\kappa({\tilde y}) \ot 
z^{-d+I(x)-p}{\tilde w}$, 
where $w= \kappa(x)$, 
${\tilde y} \ot {\tilde w} = \ol{R}^\vee(w \ot y)$ 
and $p = P_0(w,y) = P_0(y,w)$.
\end{proof}

\subsection{Classical and affine parts of the reflection equation}
\label{appsub:ca}
\begin{figure}[h]
\caption{Diagram for (\ref{eq:remigi})}\label{fig:remigi}
\vspace{1.2cm}

\unitlength 0.1in
\begin{picture}( 59.5100, 40.6500)(  17.00,-46.1000)
%
\special{pn 8}%
\special{pa 3800 830}%
\special{pa 3800 4610}%
\special{fp}%
%
\special{pn 8}%
\special{pa 3800 1910}%
\special{pa 2100 1470}%
\special{fp}%
%
\special{pn 8}%
\special{pa 3800 1910}%
\special{pa 2270 2306}%
\special{dt 0.045}%
\special{sh 1}%
\special{pa 2270 2306}%
\special{pa 2340 2310}%
\special{pa 2322 2294}%
\special{pa 2330 2270}%
\special{pa 2270 2306}%
\special{fp}%
%
\special{pn 8}%
\special{pa 2180 1100}%
\special{pa 3800 3800}%
\special{fp}%
%
\special{pn 8}%
\special{pa 3800 3810}%
\special{pa 3314 4430}%
\special{dt 0.045}%
\special{sh 1}%
\special{pa 3314 4430}%
\special{pa 3372 4390}%
\special{pa 3348 4388}%
\special{pa 3340 4366}%
\special{pa 3314 4430}%
\special{fp}%
\put(30.1900,-21.6000){\makebox(0,0)[lt]{$z^{I_1+h_1-d}x''$}}%
\put(19.9900,-9.2000){\makebox(0,0)[lt]{$z^ey$}}%
\put(18.1900,-13.7000){\makebox(0,0)[lt]{$z^dx$}}%
\put(25.4900,-18.1000){\makebox(0,0)[rt]{$z^{e+h_1}y'$}}%
\put(27.2900,-23.9000){\makebox(0,0)[rt]{$z^{I_1+h_1+h_2-d}x'''$}}%
\put(33.0900,-30.7000){\makebox(0,0)[rt]{$z^{e+h_1-h_2}y''$}}%
\put(35.0900,-45.0000){\makebox(0,0)[rt]{$z^{I_2-h_1+h_2-e}y'''$}}%
\put(29.3900,-16.1000){\makebox(0,0)[lb]{$z^{d-h_1}x'$}}%
\put(42.7000,-26.6000){\makebox(0,0){$=$}}%
\put(29.9000,-6.3000){\makebox(0,0){$K_2R^\vee K_2R$}}%
\put(57.0000,-38.3000){\makebox(0,0)[lt]{$z^{I_4+h_3-d}x^{\ast\ast}$}}%
\put(53.7000,-43.8000){\makebox(0,0)[rt]{$z^{I_3+h_3-h_4-e}y^{\ast\ast\ast}$}}%
\put(51.6000,-40.4000){\makebox(0,0)[rt]{$z^{I_4+h_3+h_4-d}x^{\ast\ast\ast}$}}%
\put(55.5000,-36.4000){\makebox(0,0)[rb]{$z^{I_3+h_3-e}y^{\ast\ast}$}}%
\put(59.7000,-33.3000){\makebox(0,0)[lb]{$z^{d-h_3}x^\ast$}}%
\put(50.2000,-31.7000){\makebox(0,0)[rb]{$z^dx$}}%
\put(61.7000,-25.8000){\makebox(0,0)[rb]{$z^{I_3-e}y^\ast$}}%
\put(62.0000,-9.1000){\makebox(0,0)[rt]{$z^ey$}}%
%
\special{pn 8}%
\special{pa 6610 3530}%
\special{pa 4874 3980}%
\special{dt 0.045}%
\special{sh 1}%
\special{pa 4874 3980}%
\special{pa 4944 3984}%
\special{pa 4926 3968}%
\special{pa 4934 3944}%
\special{pa 4874 3980}%
\special{fp}%
%
\special{pn 8}%
\special{pa 6610 3530}%
\special{pa 5090 3134}%
\special{fp}%
%
\special{pn 8}%
\special{pa 6610 1910}%
\special{pa 5134 4340}%
\special{dt 0.045}%
\special{sh 1}%
\special{pa 5134 4340}%
\special{pa 5186 4294}%
\special{pa 5162 4294}%
\special{pa 5152 4274}%
\special{pa 5134 4340}%
\special{fp}%
%
\special{pn 8}%
\special{pa 6610 1910}%
\special{pa 6134 1100}%
\special{fp}%
%
\special{pn 8}%
\special{pa 6610 830}%
\special{pa 6610 4610}%
\special{fp}%
\put(58.1000,-6.3000){\makebox(0,0){$R^{\vee\vee}K_2R^\vee K_2$}}%
\end{picture}%

\vspace{0.2cm}
\end{figure}
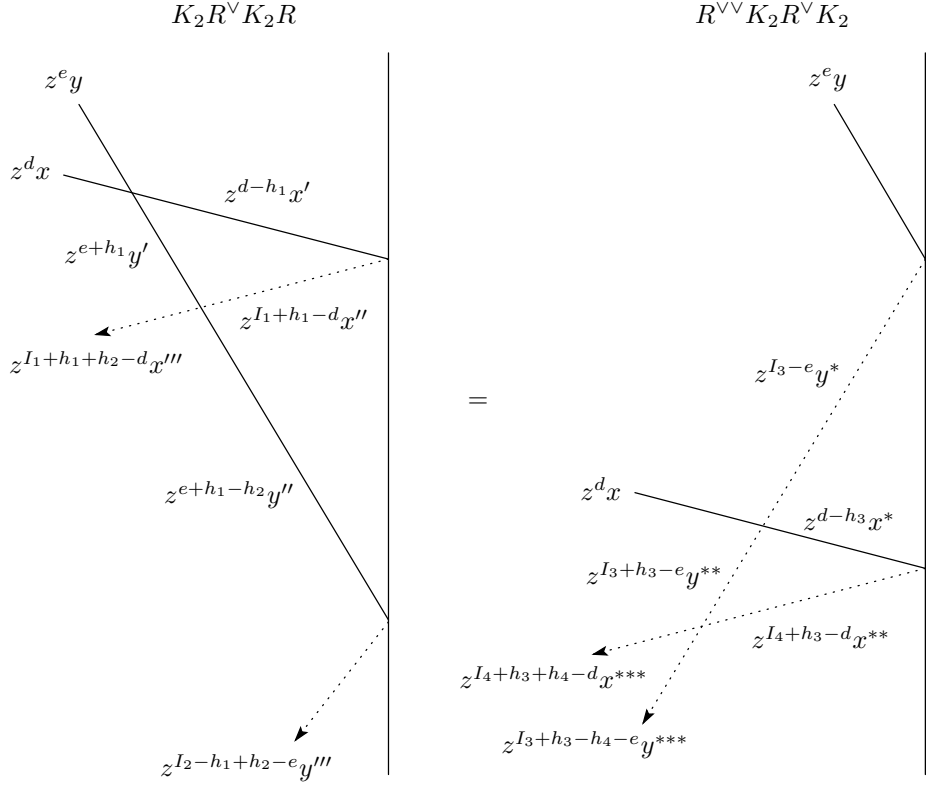

{}In the rest of the appendix, we concentrate on (\ref{eq:remigi}).
We let its two sides act on $z^dx \ot z^ey \in \aff{B_l}\ot \aff{B_m}$
and name the generated elements as
in Figure \ref{fig:remigi}.

\noindent
Here the energy $h_1, \ldots, h_4$ and $I_1, \ldots, I_4$ are given by
\begin{alignat*}{2}
&h_1 = -Q_0(x,y), &\quad &I_1 = I(x'),\\
&h_2 = -P_0(x'',y'), &\quad &I_2= I(y''),\\
&h_3 = -P_0(x, y^\ast), &\quad &I_3 = I(y),\\
&h_4 = -Q_0(x^{\ast\ast},y^{\ast\ast}), &\quad &I_4 = I(x^\ast).
\end{alignat*}
Comparing the final outputs of the two diagrams in 
Figure \ref{fig:remigi}, we find that the reflection equation 
(\ref{eq:remigi}) consists of
\begin{alignat}{2}
&\hbox{classical part:} & \nonumber \\
&\quad  x''' = x^{\ast\ast\ast},&\label{eq:x}\\
&\quad  y''' = y^{\ast\ast\ast},&\label{eq:y}\\
&\hbox{affine part:} & \nonumber \\
&\quad Q_0(x,y)+P_0(x'',y')-I(x')
=Q_0(x^{\ast\ast},y^{\ast\ast})+P_0(x,y^\ast)-I(x^\ast),&\label{eq:e1}\\
&\quad Q_0(x,y)-P_0(x'',y')+I(y'')
=Q_0(x^{\ast\ast},y^{\ast\ast})-P_0(x,y^\ast)+I(y).&\label{eq:e2}
\end{alignat}

\subsection{Switch to tropical version}\label{appsub:tropical}

There are three cases (\ref{eq:rl}), (\ref{eq:1n}) and 
(\ref{eq:12}) to treat, to which not only $I$ in the affine part but 
also the classical part is dependent.
In any case, (\ref{eq:x}) and (\ref{eq:e1}) are 
piecewise linear relations among the $2n$ coordinates 
$x=(x_1, \ldots, x_n)$ and $y = (y_1, \ldots, y_n)$ involving 
$\min$ through $Q_i$ and $P_i$ functions.
This allows us to employ the tropical analysis \cite{Ki,NoY}.
Namely, we are going to show the totally positive rational relations 
obtained by replacing all the $+, \, -$ and $\min$ 
in the original piecewise linear ones
by $\times,\, /$ and $+$, respectively.
This is justified, at a calculative level, 
by the simple identities 
$\lim_{\epsilon \rightarrow -0}\epsilon\log(X_1 \times X_2) = x_1+x_2$,
$\lim_{\epsilon \rightarrow -0}\epsilon\log(X_1/X_2) = x_1-x_2$ and 
$\lim_{\epsilon \rightarrow -0}\epsilon\log(X_1+X_2) = \min(x_1, x_2)$
under the correspondence $X_i = \exp(x_i/\epsilon)$ 
with $x_i \in \mathbb{R}$ \cite{TTMS}.
To save the notation, we let the same letter
$x_i$ to denote the tropical coordinates $X_i$.
Now the piecewise linear formulas 
(\ref{eq:R}), (\ref{eq:Rv}) and (\ref{eq:Rvv}) 
are replaced by 
rational maps $R(x,y)$ called the tropical $R$ \cite{Y,KOTY2,KOTY3}.
\begin{align}
&R(x,y) = \left(
\Bigl(y_i\frac{Q_{i-1}(x,y)}{Q_i(x,y)}\Bigr)_{i=1}^n, 
\Bigl(x_i\frac{Q_i(x,y)}{Q_{i-1}(x,y)}\Bigr)_{i=1}^n
\right),\label{eq:tR}\\
&R^\vee(x,y) = \left(
\Bigl(y_i\frac{P_i(x,y)}{P_{i-1}(x,y)}\Bigr)_{i=1}^n, 
\Bigl(x_i\frac{P_i(x,y)}{P_{i-1}(x,y)}\Bigr)_{i=1}^n
\right),\label{eq:tRv}\\
&R^{\vee\vee}(x,y) = \left(
\Bigl(y_i\frac{Q_i(y,x)}{Q_{i-1}(y,x)}\Bigr)_{i=1}^n, 
\Bigl(x_i\frac{Q_{i-1}(y,x)}{Q_i(y,x)}\Bigr)_{i=1}^n
\right),\label{eq:tRvv}\\
&Q_i(x,y) = \sum_{1 \le k \le n}
\prod_{j=1}^{k-1}x_{i+j}\prod_{j=k+1}^n y_{i+j},\quad
P_i(x,y) = x_{i+1}+y_{i+1}.\label{eq:tPQ}
\end{align}

We write $P_i(x,y)$ and $Q_i(x,y)$ without any change 
of the entry as $P_i(x,y^\ast), Q_i(y,x)$, etc. simply as 
$P_i$ and $Q_i$.
Obviously, $Q_i$ changes into $Q_{i+1}$ under the transformation 
$(x,y) \rightarrow (\hbox{Rotateleft}(x), \hbox{Rotateleft}(y))$.
One has \cite{Y}
\begin{align}
&x_{i+1}Q_{i+1} + y_iQ_{i-1} = (x_i+y_{i+1})Q_i,
\label{eq:Y1}\\
&x_i+y_{i+1} = x'_{i+1}+y'_i. \label{eq:Y2}
\end{align}
In fact, $x_iQ_i - y_iQ_{i-1} = x_1 \cdots x_n - y_1\cdots y_n$
is independent of $i$, showing (\ref{eq:Y1}).
In view of $(y',x') = R(x,y)$ (see Figure \ref{fig:remigi}),
(\ref{eq:Y2}) is equivalent to (\ref{eq:Y1}) 
upon substitution of (\ref{eq:tR}).

\subsection{$\kappa =$ Rotateleft case (\ref{eq:rl})}
\label{appsub:rl}

First we show the classical part 
$(x''',y''') = (x^{\ast\ast\ast}, y^{\ast\ast\ast})$.
{}From Figure \ref{fig:remigi}, we know 
$(x''',y'') = R^\vee(y',x'')$, $x''_i = x'_{i+1}$ and 
$y'''_i = y''_{i+1}$.
Thus $x'''$ and $y'''$ are calculated as
\begin{align*}
&x'''_i \stackrel{\rm (\ref{eq:tRv})}{=}
x''_i\frac{P_i(y',x'')}{P_{i-1}(y',x'')}
\stackrel{\rm (\ref{eq:tPQ})}{=}
x'_{i+1}\frac{x'_{i+2}+y'_{i+1}}{x'_{i+1}+y'_{i}}
\stackrel{\rm (\ref{eq:Y2}),(\ref{eq:tR})}{=}
\frac{x_{i+1}(x_{i+1}+y_{i+2})Q_{i+1}}{(x_{i}+y_{i+1})Q_i},\\
&y'''_i \stackrel{\rm (\ref{eq:tRv})}{=}
y'_{i+1}\frac{P_{i+1}(y',x'')}{P_{i}(y',x'')}
\stackrel{\rm (\ref{eq:tPQ})}{=}
y'_{i+1}\frac{x'_{i+3}+y'_{i+2}}{x'_{i+2}+y'_{i+1}}
\stackrel{\rm (\ref{eq:Y2}),(\ref{eq:tR})}{=}
\frac{y_{i+1}(x_{i+2}+y_{i+3})Q_{i}}{(x_{i+1}+y_{i+2})Q_{i+1}}.
\end{align*}
Similarly, Figure \ref{fig:remigi} tells 
$(x^{\ast\ast\ast},y^{\ast\ast\ast}) 
= R^{\vee\vee}(y^{\ast\ast},x^{\ast\ast})$, hence
\begin{equation}\label{eq:x***}
x^{\ast\ast\ast}_i = 
x^{\ast\ast}_i\frac{Q_i(x^{\ast\ast},y^{\ast\ast})}
{Q_{i-1}(x^{\ast\ast},y^{\ast\ast})},\quad 
y^{\ast\ast\ast}_i = 
y^{\ast\ast}_i\frac{Q_{i-1}(x^{\ast\ast},y^{\ast\ast})}
{Q_{i}(x^{\ast\ast},y^{\ast\ast})}.
\end{equation}
By using (\ref{eq:rl}) and (\ref{eq:tRv}) we find
\begin{equation}\label{eq:xy**}
x^{\ast\ast}_i = x_{i+1}\frac{x_{i+2}+y_{i+3}}{x_{i+1}+y_{i+2}},
\quad
y^{\ast\ast}_i = y_{i+1}\frac{x_{i+1}+y_{i+2}}{x_{i}+y_{i+1}}.
\end{equation}
{}From (\ref{eq:xy**}) it is easy to show
\begin{equation}\label{eq:Q**}
Q_i(x^{\ast\ast},y^{\ast\ast}) = 
\frac{x_{i+1}+y_{i+2}}{x_{i+2}+y_{i+3}}Q_{i+1}.
\end{equation}
Upon substitution of (\ref{eq:xy**}) and (\ref{eq:Q**}), 
$x^{\ast\ast\ast}_i$ and $y^{\ast\ast\ast}_i$ in 
(\ref{eq:x***}) coincide with $x'''_i$ and $y'''_i$ 
obtained in the above.

Next we proceed to the affine part (\ref{eq:e1}) and 
(\ref{eq:e2}), which read, in the tropical setting as
\begin{align*}
&(x''_1+y'_1)x'_1Q_0 = 
(x_1+y^\ast_1)x^\ast_1Q_0(x^{\ast\ast},y^{\ast\ast}),\\
&\frac{Q_0}{(x''_1+y'_1)y''_1} = 
\frac{Q_0(x^{\ast\ast},y^{\ast\ast})}{(x_1+y^\ast_1)y_1}.
\end{align*}
By applying $x''_1=x'_2$, $x'_1=x_1Q_1/Q_0$, 
$x^\ast_1=x_1(x_2+y_3)/(x_1+y_2)$ to the former, and 
$y''_1 = y'_1(x''_2+y'_2)/(x''_1+y'_1) = 
y_1Q_0(x'_3+y'_2)/((x''_1+y'_1)Q_1)$ to the latter along with 
$y^\ast_1 = y_2$, they are simplified into
\begin{align*}
&(x'_2+y'_1)x_1Q_1 = x_1(x_2+y_3)Q_0(x^{\ast\ast},y^{\ast\ast}),\\
&\frac{Q_1}{x'_3+y'_2} = 
\frac{Q_0(x^{\ast\ast},y^{\ast\ast})}{x_1+y_2},
\end{align*}
which are obvious from (\ref{eq:Y2}) and (\ref{eq:Q**}).

\subsection{$\kappa =$ Switch$_{1n}$, 
Switch$_{12}$ cases (\ref{eq:1n}), (\ref{eq:12})}
\label{appsub:1n2}

Here we assume $n$ is even.
Set Switch$_{12}(x) = 
(x_{\underline{1}}, x_{\underline{2}}, \ldots, x_{\underline{n}})$, 
namely, $\underline{i} = i-1\, (i+1)$ for $i$ even (odd).
A direct calculation leads to 
\begin{equation}\label{eq:alpha}
x^{\ast\ast}_i 
= x_{\underline{i}}
\frac{\alpha_{\underline{i}+1}}{\alpha_{\underline{i}}},\quad 
y^{\ast\ast}_i = y_{\underline{i}}\frac{\alpha_{i+1}}{\alpha_i},
\quad \alpha_i = x_i + y_{\underline{i}}.
\end{equation} 
\begin{lemma}\label{lem:nl3}
For $\kappa =$ Switch$_{12}$, 
\begin{equation*}
Q_i(x^{\ast\ast},y^{\ast\ast}) = 
\begin{cases}
Q_i 
& i \hbox{ even},\\
\frac{(x_{i+1}+y_i)(x_iQ_{i+1}+y_{i+1}Q_{i-1})}
{(x_i+y_{i+1})(x_{i+2}+y_{i+3})} 
& i \hbox{ odd}.
\end{cases}
\end{equation*}
For $\kappa =$ Switch$_{1n}$, the same relation 
with the opposite alternative with respect to the 
parity of $i$ holds.
\end{lemma}
\begin{proof}
It suffices to show Switch$_{12}$ case only.
Suppose $Q_i(x^{\ast\ast},y^{\ast\ast}) = Q_i$ for any even $i$.
Then in (\ref{eq:Y1}) with odd $i$, 
replacement of $(x,y)$ by $(x^{\ast\ast},y^{\ast\ast})$
and application of (\ref{eq:alpha}) lead to the sought formula
for $Q_i(x^{\ast\ast},y^{\ast\ast})$ with odd $i$.
Thus we are left with $i$ even case, which is done by checking
$Q_0(x^{\ast\ast},y^{\ast\ast}) = Q_0$ only.
We use the expression 
$Q_0 = \sum_{k=1}^{n/2}x_1x_2\cdots x_{2k-2}\alpha_{2k-1}
y_{2k+1}y_{2k+2}\cdots y_n$.
Under the replacement $(x,y)$ by $(x^{\ast\ast},y^{\ast\ast})$,
$\alpha_{2k-1}$ changes into $\alpha_{2k+1}$ while 
$x_1 \cdots x_{2k-2}$ and $y_{2k+1}y_{2k+2}\cdots y_n$ acquire the 
extra factors $\alpha_{2k-1}/\alpha_1$ and $\alpha_1/\alpha_{2k+1}$,
respectively. Hence the summand for each $k$ remains invariant.
\end{proof}
First we prove the classical part 
$x''' = x^{\ast\ast\ast}$ and 
$y''' = y^{\ast\ast\ast}$ in Switch$_{12}$ case.
By using (\ref{eq:Y2}), we have
\begin{align*}
&x'''_i = 
\frac{x_{i+1}(x_iQ_{i+1}+y_{i+1}Q_{i-1})}
{Q_{i-1}(x_i+y_{i+1})} \;\;(i \hbox{ odd}),\quad
\frac{Q_ix_{i-1}(x_{i+1}+y_{i+2})}{x_{i-1}Q_i+y_iQ_{i-2}}
\;\;(i \hbox{ even}), \\
&y'''_i = 
\frac{Q_{i-1}y_{i+1}(x_{i+2}+y_{i+3})}{x_{i}Q_{i+1}+y_{i+1}Q_{i-1}} 
\;\;(i \hbox{ odd}),\quad
\frac{y_{i-1}(x_{i-1}Q_{i}+y_{i}Q_{i-2})}
{Q_{i}(x_{i-1}+y_{i})}
\;\;(i \hbox{ even}).
\end{align*}
On the other hand,
\begin{equation*}
x^{\ast\ast\ast}_i = x^{\ast\ast}_i
\frac{Q_i(x^{\ast\ast},y^{\ast\ast})}
{Q_{i-1}(x^{\ast\ast},y^{\ast\ast})}, \quad
y^{\ast\ast\ast}_i = y^{\ast\ast}_i
\frac{Q_{i-1}(x^{\ast\ast},y^{\ast\ast})}
{Q_{i}(x^{\ast\ast},y^{\ast\ast})}.
\end{equation*}
Applying Lemma \ref{lem:nl3} and (\ref{eq:alpha}), 
one finds  $x'''_i=x^{\ast\ast\ast}_i $ and 
$y'''_i=y^{\ast\ast\ast}_i $.

Second we show $x''' = x^{\ast\ast\ast}$ and 
$y''' = y^{\ast\ast\ast}$ for Switch$_{1n}$. 
This is a corollary of Switch$_{12}$ case.
To see this, note that 
$x'''_i$ in the two cases have the same expressions
except the opposite alternative concerning the parity of $i$, 
and the same holds for 
$x^{\ast\ast\ast}_i, y'''_i$ and $y^{\ast\ast\ast}_i$ as well.

Third we prove the affine part (\ref{eq:e1})
and (\ref{eq:e2}) in Switch$_{12}$ case.
Since $I$ (\ref{eq:12}) is trivial, we are to show 
\begin{equation*}
Q_0= Q_0(x^{\ast\ast},y^{\ast\ast}),\quad 
P_0(x'',y') = P_0(x,y^\ast).
\end{equation*}
The former is due to Lemma \ref{lem:nl3}, 
and it is easy to check the both sides of the latter agree with
$x_1+y_2$ by using (\ref{eq:Y2}).

Finally we prove (\ref{eq:e1}) and (\ref{eq:e2}) in Switch$_{1n}$ case.
Their tropical version read
\begin{align*}
&Q_0P_0(x'',y')\frac{x'_1}{x'_n} = 
Q_0(x^{\ast\ast},y^{\ast\ast})P_0(x,y^\ast)
\frac{x^\ast_1}{x^\ast_n}, \\
&\frac{Q_0y''_n}{P_0(x'',y')y''_1} = 
\frac{Q_0(x^{\ast\ast},y^{\ast\ast})y_n}{P_0(x,y^\ast)y_1}.
\end{align*}
With the help of Lemma \ref{lem:nl3} and 
(\ref{eq:Y2}), these relations can be 
directly checked.
This completes the proof of Proposition \ref{pr:re}.


\section*{Acknowledgements}
A.K. and M.O. are supported by Grants-in-Aid for Scientific 
Research JSPS No.15540363 and No.14540026, respectively.

\end{document}